\def\h{${{h}_{75}}^{-1}$}

\def\subsun{\mbox{$_{\normalsize\odot}$}}
\def\deg{\hbox{$^\circ$}}
\def\sun{\hbox{$\odot$}}

\def\lesssim{\mathrel{\hbox{\rlap{\hbox{\lower4pt\hbox{$\sim$}}}\hbox{$<$}}}}
\def\gtrsim{\mathrel{\hbox{\rlap{\hbox{\lower4pt\hbox{$\sim$}}}\hbox{$>$}}}}
\def\la{\mathrel{\hbox{\rlap{\hbox{\lower4pt\hbox{$\sim$}}}\hbox{$<$}}}}
\def\ga{\mathrel{\hbox{\rlap{\hbox{\lower4pt\hbox{$\sim$}}}\hbox{$>$}}}}

\documentclass{aa}
\usepackage{graphicx}

\def\etal{{\it et al.\/}\ }
                       
\begin{document}

\title {Dynamical state and star formation properties of the merging galaxy 
cluster Abell 3921
\thanks{Based on observations collected at the
  European Southern Observatory, Chile. Proposal numbers:
67.A-0494(A), 67.A-0495(A) and 70.A-0710}}
\author{Ferrari, C.\inst{1,2}
\and Benoist, C.\inst{1}
\and Maurogordato, S.\inst{1}
\and Cappi, A.\inst{3}
\and Slezak, E.\inst{1}}
\offprints{Chiara Ferrari}  
\institute{    
  Laboratoire Cassiop\'ee, CNRS, Observatoire de la C\^ote d'Azur, BP4229, 
06304 Nice Cedex 4, France \and Institut f\"ur Astrophysik,
Technikerstra{\ss}e 25, 6020 Innsbruck, Austria \and INAF,
Osservatorio Astronomico di Bologna, via Ranzani 1, 40127 Bologna,
Italy }
\date{Received~9 August 2004; accepted~31 August 2004}  

\abstract{
  We present the analysis and results of a new VRI photometric and
spectroscopic survey of the central $\sim~1.8{\times}1.2~{\rm Mpc}^2$
region of the galaxy cluster A3921 ($z=0.094$). We detect the presence
of two dominant clumps of galaxies with a mass ratio of $\sim$5: a
main cluster centred on the BCG (A3921-A), and a NW sub-cluster
(A3921-B) hosting the second brightest cluster galaxy. The distorted
morphology of the two sub-clusters suggests that they are interacting,
while the velocity distribution of 104 confirmed cluster members does
not reveal strong signatures of merging. By applying a two-body
dynamical formalism to the two sub-clusters of A3921, and by comparing
our optical results to the X-ray analysis of A3921 based on XMM
observations (Belsole \etal2004), we conclude that A3921-B is probably
tangentially traversing the main cluster along a SW/NE direction. The
two sub-clusters are observed in the central phase of their merging
process ($\pm$ 0.3 Gyr), with a collision axis nearly perpendicular to
the line of sight. Based on the spectral features of the galaxies
belonging to A3921, we estimate the star formation properties of the
confirmed cluster members. Substantial fractions of both emission-line
($\sim$13\%) and post-star-forming objects (k+a's, $\sim$16\%) are
detected, comparable to those measured at intermediate redshifts. Our
analysis reveals a lack of bright post-star forming objects in A3921
with respect to higher redshift clusters, while the fraction of k+a's
increases towards fainter magnitudes ($M_{R_{\rm AB}}>-20$). Similar
results were obtained in Coma cluster by Poggianti \etal2004, but at
still fainter magnitudes, suggesting that the maximum mass of actively
star-forming galaxies increases with redshift (``downsizing
effect''). The spatial and velocity distributions of k+a's galaxies do
not show significant differences to that of the passive population,
and to the whole cluster. Most of these objects show relatively red
colours and moderate Balmer absorption lines, which suggest star
formation has ceased $\sim 1 - 1.5$ Gyr ago. Their presence is
therefore difficult to relate to the on-going merging event. We find
that star forming galaxies share neither the same kinematics, nor the
same projected distribution of the passive cluster members. Moreover,
most of emission-line galaxies are concentrated in A3921-B and in the
region between the two sub-clusters. We therefore suggest that the
ongoing merger may have triggered the star-formation episode in at
least a fraction of the observed emission-line galaxies.
 }
\maketitle   

\section{Introduction}

In the standard cosmological scenario of hierarchical structure formation,
bound objects form from the collapse of initial density fluctuations that grow
under the influence of gravity through merging of smaller structures that have
formed before. As optical and X-ray studies reveal that clusters of galaxies
are still forming at the present epoch (e.g. Jones \& Forman 1992, West
\etal1995, Donnelly \etal2001), merging clusters provides a unique tool to
test or analyze the physics of structure formation and evolution.

Major cluster--cluster collisions are the most energetic events that have
occurred in the Universe since the Big Bang, as they release total energies up
to $3~{\times}~10^{64}$~erg (Sarazin 2003), and their effects on all cluster
components are far from being fully understood. The intra-cluster gas
experiences compression, rarefaction and shock waves (Schindler \& M\"uller
1993), and shows characteristic features such as cold fronts (Markevitch
\etal2000). Additionally, on the side of the galaxy distribution, the velocity
dispersion of the cluster members can increase up to a factor of two during
the merger event (Schindler and B\"ohringer 1993). While the effects on the
intra-cluster medium and on cluster internal dynamics have been analyzed in
some details, the effects on the galaxies are still debated. Caldwell et al.
(1993) showed that strong Balmer-line absorption galaxies are distributed
between the main Coma cluster and the merging SW sub-cluster, suggesting star
formation is triggered by the merging event. However, various physical
mechanisms have been shown to affect the process of star formation within
clusters.  Changes of the tidal gravitational field during cluster merging
could affect star formation in galaxies (e.g. Bekki 1999). The increasing
external pressure following the infall of galaxies into the dense
intra-cluster gas may trigger star formation (Dressler \& Gunn 1983, Evrard
1991), while gas stripping in galaxies due to ram-pressure exerted by the ICM
could weaken the starburst phenomenon during cluster--cluster collisions
(Fujita \etal1999). So far it is not clear which one of these competing
effects is the dominant one, since results in favour of the former (e.g.
Abraham \etal1996, Wang, Ulmer \& Lavery 1997, Moss \& Whittle 2000, Flores
\etal2000, Gavazzi \etal2003, Poggianti \etal2004) and the latter (e.g.
Tomita \etal1996, Balogh \etal1997,~1998, Baldi, Bardelli \& Zucca
2001) were obtained.

Combined optical and X-ray studies have been particularly successful
in revealing the complex dynamics of merging clusters (e.g. Davis
\etal1995, Lemonon \etal1997, Roettiger \etal1998, Durret \etal1998,
Arnaud \etal2000, Donnelly \etal2001, Ferrari \etal2003). The X-ray
properties of merging clusters are currently being investigated
through a set of XMM observations (XMM guaranteed time, Sauvageot
\etal2001), together with the properties of their galaxy distribution
through optical observations by our group, hence providing a unique
combined analysis. In this paper we will concentrate on the optical
analysis of the merging cluster Abell 3921, and our results will be
compared to conclusions obtained from the X-ray analysis of XMM data
by Belsole \etal2004.

Abell 3921 is a R=2, BM II Abell cluster at $z=0.094$ (Katgert
\etal1998). Previous ROSAT and Ginga observations revealed the
presence of a main cluster and a substructure with a very perturbed
morphology, interpreted as falling onto the main component (Arnaud
\etal1996). Following XMM-Newton/EPIC observations (Sauvageot
\etal2001) confirmed the presence of a two component structure.  This
motivated new optical observations in order to better characterize the
merger scenario of this complex cluster. In this paper, we analyze the
spectroscopic and photometric galaxy catalogues of the central
$\sim~1.8{\times}1.2~{\rm Mpc}^2$ region of the cluster, with new data
of multi-object spectroscopy (239 new spectra) and VRI-bands imaging,
obtained with EFOSC2 at the 3.6m ESO telescope and with WFI at the
2.2m ESO telescope respectively. Sect. 2 briefly describes the
observations, the data processing technique and the completeness level
achieved by the new spectroscopic sample. In Sect. 3 cluster members
are identified and in Sect. 4 the optical morphology of the cluster is
studied. In Sect. 5 we perform a kinematical and dynamical analysis of
the cluster. The optical masses of the two main subclusters are
estimated in Sect. 6, and we solve the two-body problem for these two
systems. The photometric and spectral properties of the cluster
members are investigated in Sect. 7, allowing to define several
subsamples whose spatial and velocity distributions are analyzed. The
main results and their interpretation are summarized in the final
Sect. 8.  All numbers are expressed as a function of $h_{75}$, the
Hubble constant in units of 75~km/s/Mpc. We have used the $\Lambda$CDM
model with $\Omega_m=0.3$ and $\Omega_{\Lambda}=0.7$, then 1~arcmin
corresponds to $\sim$0.097~\h~Mpc in the following.

\section{The data}

\subsection {Imaging}

The optical observations of Abell 3921 were carried out using the Wide
Field Imager (WFI) mounted at the Cassegrain focus of the MPG/ESO 2.2m
telescope at La Silla observatory. WFI is a mosaic camera with
4$\times$2 CCD chips covering a total area of 34$\times$33 {\rm
arcmin}$^2$. To cover the gaps between the eight individual chips of
the camera we adopted a standard dithering sequence. The field centred
on $\alpha = 22^h49^m44^s$ $\delta = -64\deg22^m15^s$ was observed in
service mode in the V,R and I passbands.  The data, including
photometric calibration images, were obtained between August 20th and
23rd, 2001. In Table~\ref{tab:obs}, we summarize the observations.

\begin{table*}
\begin{center}   
\begin{tabular}{cccccc}  
\hline
\hline
Filter & ESO id  & Nr. of Exp. & Total Exp. & Seeing & Mag.limit (AB)\\
 &  &  & (s) & (arcsec)  & (5$\sigma$,2$\times$FWHM) \\
\hline
V & V/89   & 5  & 750  & 1.35 & 22.5 \\
R & Rc/162 & 5  & 750  & 1.15 & 22.7 \\
I & Ic/Iwp & 10 & 1800 & 1.25 & 21.5 \\
\hline
\end{tabular}
\caption{Imaging: summary of the observations}
\label{tab:obs}
\end{center}
\end{table*}

The data reduction was performed using the package "alambic" developed
by B.Vandame based on tools available from the multi-resolution visual
model package (MVM) by Bijaoui and collaborators (Bijaoui \& Ru\'e
1995, Ru\'e \& Bijaoui 1997). The photometric calibration was obtained
using several standard stars from Landolt (1992) over a large range of
airmasses, leading to an accuracy of the photometry of $\sim 0.05$mag.

The SExtractor software package (Bertin \& Arnouts 1996) was used to identify
all sources in the field as well as to classify them and measure their
magnitudes.  The latter were corrected for galactic absorption using
$E(B-V)=0.027$ as derived from Schlegel \etal(1998), yielding $A_V=0.09$mag,
$A_R=0.07$mag and $A_I=0.05$mag. They were also transformed to the AB system
given by the following relations: $V_{\rm AB}=V-0.01$, $R_{\rm AB}=R+0.19$,
$I_{\rm AB}=I+0.49$.

\subsection {Spectroscopy}

In this paper we present the results of the analysis of new
spectroscopic data obtained through two sessions of observations at
the ESO 3.6 m telescope (2 nights in September 2001, 2 nights in
October 2002). We used the ESO Faint Object and Camera (EFOSC2) with
grism\#03 and a punching head of 1.35'', obtaining a spectral
resolution of FWHM$\sim$7.5~\AA~over the wavelength range
3050-6100~\AA.  For each frame we made at least two science exposures,
in order to eliminate cosmic rays, with an integrated exposure time of
5400~s for brighter objects ($R_{\rm AB}<$18) and of 7200~s for
fainter ones (${\rm R}_{\rm AB}<$19).  We made a standard
spectroscopic reduction using our automated package for multi--object
spectroscopy based on the task ``apall'' in IRAF\footnote{IRAF is
distributed by the National Optical Astronomy Observatories, which are
operated by the Association of Universities for Research in Astronomy,
Inc., under cooperative agreement with the National Science
Foundation}. Spectra were wavelength calibrated using the
Helium--Argon lamp spectra taken after each science exposure.  We
determined redshifts using the cross--correlation technique (Tonry \&
Davis, 1981) implemented in the task ``xcsao'' of the RVSAO
package. Spectra of late--type stars were used as radial velocity
standards.

In Table~\ref{MAINTAB} (available in the electronic version of the paper) we
list the results of our spectroscopic observations in the following way:
Col.~1: identification number of each target galaxy; Cols.~2 and 3: right
ascension and declination (J2000.0) of the target galaxy; Cols.~4, 5 and 6:
the V, R and I band magnitudes in the AB system; Cols.~7 and 8: best estimate
of the heliocentric redshift (expressed as $cz$) and associated error from the
cross-correlation technique (those values have been set to ``-2'' if the
object is a star and to ``-1'' if we have no redshift information); Col.~9:
run of observations; Col.~10: a quality flag for the redshift determination:
1=good determination (R parameter of Tonry \& Davis $\geq$~3), 2=uncertain
determination, 3=no determination; Cols.~11 and 12: the equivalent widths
(negative or positive in the case of emission or absorption features
respectively) for [OII]$\lambda$3727 and ${\rm H}_\delta$ ($\lambda$=4101~\AA)
lines, measured only for the cluster members; Col.~13: the spectral
classification of the galaxy members.

A total of 239 new spectra have been obtained, among which 56 are
stars, 100 are galaxies with a very good redshift determination and 83
are spectra with $R<3$.  In the following analysis, we also consider
the redshifts measured in the region of A3921 by other authors (ENACS
collaboration, Katgert \etal1996, Mazure \etal1996). We compare the
values of the redshifts of 11 galaxies obtained from the common sample
to those published (35) and we obtain a mean difference of
-12.7$\pm$75.8~km/s, which shows a good consistency between the two
datasets.  The final sample includes our 239 redshifts and 24
available redshifts from the ENACS catalogue (Katgert
\etal1996, Mazure \etal1996), among which 22 with quality flag=1 following our
criterion. It covers a field of ${\sim}25.5'{\times}25'$, slightly
larger than the size of our own spectroscopic follow-up
(${\sim}18'{\times}12.5'$).

\begin{figure}
\resizebox{8cm}{!}  
{\includegraphics{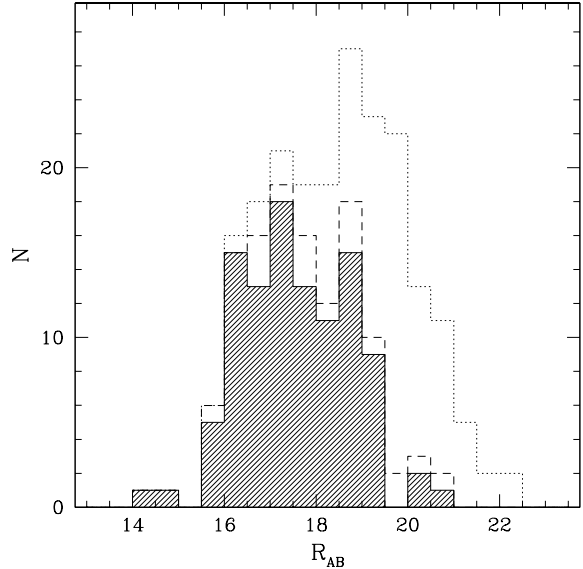}}  
\hfill  
\parbox[b]{8cm}{
 \caption{R-band magnitude distributions of the galaxies of our 
   spectroscopic sample (207 objects - doted line), of all the galaxies with
   good velocity determination (122 - dashed line), and, among them, of those
   belonging to A3921 (104 - shaded area).}
\label{histoI}}  
\end{figure}

The $R$--band magnitude distribution of the whole spectroscopic sample
of galaxies is presented in Fig.~\ref{histoI} (207 galaxies), and
compared to the distribution of galaxies with a good redshift
determination, i.e.  quality flag=1 (122 galaxies), and among them,
those belonging to A3921 (104 galaxies, see Sect.~\ref{membership}).

\begin{figure}
\resizebox{8cm}{!}
{\includegraphics{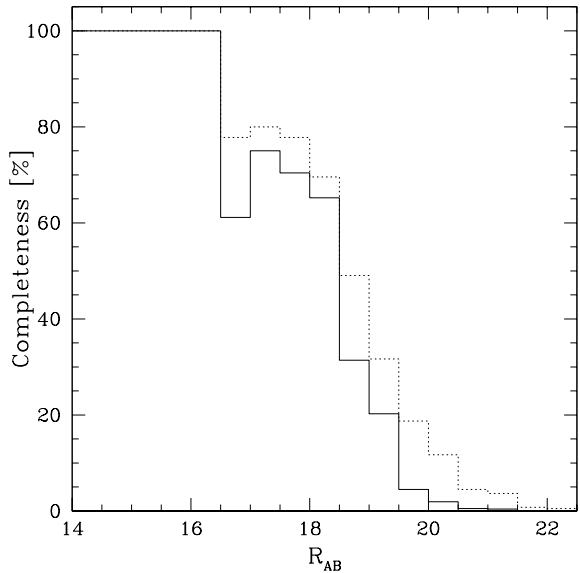}}
\hfill
\parbox[b]{8cm}{
\caption{Completeness of the spectroscopic sample as a function of the
  R-band magnitude in the central $18{\times}12.5~{\rm arcmin}^2$ field for
  all galaxies observed (dashed line) and for those that led to a Quality
  Flag=1 (solid line). }
\label{ComplI}}
\end{figure}

The ratio of the number of galaxies with measured redshift to the
total number of galaxies detected within the central field of
$18{\times}12.5~{\rm arcmin}^2$ covered by our last observations is
plotted in Fig.~\ref{ComplI} as a function of the $R$--band
magnitude. The spectroscopic catalogue is complete at more than 50\%
level up to $R_{\rm AB}=18.5$ ($R_{\rm AB}^*$+2.1).

Among the 263 objects of our spectroscopic sample, we will consider in
the following only the 122 galaxies with redshifts determined with a
quality flag=1 (i.e. our new 100 high quality spectra plus 22
published previously).

\section {Cluster membership}\label{membership}

\begin{figure}  
\resizebox{8cm}{!}{\includegraphics{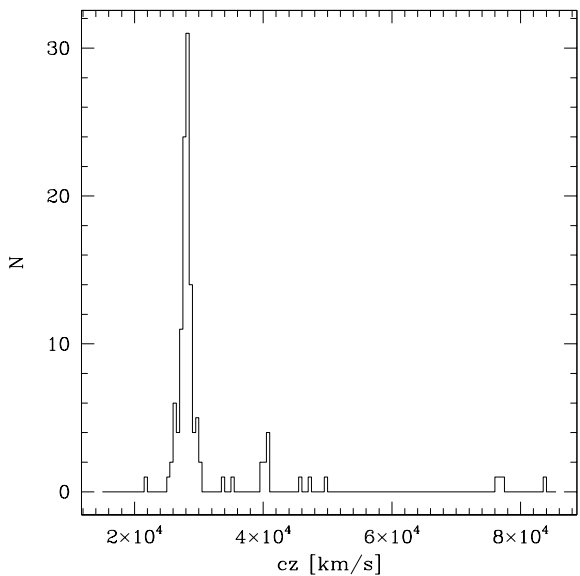}}  
\hfill  
\parbox[b]{8cm}{  
\caption{Apparent radial velocity ($cz$) histogram of the galaxies in the
  central field of the cluster A3921, with a binning of 500~km/s.}
\label{histoTOT}}  
\end{figure}  

Fig.~\ref{histoTOT} shows the radial velocity ($cz$) distribution of
this dataset in bins of 500~km/s.  The bulk of the cluster is
concentrated between 25400 and 30400~km/s. Four objects have $cz$
higher than 75000, one galaxy appears to be in the foreground, while
13 are located between 31000 and 50000~km/s and most of them (12) are
also spatially concentrated in the region around the two brightest
objects of the West side of the cluster, BG2 and BG3. If we consider
only the 8 galaxies around 40000~km/s, they correspond to a peak
located at ${C}_{BI}=40378$~km/s ($\bar{z}=0.135$) with a velocity
dispersion $S_{BI}=427$~km/s, indicating that they could be a
background group.

\begin{figure*}
\resizebox{18cm}{!}{\includegraphics*{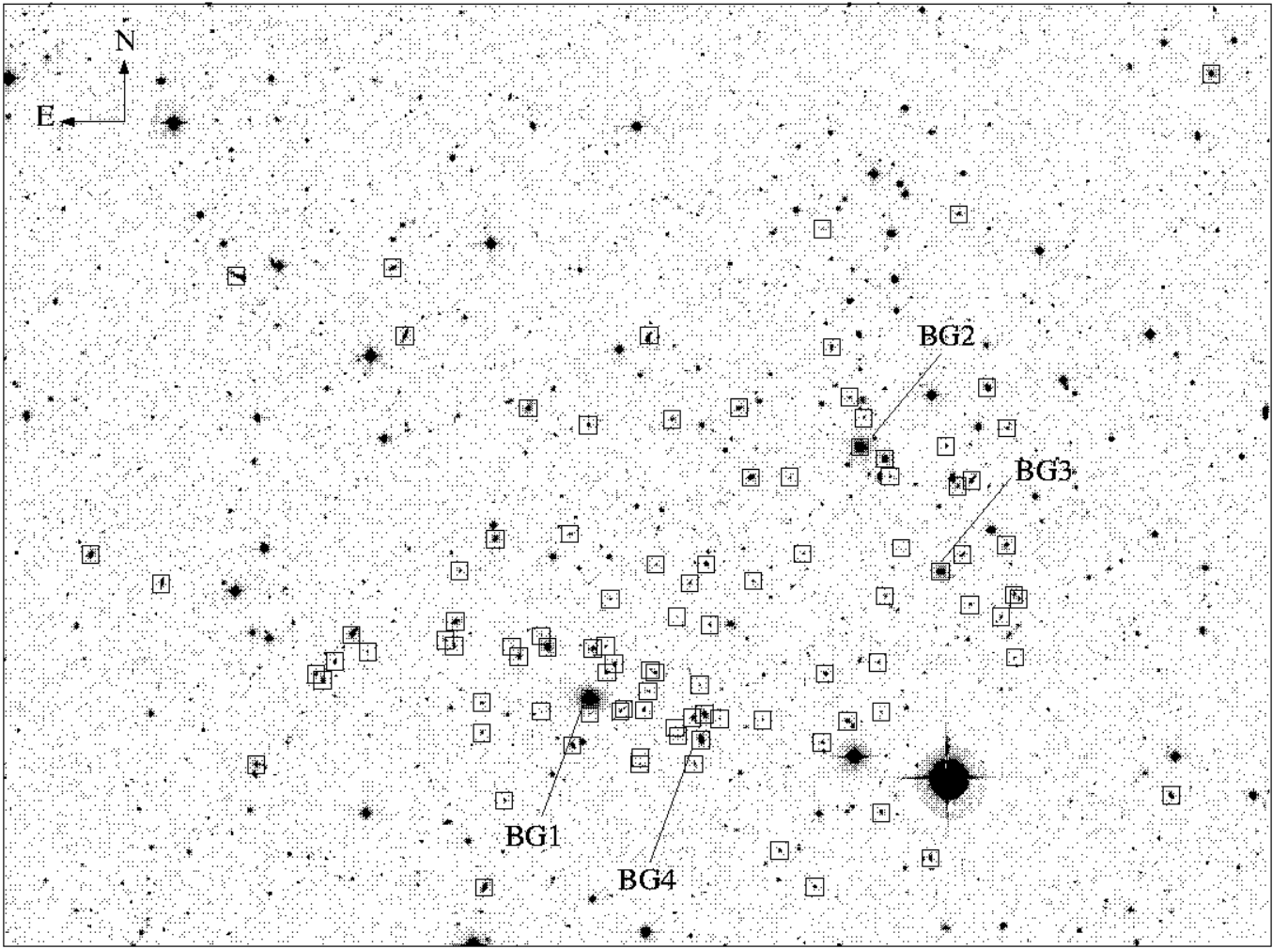}}  
\hfill
\parbox[b]{18cm}{
\caption{
 Galaxies identified as members of A3921 (squares),
  superimposed to a fraction of the R-band image (27'${\times}20'$).}
\label{FC}}
\end{figure*}

In order to eliminate the galaxies not belonging to the cluster, we
apply the standard iterative 3$\sigma$ clipping (Yahil \& Vidal 1977).
104 cluster members with $cz$ between 25400 and 30400~km/s are
selected. Note that the background peak being located $\sim$10000~km/s
behind the mean velocity of the cluster is excluded as a subgroup of
the cluster.  Fig.~\ref{FC} shows the location of the 104 identified
members superimposed to a fraction of the R-band image.

\begin{figure*}  
\resizebox{18cm}{!}{\includegraphics*{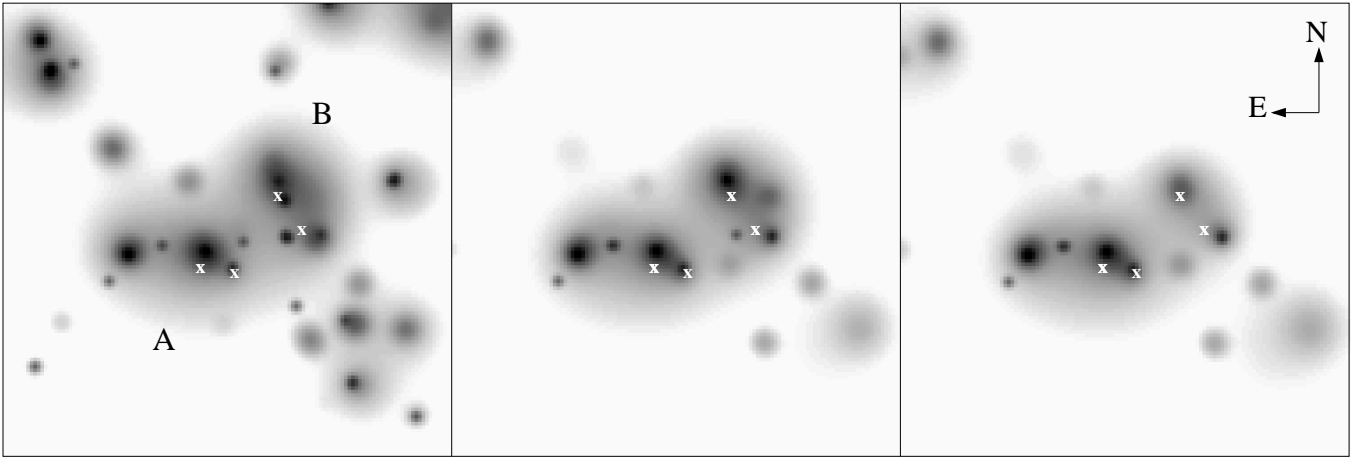}}  
\hfill  
\parbox[b]{18cm}{  
\caption{  
    Projected galaxy density maps (with $R_{\rm AB}\le 19$) on a 34${\times}34
  {\rm arcmin}^2$ field centred on A3921.  {\bf Left panel:} All galaxies;
  {\bf central panel:} the red sequence galaxies, and {\bf right panel:} same
  as central panel but removing galaxies known not to be cluster members from
  spectroscopy. The white crosses show the positions of the four brightest
  galaxies indicated in Fig.~\ref{FC}.}
\label{IsoDMaps}}  
\end{figure*}  

\begin{figure*}  
\resizebox{16.cm}{!}{\includegraphics*{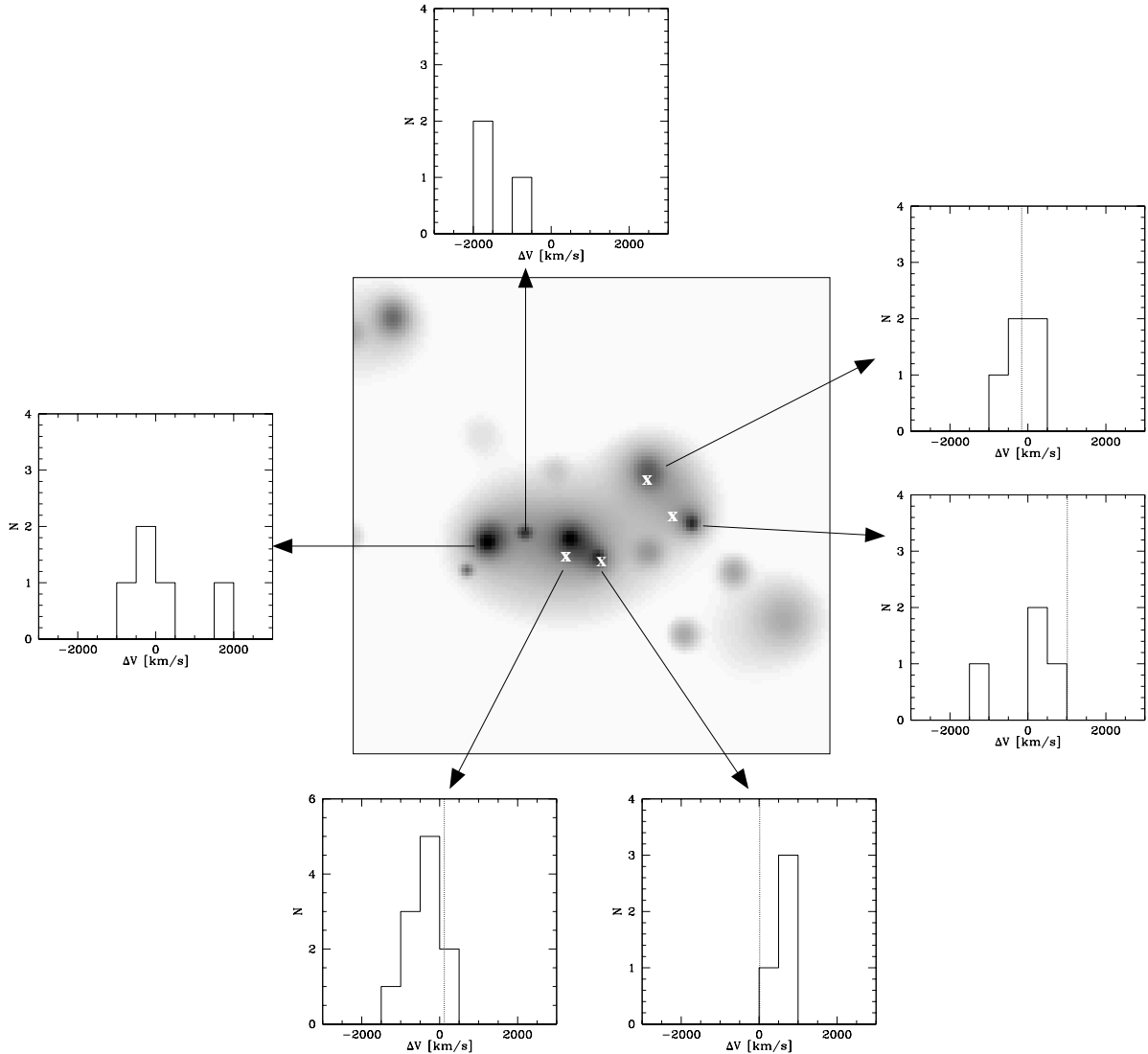}}  
\hfill  
\parbox[b]{18cm}{  
\caption{  
  Velocity distributions of the main structures of the projected
  galaxy density map (same as the right panel of
  Fig.~\ref{IsoDMaps}). The dotted line in the histogram indicate
  the relative velocity of the neighbouring brightest galaxy. At the
  level of $R_{\rm AB}=19$ all the selected clumps have a 100\%
  coverage except the eastern group with 70\% completeness.}
\label{IsoDMapsGR}}  
\end{figure*}

\begin{figure*}  
\centering
\resizebox{16cm}{!}{\includegraphics{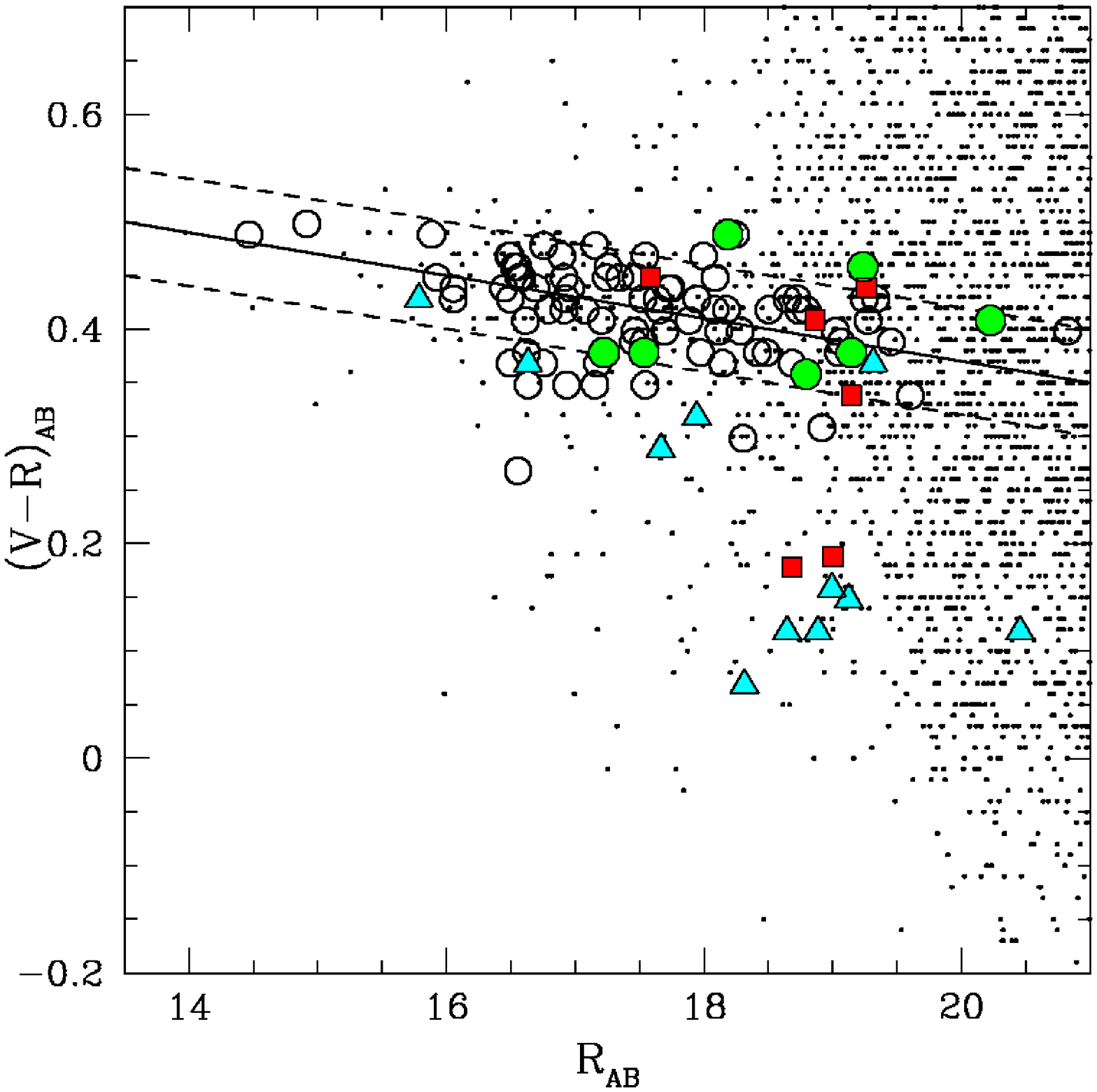}{\includegraphics{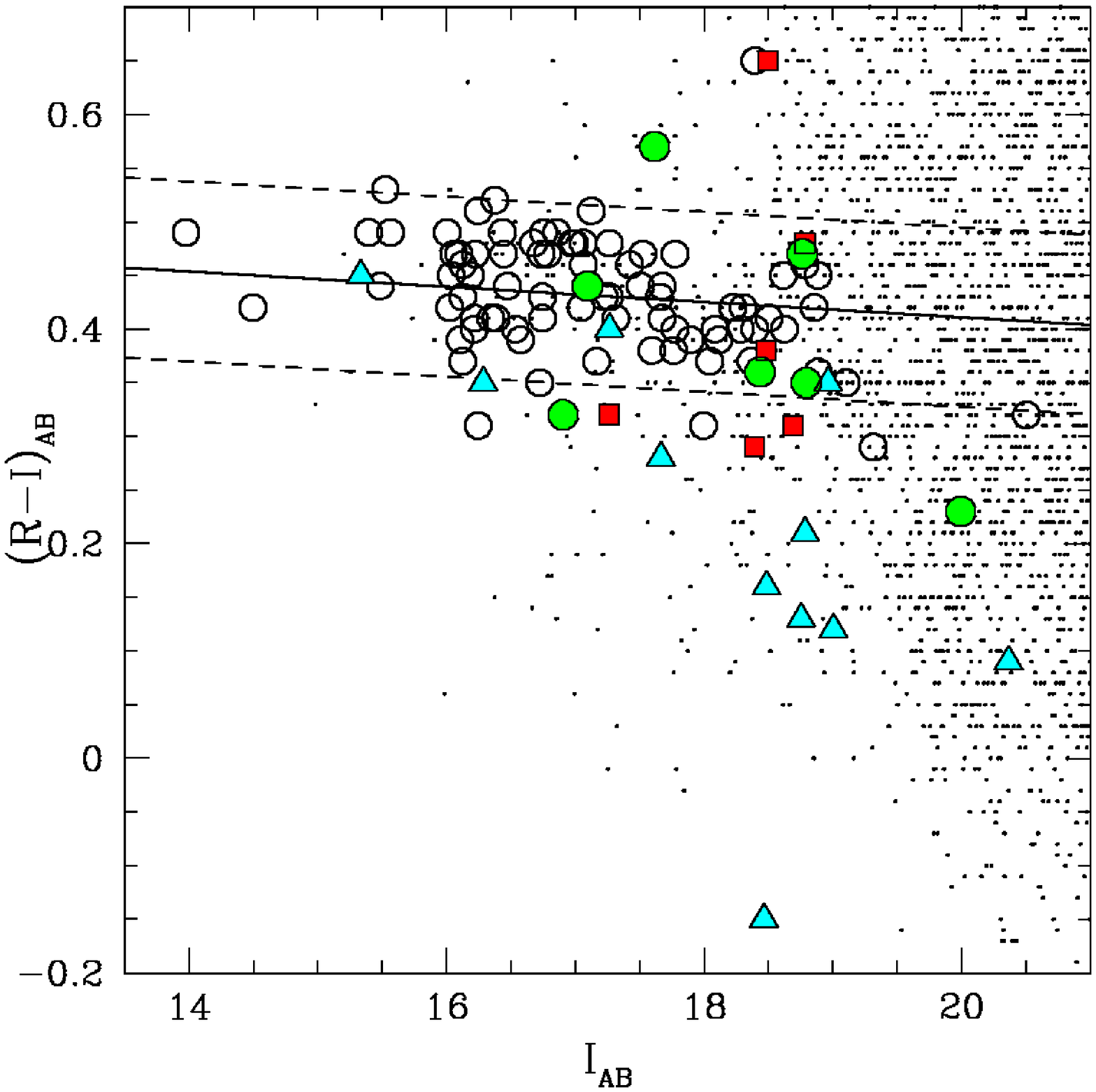}}}
\parbox[b]{16cm}{
\caption{     
  $(V-R)_{\rm AB}$ {\it vs} $R_{\rm AB}$ (left) and $(R-I)_{\rm AB}$
  {\it vs} $I_{\rm AB}$ (right) colour-magnitude diagrams.  All
  galaxies within 34$\times$34 ${\rm arcmin}^2$ are shown. Big symbols
  correspond to confirmed cluster members whereas dots correspond to
  galaxies without spectroscopic information.  Triangles correspond to
  emission line galaxies, squares to k+a type, circles to galaxies
  presenting an H-K inversion following the classification described
  in the text. The solid line is the best linear-fit to the red
  sequence of the cluster, while the dotted line is at $\pm
  1\sigma_{\rm RS}$.}
\label{cmd}}
\end{figure*}

\begin{figure*}  
\resizebox{18cm}{!}{\includegraphics*{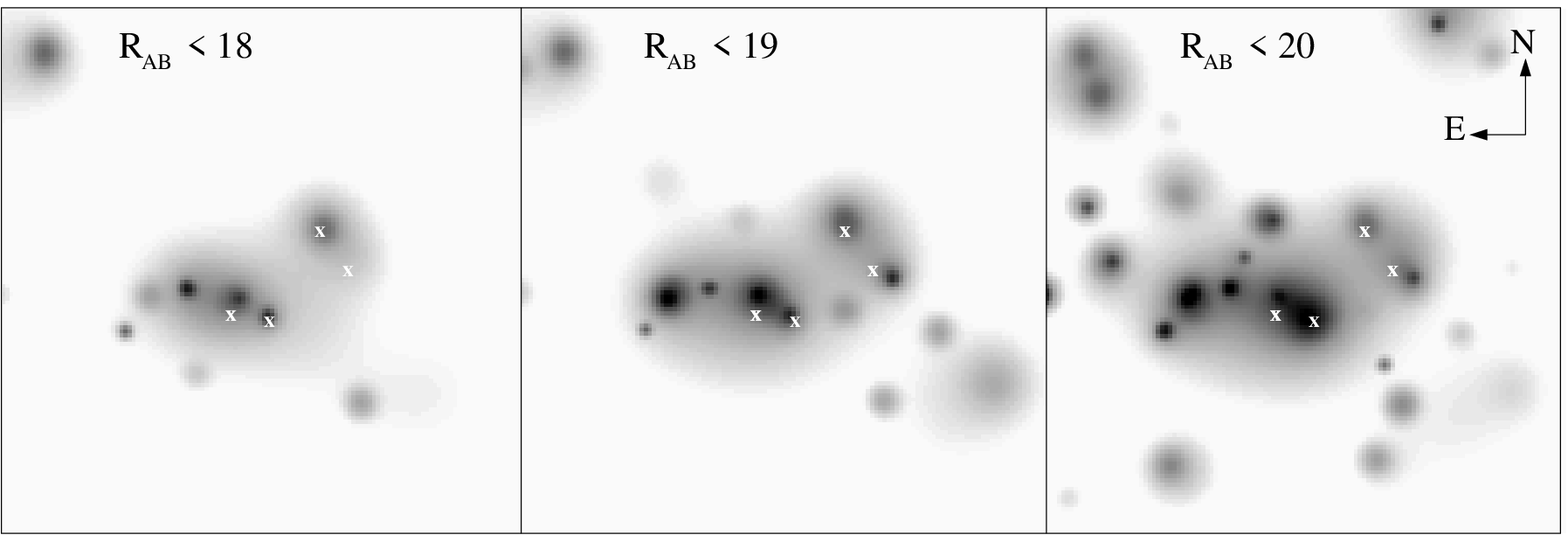}}  
\hfill  
\parbox[b]{18cm}{  
\caption{  
  Projected galaxy density maps (using red sequence galaxies and excluding
  galaxies known not to be cluster members from spectroscopy, as in
  the right panel of Fig.~\ref{IsoDMaps}) on a 34${\times}34 {\rm
  arcmin}^2$ field centred on A3921 and for three magnitude cuts.  The
  white crosses indicate the positions of the four brightest galaxies
  indicated in Fig.~\ref{FC}.}
\label{IsoDMapsMag}}  
\end{figure*}  

\section{Colour properties and spatial morphology of A3921}\label{morph}

In order to investigate the projected spatial morphology of A3921
several density maps of the galaxy distribution are built on the basis
of a multi-scale approach. The adopted algorithm is a 2D
generalization of the algorithm presented in Fadda \etal(1998). It
involves a wavelet decomposition of the galaxy catalogue performed on
five successive scales from which the significant structures are
recombined into the final map (following the equation [C7] of Fadda
\etal 1998). These significant structures are obtained by thresholding
each wavelet plane at a level of three times the variance of the
coefficients of each plane except for the two smallest scales for
which the threshold is increased to four and five times the variance
in order to reduce false detections due to the very low mean density
of the Poisson process at these scales (0.01 for a chosen grid of
128$\times$128 ${\rm pixel}^2$).

Such density maps are presented in Fig.~\ref{IsoDMaps} for three input
catalogues. In the left panel, all galaxies with $R_{\rm AB}<19$
(${\rm R}_{\rm AB}^*$+2.6) are used, leading to a map revealing two
dominant clumps in the central part of the field (hereafter A3921-A in
the centre and A3921-B to the NW) in the central
$\sim$~34$\times$34~${\rm arcmin}^2$.  In order to avoid possible
projection effects, we isolate galaxies likely to be early-types at
the same redshift on the basis of their colour properties.  Indeed,
despite the bimodal structure of A3921 one can notice in the two
colour magnitude diagrams of Fig.~\ref{cmd} a well defined red
sequence, the characteristic linear structure defined by the bulk of
early-type galaxies in a cluster.  The determination of the slope,
intercept and width of these "red sequences" is performed using the
technique described in Appendix~\ref{RS}.  The central panel of
Fig.~\ref{IsoDMaps} shows the resulting red sequence density map
(keeping only galaxies at $\pm 1 \sigma$ around the red-sequence),
where one can notice that several small-scale clumps in the cluster
core disappear as well as in its surroundings, leaving the
larger-scale structure of the two main clumps unaffected.

In order to go one step further in avoiding projection effects we take
advantage of our spectroscopic follow up. Even if it is complete down
to a level of 50\% at $R_{\rm AB}=19$, most of the clumps seen
visually have been spectroscopically sampled allowing us to discard
possible external groups. For example, as mentioned in the previous
section, the background redshift-peak group may lead to an
over-density unrelated to the cluster in the iso-density
map. Therefore, from the sample of red sequence galaxies we
additionally exclude galaxies known from spectroscopy not to be
cluster members. The result is shown in the right panel of
Fig.~\ref{IsoDMaps}, which is quite similar to the map in the central
panel except that the location of clump B is shifted towards South and
its central peak is shifted close to the position of the BG2. If we
can trust in the high confidence level that the sub-structures still
visible are part of A3921, it is still not clear whether or not they
are projection effects within the cluster. In Fig.~\ref{IsoDMapsGR}
the relative velocity distributions of these various sub-structures
are presented using the available redshifts. At the level of ${\rm
R}_{\rm AB}=19$ these clumps actually have full spectroscopic coverage
except for the eastern group with 5 confirmed cluster members out of 7
galaxies belonging to the red sequence. The strong clustering in
redshift space for all the sub-structures confirms the accuracy of the
picture reflected by this density map.

In Fig.~\ref{IsoDMapsMag} the same maps are presented for three
magnitude cuts. The global bi-modal structure of the cluster remains
unchanged with luminosity. This is particularly true for the central
parts of clumps A and B.  However, note that the eastern group, more
compact than the clump B, is stronger at fainter levels.

\begin{figure}  
\centering
\resizebox{8cm}{!}{\includegraphics{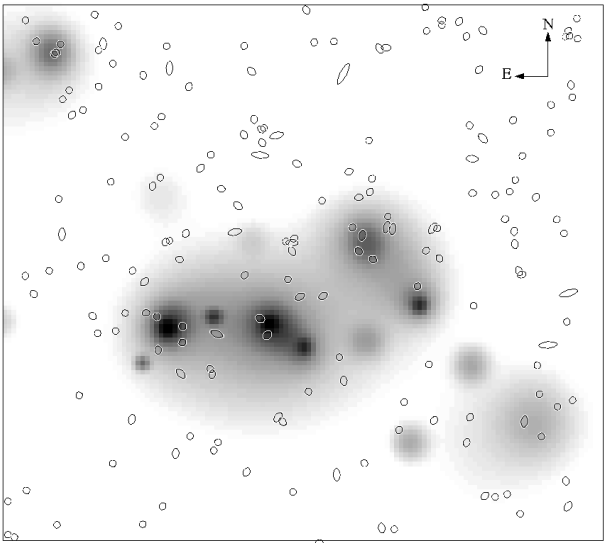}}  
\parbox[b]{8cm}{  
\caption{  
  Projected galaxy density map (34$\times$34 ${\rm arcmin}^2$) of the
  red-sequence galaxies (see Fig.\ref{IsoDMapsGR}). The symbols
  represent the galaxies with $R_{\rm AB}<20$ and bluer than the red
  sequence galaxies.}
\label{IsoDblue}}
\end{figure}

The results given above are based on the only red sequence
galaxies. In Fig.~\ref{IsoDblue} the galaxies bluer than the red
sequence are presented, superimposed on the red sequence iso-density
map. These blue galaxies appear much less clustered than the redder
ones over the whole field and show only little correlation with the
red density peaks except in the case of clump B showing an excess of
blue galaxies relative to the other clumps and in particular to the
center of the clump A.  This asymmetry is also present in the
distribution of emission line galaxies as it will be shown and
discussed in Sect.~\ref{eml_dist}.

To summarize, A3921 is characterized by a) a bimodal morphology, b) the
presence of several substructures inside each of the two main clumps, c) an
offset of the BCG from the main density peak of clump A, and d) an excess of
blue galaxies around the clump B. These results suggest that this system is
out of dynamical equilibrium and that it is probably composed of a main cluster
interacting with at least two groups, one to the North-East (clump B) and one
to the East, the latter being significantly less luminous than the former.  In
the following section, we will analyze the dynamical properties of A3921 in
order to understand which phase of the merging process we are witnessing.

\section {Cluster Kinematics}

\subsection{Velocity distribution of the whole cluster}\label{ROSTAT}

\begin{figure}
\centering
\resizebox{8cm}{!}{\includegraphics{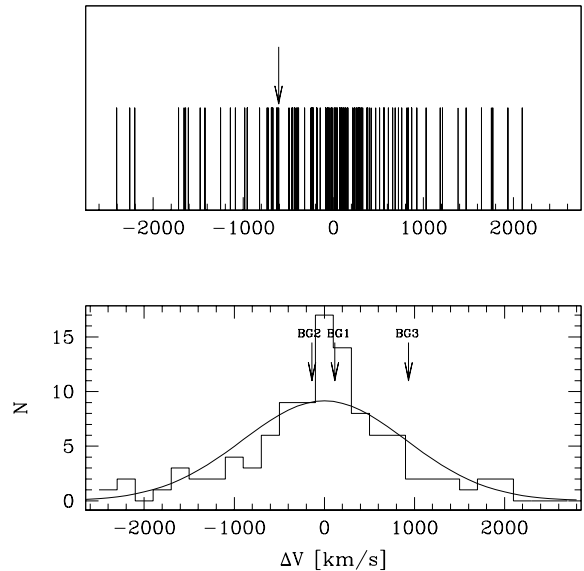}}
\parbox[b]{8cm}{
\caption{Top: Stripe density plot of the radial velocity offsets of the 104
  A3921 members from $\bar{z}$ (corrected for cosmological and relativistic
  effects). The position of the gap detected in the cluster velocity
  distribution is indicated by an arrow. Bottom: Velocity histogram of the
  confirmed cluster galaxies in bins of 200~km/s.  The Gaussian best--fit to
  the velocity distribution is superimposed.}
\label{histoA3921}}  
\end{figure}

Using the biweight estimators for location and scale (Beers
\etal1990), we find a mean apparent velocity of $C_{BI}=28047 ^{+76}
_{-77}$~km/s, corresponding to a mean redshift of
$\bar{z}{\simeq}0.0936$, and a velocity dispersion of $S_{BI}=831
^{+100} _{-76}$ km/s (at $1\sigma$ significance level, see
Table~\ref{ROST_Tab}). Fig.~\ref{histoA3921} shows the histogram of
the cosmologically and relativistically corrected velocity offsets
from the mean cluster redshift ($\Delta{v}=c(z-\bar{z})/(1+\bar{z})$)
of the 104 cluster members. Contrary to the projected morphology on
the sky we do not find any bi-modal structure in the velocity
distribution.  In dissipationless systems, gravitational interactions
of cluster galaxies over a relaxation time generate a Gaussian
distribution of their radial velocities; possible deviations from
Gaussianity could provide important indications of ongoing dynamical
processes.  In the following, we are therefore interested in measuring
a possible departure of the observed cluster velocity distribution
from a Gaussian.

\begin{table*}
\begin{center}
\begin{tabular}{cccccccc}
\hline
\hline    
Subsample & $N_{gal}$ & ${C}_{BI}$ & ${S}_{BI}$ & Skewness & AI & 
Kurtosis &TI \\
 & & [km/s] &  [km/s]  & \% & \% & \% & \%\\
\hline
\multicolumn{8}{c}{ }\\  
Whole sample & 104 & $28047 ^{+76} _{-77}$ & $831 ^{+100} _{-76}$ & 
$\geq$10 & $>$20 & $\leq$10 & $\leq$1   \\
\multicolumn{8}{c}{ }\\
A3921-A & 41 & $28017 ^{+145} _{-173}$ & $1008 ^{+156} _{-106}$ & $>$10 
& $>$20 & $>$20 &$>$20\\
\multicolumn{8}{c}{ }\\
A3921-B & 20 & $27920  ^{+88} _{-86}$ & $451 ^{+215} _{-80}$ & $<$10 & 
$<$5 & $>$20 & $>$10 \\
\multicolumn{4}{c}{ }\\
\hline
\end{tabular}
\caption{Properties of the $cz$ distribution and significance levels of
  shape estimators for various subsamples of A3921. ${C}_{BI}$ and
  ${S}_{BI}$ are the mean velocity and the velocity dispersion of the
  different distributions (biweight estimators for location and scale,
  Beers \etal1990) }
\label{ROST_Tab}
\end{center}
\end{table*}

A light--tailed velocity distribution can indicate the presence of two
or more overlapping Gaussian components in the whole velocity
histogram, while an asymmetric distribution can be the result of the
contemporary presence of subclusters with different numbers of
galaxies (Ashman \etal1994). We use two kinds of shape estimators: the
traditional third and fourth moments, i.e. skewness and kurtosis, and
the asymmetry and tail indices (Bird \& Beers, 1993). In
Table~\ref{ROST_Tab} we report the corresponding values and
significance levels, estimated from Table~\ref{ROST_Tab} in Bird \&
Beers 1993 under the null hypothesis of a Gaussian distribution.

The values obtained for Skewness and AI show that the velocity
histogram is quite symmetric (significance level $>$10\%). Values
observed for kurtosis and TI indicate a heavy--tailed distribution,
rejecting the Gaussian hypothesis at $\sim$10\% significance level the
first, even better than 1\% the second.  While light-tailed
distributions indicate multi-modality, heavy populated tails could be
due to contamination by non-cluster galaxies. We think that the
analysis of Sect.~\ref{membership} excludes this possibility, but we
apply an additional test in order to definitively reject the presence
of outliers in our velocity sample of 104 galaxies.  Extensive data
available for low-$z$ clusters show that most ($\geq$95\%) of the
galaxies in the central region of the clusters have radial velocities
within $\pm$3500~km/s of the mean cluster redshift (Postman \etal
1998). We thus verify that all the 104 galaxies of our final sample
have $|\Delta{v}|<$3500~km/s (with $\Delta{v}$ defined at the
beginning of this section). In any case, the shape parameters do not
show strong evidence of subcomponents in velocity space.

7 of the 13 normality tests performed by ROSTAT reject the Gaussian
hypothesis at better than 10\% significance level (see
Table~\ref{Test1D}).  Moreover, we find one significant weighted gap
in our dataset (Beers \etal1990) at $cz=27381.9$ km/s (shown in
Fig.~\ref{histoA3921}) and characterized by a normalized size of 2.47
and a significance of 3\%.  On the contrary, among the 6 tests
performed by ROSTAT that do not reject the Gaussian hypothesis, the
DIP test (Hartigan \& Hartigan 1985) accepts the unimodal hypothesis
at better than 99\% level, in agreement with the conclusions obtained
with the symmetry tests (i.e. Skewness and AI).

Classical tests of Gaussianity give therefore controversial results
about the kinematical properties of A3921.

\begin{table}  
\begin{center}  
\begin{tabular}{ccr}  
\hline
\hline
 & Whole cluster & \\
\hline
Statistical Test & Value & Significance\\
\hline
A & 0.735 & $<$1\%\\
B2 & 3.567 & 8.8\%\\
I & 1.099 & $<$5\% \\  
KS & 0.901 & 5.0\% \\  
V & 1.621 & 2.5 \% \\  
${\rm W}^2$ & 0.163 & 1.6\% \\  
${\rm U}^2$ & 0.159 & 1.2\% \\  
${\rm A}^2$ & 0.907 & 2.1\% \\
\hline  
 & A3921-A & \\
\hline
Statistical Test & Value & Significance\\
None & - & - \\
\hline
 & A3921-B & \\
\hline
Statistical Test & Value & Significance\\
A & 0.684 & $<5\%$ \\
U & 4.822 & $\sim$1\% \\
B1 & -0.657 & 7.8\% \\
B2 & 4.738 & 2.1\% \\
B1 \& B2 & 6.123 & 4.7\% \\
\hline
\end{tabular}  
\caption{1-D statistical tests performed in ROSTAT package that exclude
the hypothesis of a single Gaussian distribution for the different
velocity datasets considered in the paper. In Cols.~1 and 2 we report
the name and the value of the statistics, while Col.~3 indicate their
significance levels.}
\label{Test1D}
\end{center}  
\end{table}

\subsection{Analysis of a possible partition in redshift space and 
kinematical
indicators of subclustering}

\begin{figure*}
\centering

\resizebox{16cm}{!}{\includegraphics{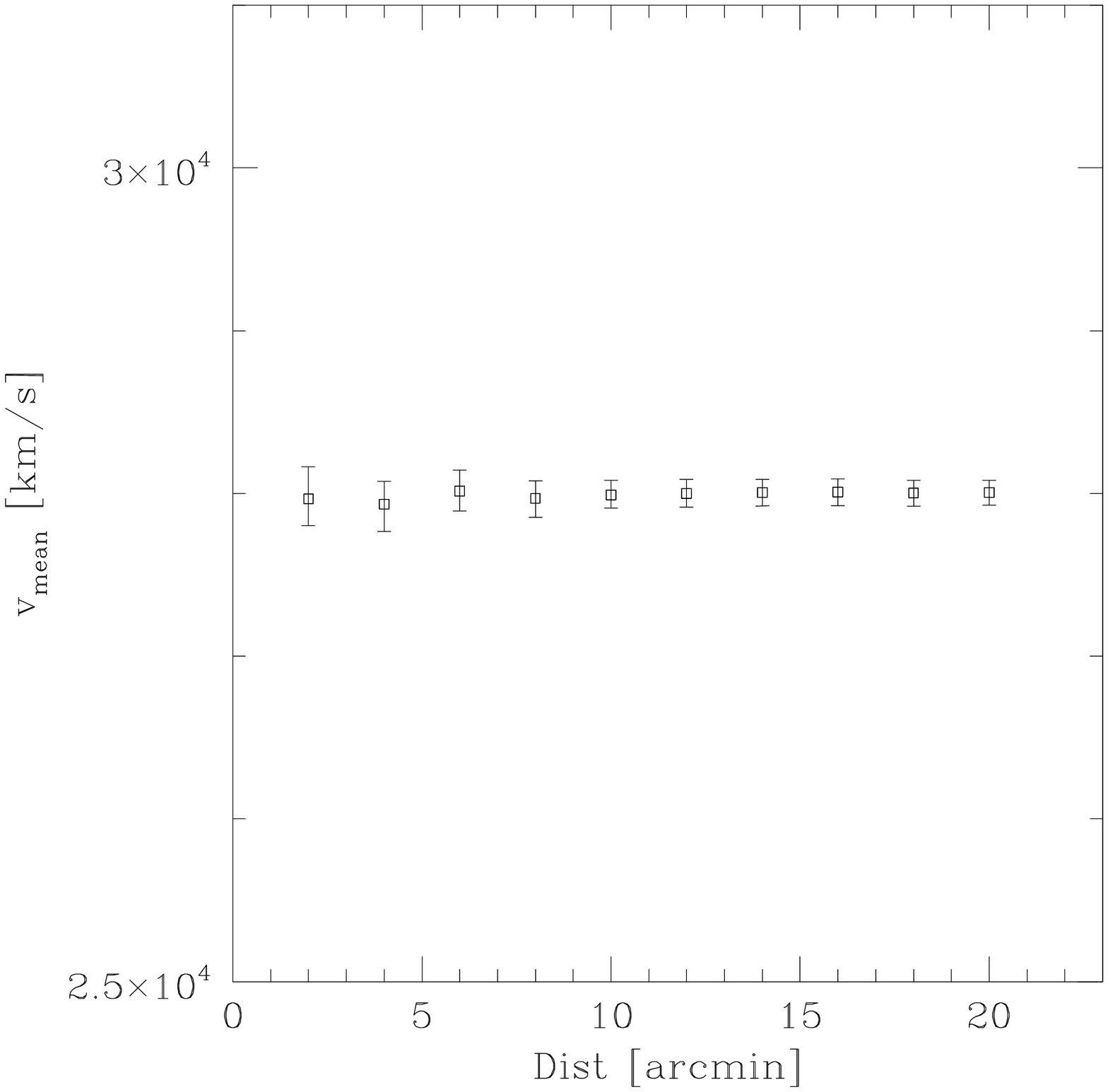}
\includegraphics{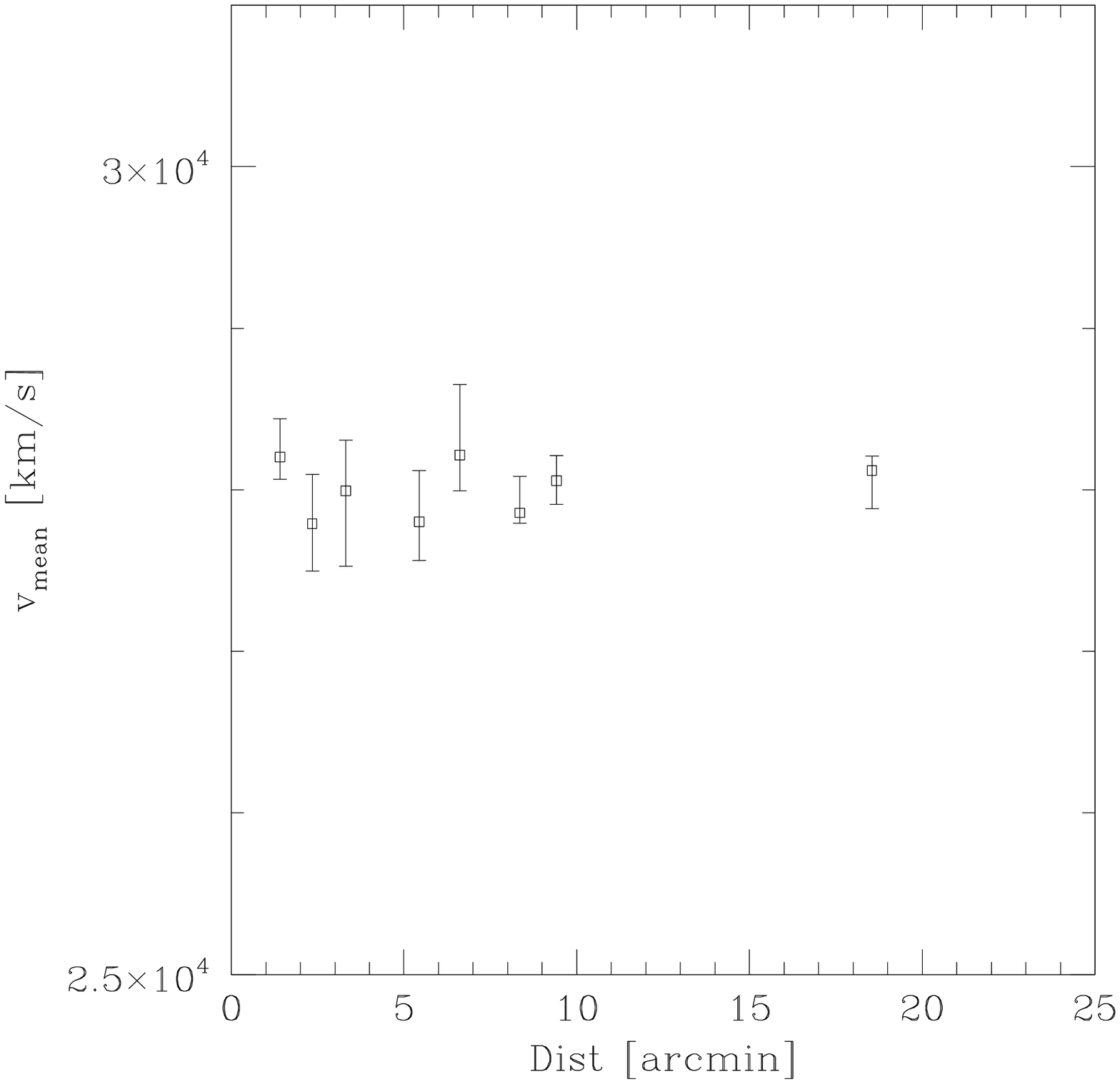}}

\resizebox{16cm}{!}{\includegraphics{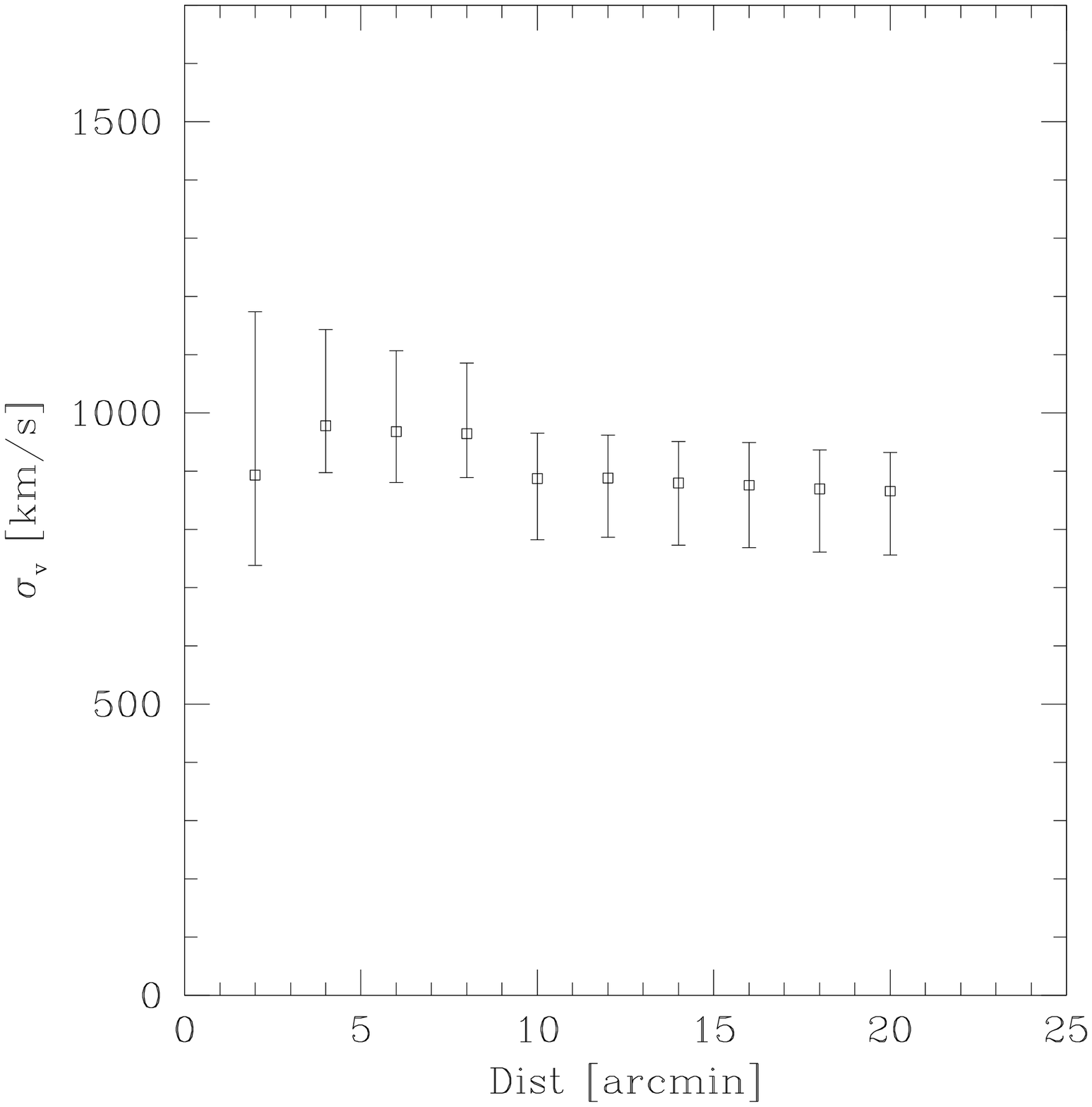}
\includegraphics{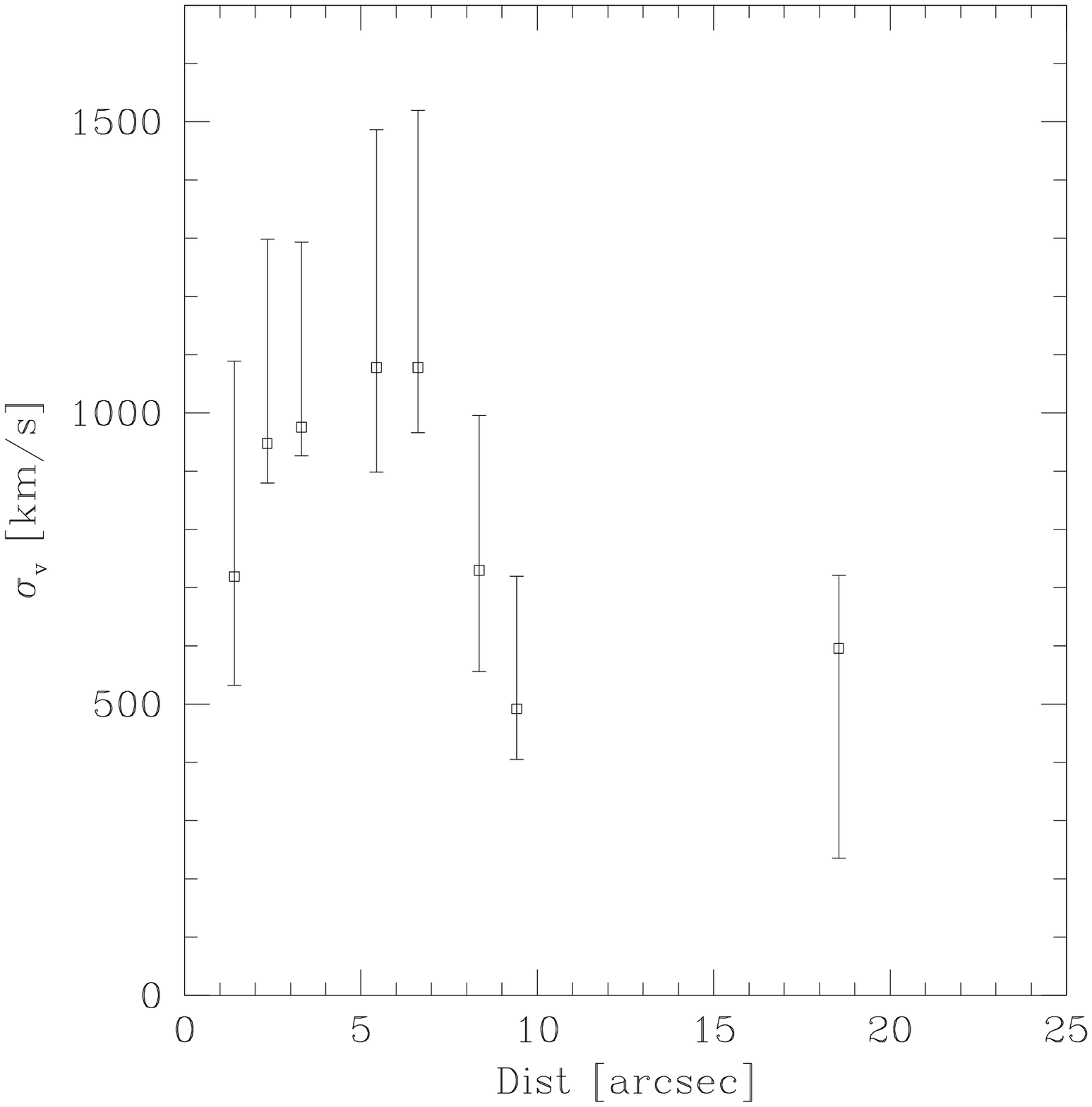}}

\parbox[b]{16cm}{  
\caption{Profile of the velocity means (top) and dispersions (bottom) in the
  cluster: left: as a cumulative function of galaxies centred on BG1 
position
  - right: differential values as a function of radius in rings with 13
  galaxies; the last point is far from the others due to the lower
  degree of completeness reached by spectroscopic data in the external 
regions
  of the cluster (out of the central $\sim$12').}
\label{VDP}}
\end{figure*}  

Due to the uncertain results obtained in previous section, we apply
the KMM mixture-modeling algorithm of McLachlan \& Basford (1988)
which, using a maximum-likelihood technique, assigns each galaxy to a
possible parent population, and evaluates the improvement in fitting a
multiple--component model over a single--one. First of all, we try to
allocate the 104 galaxies of our dataset into two possible
subclusters, respectively with mean velocity lower and higher than the
gap position. Several tests are then performed by changing both the
estimated mean velocities and the estimated mixing proportion for each
group, but we always obtain the same result: the KMM algorithm tries
to fit a 2--group partition from these guesses, obtaining intermediate
confidence levels that cannot reject the null hypothesis of unimodal
distribution. This result actually agrees both with the DIP test and
with what is observed from the shape parameters of the velocity
distribution, the absence of light--populated tails and of a
significant asymmetry in the histogram excluding the possible presence
of several overlapping subunits populated differently.

Classical statistical tests for sub--clustering are then applied to
the 104 cluster members in order to look for the existence of
correlated substructures in velocity and spatial distributions
(Dressler \& Shectman, 1988; Bird, 1994; West \& Bothun, 1990).  The
results, obtained by the bootstrap technique and 1000 Monte Carlo
models, are presented in Table~\ref{3D}.

\begin{table}  
\begin{center}  
\begin{tabular}{ccc}  
\hline
\hline  
Indicator & Value & Significance\\  
\hline
$\Delta$ & 109.284 & \bf {0.680}  \\  
$\epsilon$ & ${0.991}{\times}{10}^{+27}~{\rm kg}$ & 0.353 \\  
$\alpha$ & 0.129~${{h}_{75}}^{-1}$~Mpc & \bf {0.914} \\  
\hline  
\end{tabular}
\caption{3-D substructure indicators for the sample of 104 objects with
  quality flag=1 in the velocity range 25400$\div$30400~km/s}
\label{3D}
\end{center}  
\end{table}  
 
\begin{figure}
\centering
\resizebox{8cm}{!}{\includegraphics{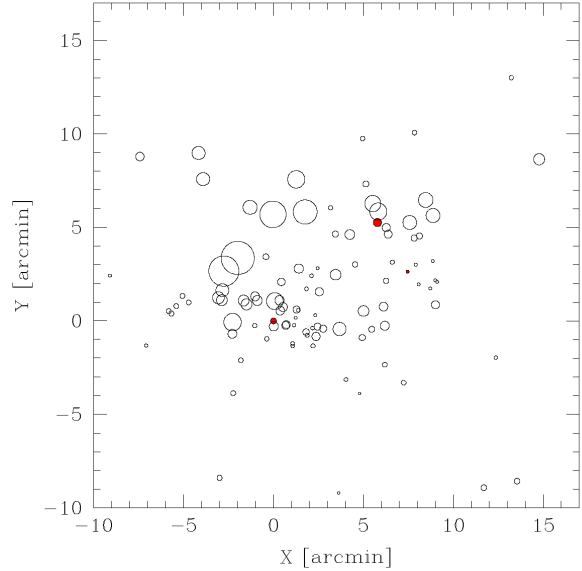}}
\parbox[b]{8cm}{  
\caption{Projected positions, centred on BG1 coordinates, of the galaxies
  with $v_r=25400{\div}30400$. Each galaxy is represented by a circle, whose
  dimension is weighted by the estimator of Dressler \& Shectman (1988).
  Filled circles (red in the electronic version of the paper) are 
centred on
  the three BGs positions.}\label{Dressler}}
\end{figure}

Both $\Delta$ and $\alpha$ parameters do not find evidence of
substructures with a high significance level, while $\epsilon$ test
has an intermediate and not conclusive value. Fig.~\ref{Dressler}
shows graphically the results obtained with the $\Delta$ test.

In conclusion, while no bi-modality is shown in the velocity distribution, the
various tests of gaussianity and subclustering do not reach a consensus, but
do not show "extreme" departures from gaussianity.

\subsection{Radial profile of the velocity dispersion}

The analysis of the mean velocity and particularly of the radial
profile of the velocity dispersion provides a useful tool for
investigating the dynamics of galaxy clusters, as they can reveal
signs of subclustering and ongoing merging (Quintana \etal1996, Muriel
\etal2002).  Such integrated measurements are therefore performed with
the 104 members of A3921 up to a distance of $\sim$~2~\h~Mpc from the
cluster centre (taken to be at the position of BG1), and are presented
in the left panels of Fig.~\ref{VDP}.  The mean velocity has a very
constant value ($\sim$28000~km/s) from the centre of the cluster to
its outer edges.  Following the classification of the radial profiles
of the velocity dispersions of den Hartog \& Katgert (1996), A3921
presents an "inverted" shape: the cumulative velocity dispersion shows
an initial increase with radius up to $\sim$4~arcmin apart from the
cluster centre and then decreases nearly to its first bin value
($\sim$900~km/s). Finally, the velocity dispersion becomes flat in the
external regions of the cluster ($\geq$1~\h~Mpc). This result suggests
that this final value is representative of the total kinetic energy of
the cluster members (Fadda \etal1996). A similar shape was detected by
den Hartog \& Katgert (1996) and Nikogossyan \etal(1999) in the case
of the galaxy cluster A194, and these authors interpreted such a
profile as originated in a nearly relaxed region.  Therefore, both the
integrated radial profiles of the velocity dispersion and of the mean
velocity do not reveal the presence of significant velocity gradients
that could be produced by the presence of internal
sub-structures. These results are confirmed by the right panels of
Fig.~\ref{VDP}, where mean velocities and velocity dispersions are
estimated in rings containing the same number of galaxies (13). The
shapes of these differential profiles agree with the integrated ones,
presenting again an inverted shape. Its minimum velocity dispersion
corresponds to the ring containing the group of galaxies located
around BG2.  Within error bars, mean velocities are nearly constant
around 28000~km/s, as for the integrated profile.

\subsection{Velocity distributions of A3921-A and A3921-B}\label{dyn2}

\begin{figure}
\centering
\resizebox{8cm}{!}{
\includegraphics{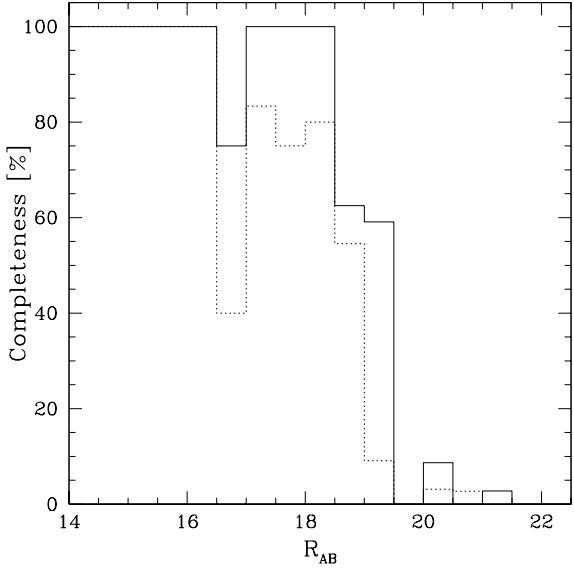}}
\parbox[b]{8cm}{
\caption{Completeness of the spectroscopic sample for the two clumps A3921-A 
(solid line) and in A3921-B (dashed line).}
\label{ComplIclumps}}
\end{figure}

\begin{figure}  
  \centering \resizebox{8cm}{!}{\includegraphics{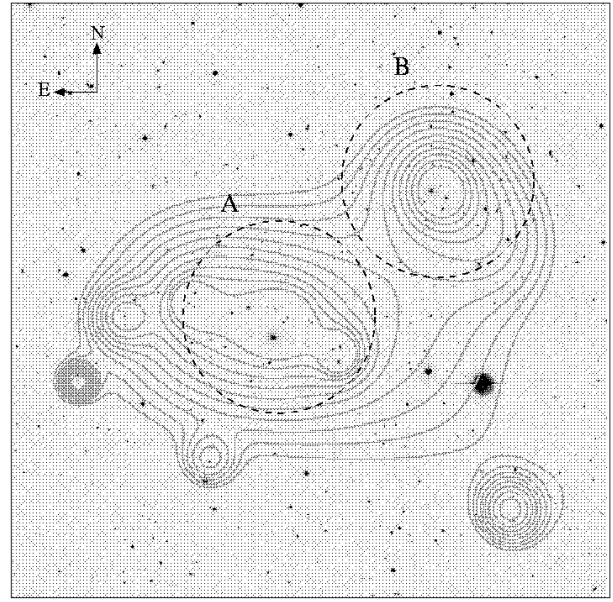}}
  \parbox[b]{8cm}{
\caption{  
  Iso-density contours of the projected distribution of the red-sequence
  galaxies (with $R_{\rm AB}\leq 18$ and after removing galaxies known 
not to
  be cluster members from spectroscopic data) superimposed to the central
  22$\times$22 ${\rm arcmin}^2$ of the R--band image of A3921.  The 
dynamical
  analysis of the two clumps A and B is performed based on the galaxies
  selected inside the two dashed circles (R$\simeq$ 0.34 Mpc).}
\label{IsoDf}}
\end{figure}

As shown above, the dynamical and kinematical properties of the whole
cluster do not reveal strong signatures of merging.  More indications
on the dynamical state of the cluster could come from the analysis of
the velocity distribution of the two main subclusters detected on the
iso-density maps separately. For this, our velocity sample is divided
into two datasets, containing the confirmed cluster members in the two
circles displayed in Fig.~\ref{IsoDf}, chosen as the largest
without intersecting. The radius of the two circles is $\simeq$ 0.34
Mpc and they are centred on the density peaks previously detected
(Sect.~\ref{morph}). We plot in Fig.~\ref{ComplIclumps} the ratio of
the number of galaxies with very good redshift determination (Q.F.=1)
to the total number of objects detected in the two circles of
Fig.~\ref{IsoDf} as a function of $R$--band magnitude; the
spectroscopic sampling of these two regions, and in particular of
A3921-A, appears quite good, with completeness levels $\geq$50\% for
$R_{\rm AB}\leq$19.5 ($\simeq R_{\rm AB}^*$+3.1).

\begin{figure}  
\resizebox{8cm}{!}{\includegraphics{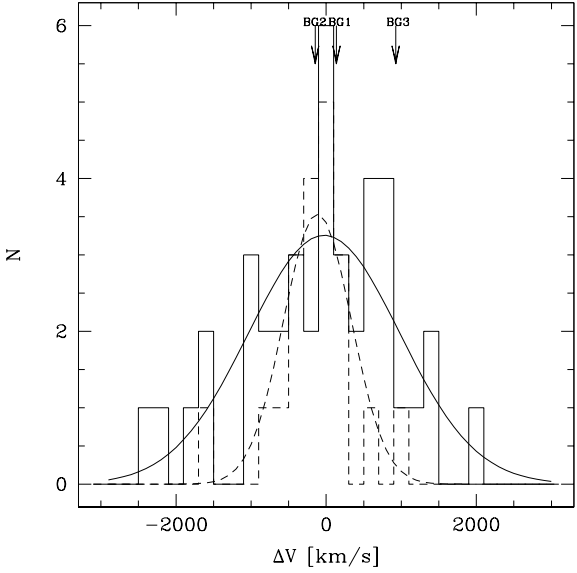}}  
\hfill  
\parbox[b]{8cm}{  
\caption{Velocity histogram, with a binning of 200~km/s, of the galaxies in
  A3921-A (solid line) and in A3921-B (dotted line), according to the
  division showed in Fig.~\ref{IsoDf}. The relative Gaussian best--fits
  to the velocity distributions are superimposed. Arrows show the
  radial velocities of the three brightest cluster galaxies. }
\label{histoMG}}  
\end{figure}

Fig.~\ref{histoMG} shows the velocity distributions of the two
datasets and in Table~\ref{ROST_Tab} we quote their velocity means and
dispersions.  Both datasets show a velocity location very close to
each other and to the whole cluster value (the velocity
offset\footnote{As usual cosmologically and relativistically
corrected} $\Delta{v}$ between A3921-A and A3921-B mean velocities is
of only $89 ^{+155} _{-177}$ km/s); the clump A is characterized by
the highest velocity dispersion.

Most (8) of the 13 normality tests contained in ROSTAT package accept the
hypothesis of Gaussianity for the velocity distribution of the A3921-B
component, and actually all of them do not reject the null hypothesis for the
radial velocities of the clump A3921-A (Table~\ref{Test1D}). Shape parameters
(Table~\ref{ROST_Tab}) accept the Gaussian hypothesis at more than 10\%
significance level, with the exception of the skewness and of the kurtosis,
that indicates the possible presence of an asymmetric velocity distribution
(significance level lower than 10\%) and of heavy populated tails
(significance level lower than 5\%) in the A3921-B velocity histogram.

Summarizing, Gaussianity tests suggest that the merging event has not
strongly affected the internal dynamics of the two sub-clusters.

\section{Dynamical study of the system}

\subsection{Mass estimate of the two main clumps}\label{MassE}

Determining the respective masses of A3921-A and A3921-B
requires idealized assumptions which cannot be
strictly valid for an interacting system. However,
as seen in section ~\ref{dyn2}, the dynamics of the central regions of the
two clumps appears to be relatively unaffected by the merging event.

We have therefore assumed that each subcluster is virialized. 
As in the case of velocity dispersions, the virial radius of each clump
was estimated selecting only those
galaxies within a projected distance of $\sim$ 0.34 \h~ Mpc 
from its center (see Fig.~\ref{IsoDf}). In
order to have a better statistics and at the same
time to minimize biases due to spectral incompleteness and 
background contamination, we included all the galaxies
belonging to the red sequence (thus assuming that early--type galaxies
 trace the mass profile;
see Katgert, Biviano \& Mazure 2004), except for those with a measured 
redshift and 
identified as outliers on the basis of our previous analysis.

The mass was calculated with the classical virial equation:

\begin{equation}
 M_{vir} = \frac{r_{vir} \sigma_{vir} ^2}{G}
 \label{eq:vir}
\end{equation}

\noindent where $\sigma_{vir}$ is 
the three--dimensional velocity dispersion of the system, and
$r_{vir}$ is the virial radius:

\begin{equation}
\label{eq:rvir}
r_{vir} \equiv -\frac{G M_{vir}^2}{U} = 
N^2  \left( \sum _{i<j} ^{N-1} \sum _{j=i+1} 
^{N} \frac{1}{r_{ij}} \right) ^{-1} =
\frac{2 N}{N-1}  r_h
 \label{eq:radii}
\end{equation}

\noindent In equation (\ref{eq:radii}) $U$ is the potential energy 
of the system and $r_h$ is the mean harmonic radius, defined as:

\begin{equation}
\label{eq:rh}
r_h = \frac{N(N-1)}{2} \left( \sum _{i<j} ^{N-1} \sum _{j=i+1} ^{N}
\frac{1}{r_{ij}} \right) ^{-1}
 \label{eq:rhpw}
\end{equation}

\noindent where $r_{ij}$ is the separation between the $i$th and $j$th galaxies, and $N$
is the total number of objects in the system).

We have to apply the above relations to projected separations, so that we can 
estimate the projected mean harmonic radius $R_H$ and obtain
the corresponding projected virial radius $R_{vir}$.
Assuming spherical symmetry, $r_{vir} = (\pi/2) R_{vir}$ 
(see Limber \& Mathews 1960) and $\sigma_{vir} = 3 \sigma_r$, where
$\sigma_r$ is the radial velocity dispersion of the system. 

The above method is based on the pairwise estimator of
the harmonic radius: an alternative is the 
so--called ringwise estimator (see Carlberg et al. 1996). 
We have applied both methods,
finding no significant difference (notice that this is not true in general).

\begin{table*} 
\begin{center}  
\begin{tabular}{ccccc}  
\hline 
\hline 
$R_{vir}$ (A) & $R_{vir}$ (B) & $M_{vir}$ (A) & 
$M_{vir}$ (B) & $M_{vir}$ (A) / $M_{vir}$ (B) \\
0.39$\pm$0.02 & 0.38$\pm$0.02 & $4.3 ^{+1.4} _{-1.0} $ & 
$0.8 ^{+0.7} _{-0.3} $ & $5.2 ^{+4.4} _{-2.5}$ \\

\hline  
\end{tabular}  
\caption{Columns 1 \& 2: projected virial radii 
of A3921-A and A3921-B; columns 3 \& 4: virial mass estimates for A3921-A and 
A3921-B; column 5: mass ratio.
Masses are in $ 10^{14} {M}_{\odot}$ units and radiii in $h^{-1}$~Mpc units.}
\label{Mass}
\end{center}  
\end{table*}

The results obtained applying the pairwise method along with the
corresponding 1$\sigma$ errors, are shown in 
Table~\ref{Mass}. The values of the virial radii are relatively small
(only slightly larger than the window radius): we have to take into account the
the possibility that they are underestimated, although they could be 
appropriate for subcluster systems. 

Given our cosmology with $\Omega_M + \Omega_\Lambda = 1$, we would expect
$\bar{\rho}(r_{vir}) \sim 100 \rho_c$ 
(see e.g. Eke et al. 1996), where $\bar{\rho}(r_{vir})$ is the mean density 
within
$r_{vir}$ and  $\rho_c$ is the critical density. 
The ratio between these two quantities 
is related to the virial radius according to the following equation 
(always derived for our flat cosmology):

\begin{equation}
 \frac{\bar{\rho}(r_{vir})}{\rho_c(z)} = 
 \frac{1}{\rho_c(z)} \frac{3 M_{vir}}{4 \pi r_{vir} ^3} =
 \frac{\sigma_r ^2}{r_{vir} ^2} \frac{6}{H_0 ^2 [\Omega_M (1+z)^3 + 
\Omega_\Lambda]}
\label{eq:ratio}
\end{equation} 

\noindent Putting into the above equation the mean cluster redshift and 
the estimated values of the three-dimensional virial radii, $r_{vir} =
\pi R_{vir} / 2$, we find $\bar{\rho}(r_{vir})$ $\sim 2670 \rho_c$ and
$\sim 560 \rho_c$ respectively for A3921-A and A3921-B, which
indicates that we are probably underestimating the virial radii.  A
further uncertainty is associated to the surface pressure term, which
acts in the opposite direction with respect to the previous correction
for the underestimate of the virial radius (Carlberg et al. 1996 and
references therein). However, while the exact values of the virial
masses might be larger, it is clear that A3921 is made up by two
unequal mass systems, with $M_A/M_B \sim 5$, a value which essentially
reflects the factor 2 ratio between the velocity dispersion of the two
clumps.

\subsection{Two-body dynamical model}

In this section, we apply the two-body dynamical formalism (Gregory
and Thomson 1984, Beers \etal1992) to the two sub-clusters of
A3921. By assuming that at $t$=0 the two subclusters were at zero
separation, parametric solutions to the equations of motion can be
derived both in the bound:

\begin{eqnarray}
R & = & \frac{R_m}{2}~(1-{\rm cos}{\chi}) \label{eq:Rb} \\
t & = & \left(\frac{{R_m}^3}{8GM}\right)^{1/2}~(\chi-{\rm sin}{\chi})
\label{eq:Tb}\\
V & = & \left(\frac{2GM}{R_m}\right)^{1/2}~\frac{{\rm sin}{\chi}}{1-{\rm
cos}{\chi}}, \label{eq:Vb}
\end{eqnarray}

\noindent and in the unbound case:

\begin{eqnarray}
R & = & \frac{GM}{{V_{\infty}}^2}~({\rm cosh}{\chi}-1) \label{eq:Ru} \\
t & = & \frac{GM}{{V_{\infty}}^3}~({\rm sinh}{\chi}-\chi) \label{eq:Tu} \\
V & = & V_{\infty}~\frac{{\rm sinh}{\chi}}{{\rm cosh}{\chi}-1}, 
\label{eq:Vu}
\end{eqnarray}

\noindent where $R_m$ is the separation of the two subclusters at maximum
expansion, $M$ is the total mass of the system, $\chi$ is the developmental
angle and $V_{\infty}$ is the asymptotic expansion velocity.  The relative
velocity $V$ and the spatial separation $R$ between the substructures are
respectively related to their radial ($\Delta{v}$) and projected components
($R_p$) through the following relations:

$$
V=\Delta{v}/{\rm sin}{\alpha};~~~R=R_p/{\rm cos}{\alpha}
$$

\noindent where $\alpha$ is the angle between the plane of the sky and 
the line connecting the centres of the two clumps.

\begin{figure*}
\centering
\resizebox{16cm}{!}{\includegraphics{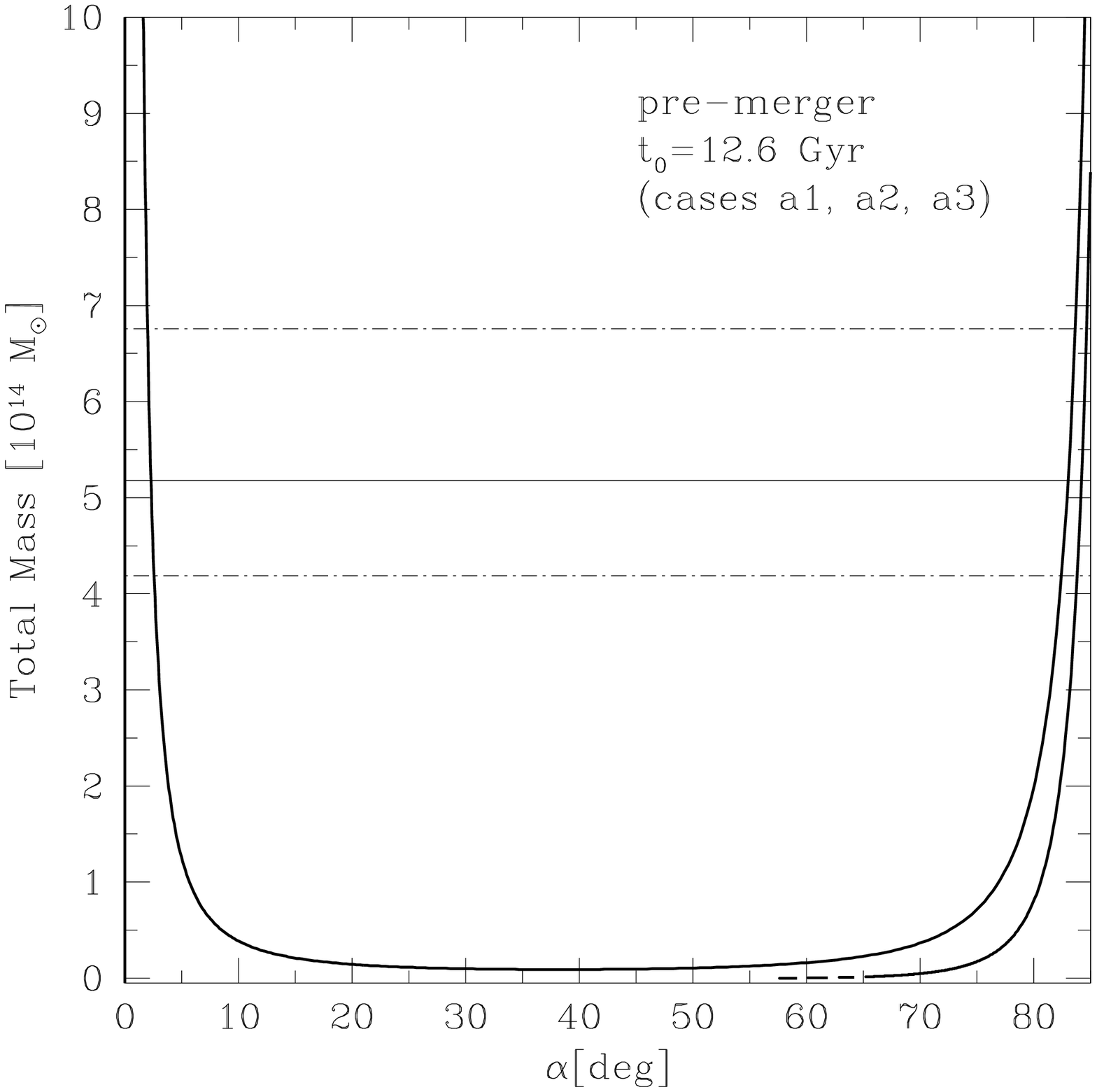}
\includegraphics{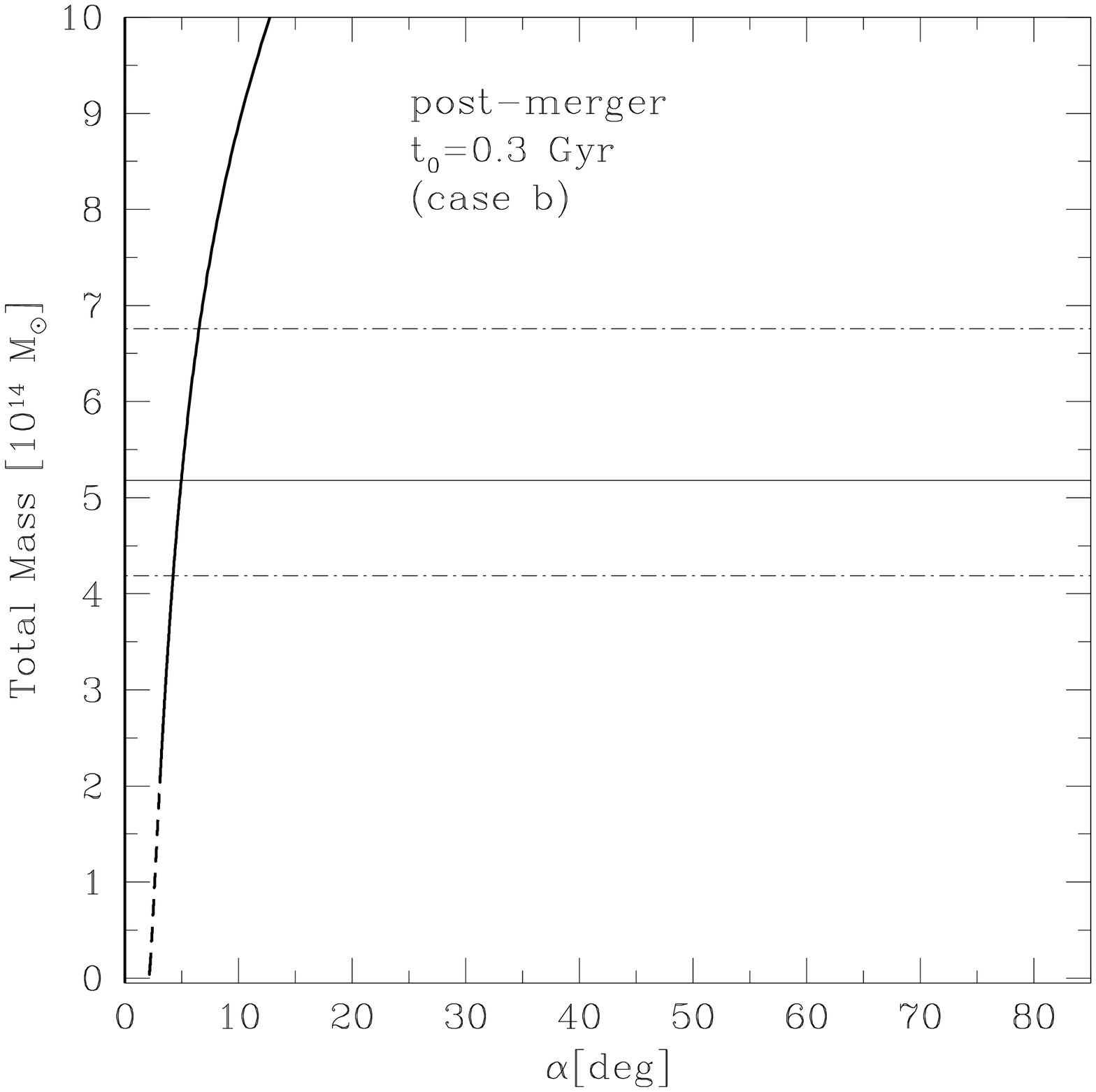}}

\resizebox{16cm}{!}{\includegraphics{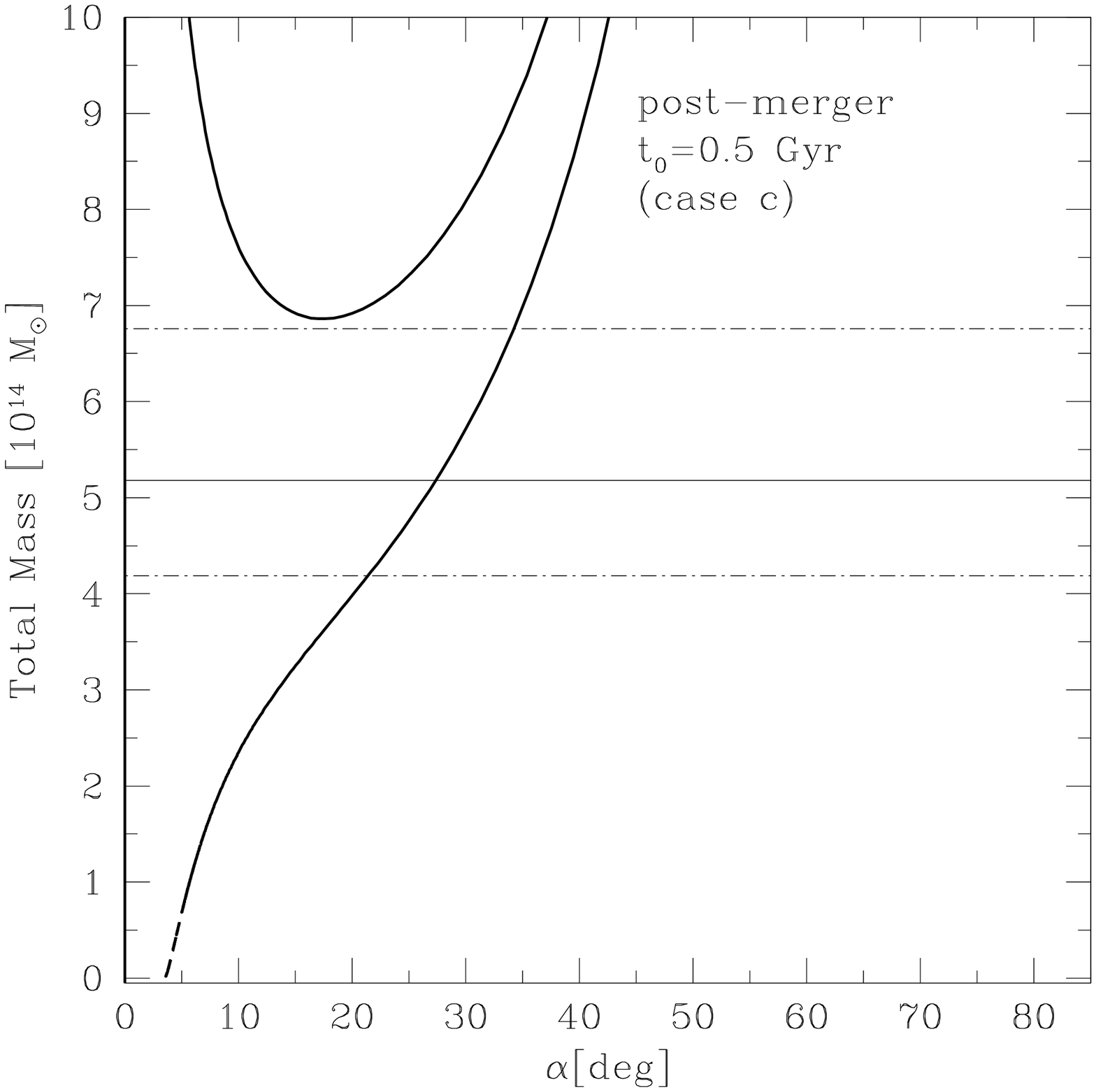}
\includegraphics{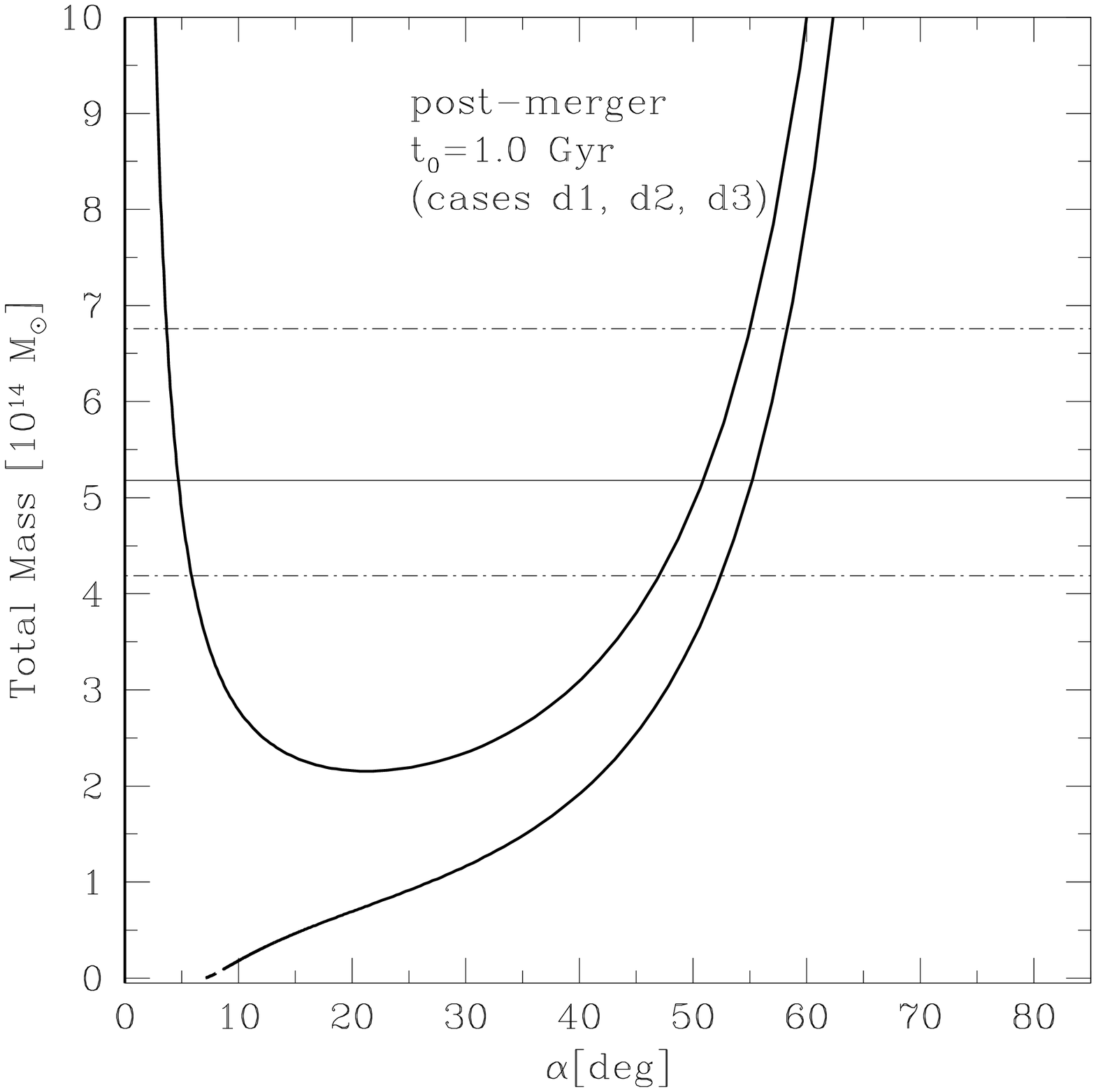}}

\parbox[b]{16cm}{
\caption{The total mass of the two clumps as a function of the angle $\alpha$
  between the plane of the sky and the line connecting the centres of
  the two sub-clusters. The horizontal solid line represents the sum
  of the pairwise projected masses as estimated in
  section~\ref{MassE}; the associated 1$\sigma$ confidence level
  domain is limited by the two dot-dashed lines.  {\bf Top left (cases
  a1, a2, a3 in Table~\ref{twobody}):} the solid line shows the
  bound-incoming and -outgoing solutions, while the dashed line
  corresponds to the unbound solutions. These solutions have been
  estimated under the hypothesis that the two systems were at zero
  separation 12.6~Gyr ago (i.e. the age of the Universe in our
  cosmology). {\bf Top right (case b):} the same but assuming that the
  two systems were at zero separation 0.3~Gyr ago. The only possible
  configuration shows that the two clumps are bound and the system is
  expanding. {\bf Bottom (cases c and d1, d2, d3):} same as in the top
  right panel, but by assuming that the collision took place 0.5
  (left) and 1.0 (right) Gyr ago respectively.}
\label{BU}}
\end{figure*}

We close the system of equations taking our measured values of
$R_p{\simeq}$0.74~Mpc and $\Delta{v}$=89~km/s, and assuming different values
for $t_0$, i.e. the epoch of the last encounter between the two clumps
(Gregory \& Thompson 1984 and Barrena \etal2002). This implies in each case a
relationship between the total mass of the system $M$ and the angle $\alpha$
which is displayed in Fig.\ref{BU}.

In the first case, we set $t_0 = 12.6$ Gyr, i.e. the age of the Universe in
our cosmology. This suggests that subclusters are moving apart or together for
the first time. We are then in the pre-merger scenario. In the case of A3921,
the measured total mass of the two sub-clusters is $M=5.2 ^{+1.6} _{-0.1}
{\times} 10^{14}~h_{75}^{-1}~{M}_{\odot}$ (Sect.~\ref{MassE}). The top, left
panel of Fig.~\ref{BU} shows that the possible solutions for this value are
bound.  In the incoming case, two solutions are possible, with respectively a
very high and a very low value for the projection angle $\alpha$, while in the
outgoing case only a high value of $\alpha$ is allowed (cases (a)
Table~\ref{twobody}).

We then suppose we are witnessing the sub-clusters after the collision,
testing the possible solutions for $t_0 = $0.3 Gyr (case (b)), 0.5 Gyr(case
(c)), and 1 Gyr (cases (d)). In all the cases, the possible solutions are
bound as shown in Fig.~\ref{BU} (top, right and bottom panels). For both the
smaller values of $t_0$ (0.3 and 0.5 Gyr), there is only one outgoing solution
possible, leading to increasing values of $\alpha$ for higher $t_0$ (see
Table~\ref{twobody}). At larger times after merging ($t_0=1$ Gyr), three
values of $\alpha$ are possible, two for incoming solutions and one in the
outgoing configuration (see Table~\ref{twobody}). At this stage, apart from
the fact that the sub-clusters are bound with a very high probability, it is
difficult to discriminate between the various merging scenarios from the
dynamical analysis of the galaxies alone.

\begin{table*}
\begin{center}
\begin{tabular}{lccccc}
\hline
\hline
Scenario & ${\rm T}_0$ & $\alpha$ &  V & R & Solutions\\
 & [Gyr] & [deg]  & [km/s] & [Mpc] &  \\
\hline
\multicolumn{6}{c}{ }\\
a1 &12.6 & 84.3 & 89.4 & 7.5 & outgoing \\
a2 &12.6  & 83.5 & 89.6 &  6.5 &  incoming \\
a3 &12.6   & 2.2 & 2318.4 & 0.7 & incoming \\
b & 0.3 & 4.9 & 1041.9 & 0.7 & outgoing \\
c & 0.5 & 27.8 & 190.8 & 0.8 & outgoing \\
d1 & 1.0 &  55.2 & 108.4 & 1.3 & outgoing \\
d2 &  1.0   & 50.6 & 115.2 & 1.2 & incoming \\
d3 &  1.0  & 4.7 & 1086.2 & 0.7 & incoming \\
\multicolumn{6}{c}{ }\\
\hline
\end{tabular}
\caption{Col.1: name of the scenario in the text -- Col.2: time since last
interaction between the two clumps -- Col.3: angle between the plane
of the sky and the line connecting the centres of the two clumps --
Col.4: relative velocity between the two clumps -- Col.5: spatial
separation between the two systems -- Col.6: state of the systems for the
possible solutions}
\label{twobody}
\end{center}
\end{table*}

\section{Spectral properties of the cluster members}\label{Sect_EL}

\subsection{Measurements of the equivalent widths}

In the following, we wish to assess the star formation properties of
the galaxies belonging to A3921, and in particular to identify
galaxies with signatures of recent star activity from an older
population, as well as galaxies with emission lines, reflecting
present star formation. This analysis is based on our high S/N spectra
catalog, therefore restricted to 83 A3921 members.  To reach this
goal, the first step is to classify objects of our spectroscopic
catalog of galaxies.  Various methods have been used in the
literature, using generally the presence and strength of the [OII]
($\lambda$=3727~\AA) line and of Balmer lines (typically one or a
combination of ${\rm H}_\delta$, ${\rm H}_\gamma$, and ${\rm H}_\beta$
lines).  In principle, as shown by Newberry \etal1990, the most robust
approach to detect post-starburst galaxies would make full use of all
three Balmer lines.  However, our limited spectral range does not
allow us to include ${\rm H}_\beta$ in numerous cases, and the S/N
ratio of the ${\rm H}_\gamma$ line is generally poor. Therefore, we
use the combination of [OII] and of the ${\rm H}_\delta$ line to
establish our spectral classification, as in Dressler \etal(1999).

We apply two methods for measuring the equivalent widths of these
lines, using respectively the tasks ``splot'' and ``sbands'' in IRAF
on our continuum normalized spectra. The continuum is fitted by a
spline function of high order (generally 9), and the quality of the
best fit verified for each spectrum. In particular, special attention
is paid to the good quality of the fit in the vicinity of the line to
measure, which is crucial for a reliable measurement of the EW. While
``splot'' fits a gaussian for the line, ``sbands'' measures the
decrement or increment in signal relative to the fitted continuum in
an interval containing the line to be measured. We check the
consistency of the measures with the two methods, and choose the
Gaussian fitting technique because it gives more accurate measurements
for the following analysis.  These measurements are listed in
Table~\ref{TabTOT}, with equivalent widths of absorption and emission
features defined as positive and negative respectively.  We estimate
the minimum measurable EW of each spectrum as in Barrena
\etal(2002), taking the width of a line spanning 3.1~\AA~(our
dispersion) in wavelength, and with an intensity three times the noise
r.m.s. in the adjacent continuum. A minimum (maximum) measurable EW of
$\sim$~2.8(-2.8)~\AA~results in the case of an absorption (emission)
lines.

\subsection{E+A galaxies}\label{CompSpec}

The fraction of poststarburst galaxies in galaxy clusters, (the so
called E+A's objects; Dressler \& Gunn 1983, also referred more
recently as ``k+a'' and ``a+k'' types by Franx 1993, Dressler
\etal1999, or ${\rm H}\delta$-strong galaxies by Couch \& Sharples 1987) has
been a quite controversial topic in the last decade.  While the
general property of these objects is the presence of a young stellar
population (A type) superimposed to an older one, a strict consensus
does not yet exist between the authors on the criteria used for their
definition as shown in Table 3 of Tran \etal2003. This can lead to
significant differences when estimating and comparing the fraction of
these objects in various samples of clusters.  The high frequency of
these objects at high redshift (up to $\sim$26\% in the MORPHS survey
Dressler \etal1999) is at contrast with the very low fraction obtained
at low redshift (around $1\%$, Dressler 1987).  These objects are
considered to have undergone a recent (younger than 1.5 Gyr) activity
of star formation, followed by a quiescent phase.  The situation at
intermediate redshifts is still debated, with low fractions obtained
by Balogh \etal1999 on the CNOC1 survey ($\leq 5\%$ for
0.18${\leq}z{\leq}$0.55), to higher values of $\sim 7\%-13\%$ obtained
by Tran \etal2003. However these discrepancies could be explained by
differences in the selection criteria and in correction for
incompleteness.  This population of E+A galaxies was suggested to be
recently accreted field galaxies, with suppression of the star
formation activity by ram pressure stripping by the ICM.  The
evolution of the fraction of such objects with redshift has often been
interpretated as a consequence of both a higher star formation density
and infall rate (Kauffmann 1995) at high redshift.  Recent analyses
also show a luminosity/mass effect, with an observed decrease in the
characteristic E+A mass (see Poggianti \etal2004, and Tran
\etal2003). We will present hereafter the analysis of the E+A galaxies
in A3921.

\subsubsection{Selection criteria and fractions}

In a first step, we adopt the criteria defined by Dressler \etal1999,
dividing E+A galaxies in k+a's and a+k's, and defining k+a and a+k
objects from spectra presenting an intermediate and strong ${\rm
H}_\delta$ line in absorption respectively, with no detectable [OII]
line in emission (with the criteria ${\rm EW([OII])}>-5 \AA$, and $3
\AA < {\rm EW({\rm H}_\delta)} < 8 
\AA$ for k+a, ${\rm EW({\rm H}_\delta)} > 8 \AA$ for a+k). We detect 6 k+a
objects in our sample, corresponding to a fraction of $7.2\%$ of the
sample, and no a+k galaxies. As we are interested in all objects with
signs of recent star formation activity, we also consider spectra
presenting a clear inversion of the intensities of the K and H calcium
lines, due to the presence of a blend of the H line with the Balmer
line ${\rm H}\epsilon$.  We measured the ratio of the line intensities
${\rm H}+{\rm H}\epsilon/{\rm K}$, shown by Rose {\it et al.}  (1985)
to have a low value for young populations and to reach a plateau at $\sim
1.2$ for older ones. We select a sample of objects with a ratio of
${\rm H}+{\rm H}\epsilon/{\rm K} < 1$ as candidates of k+a galaxies
(coded in our Table~\ref{TabTOT} with k+a?). We detect 11 objects,
with four of them included in the previous sample (${\rm EW({\rm
H}_\delta)} > 3$), and the remaining ones spreading within the range
($2.5 <{\rm EW({\rm H}_\delta)} < 3$).  When taking altogether these
objects as candidates of galaxies having recently undergone star
formation, we obtain a fraction of 15.6\% of our sample.  A
quantitative comparison of the observed fraction of active galaxies in
A3921 to those measured in other clusters is however difficult as it
can be affected by several factors like the size of the field where
spectroscopy was performed, the spatial sampling of the galaxies or
the way active galaxies were defined.  Nevertheless, the measured
fraction of E+A galaxies seems to be relatively large for a cluster at
low redshift.

\subsubsection{Luminosity effect}

In order to compare our results to those of Dressler \etal1999, we
convert their spectroscopic limit $M_V=-19.0+5logh$ to our $R_{AB}$
band, taking into account the different cosmological models used. This
leads to $M_{R_{AB}} = -20.0$ (using a mean $V-R =0.5$ as in Poggianti
\etal1999, and $R_{AB}-R =0.193$). Applying this magnitude cutoff, we
obtain respectively 1 secure k+a galaxy, and 3 k+a candidates.

\begin{figure}
\resizebox{8cm}{!}{\includegraphics{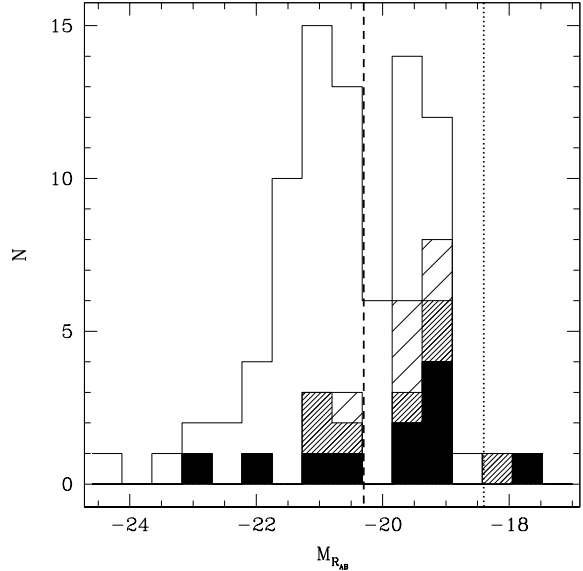}}
\hfill
\parbox[b]{8cm}{
\caption{The histogram in absolute magnitude in $R_{AB}$ band of our high
  S/N spectroscopic sample (83 objects). The various fractions of
  galaxies of different spectral types in each bin are encoded as
  follows: passive (white), k+a (less shaded), k+a? (more shaded) and
  emission-line galaxies (black). The dashed and dotted lines shows
  the magnitude limits corresponding to the MORPHS (Dressler
  \etal1999) and "giants/dwarfs" (Poggianti \etal2004) cuts
  respectively. Note that the contribution of active (k+a, k+a? and
  e=emission line) galaxies creates a second peak at fainter
  magnitudes. }
\label{dwarfs}}
\end{figure}

As seen from the histogram displayed in Fig.~\ref{dwarfs}, our secure
k+a and "k+a candidates" (k+a?) are spread in the magnitude range
$-21.1 {\leq} M_{R_{AB}} {\leq} -18.0$, with the majority of objects
lying between $-20.0 < M_{R_{AB}} < -19.0$. Taking $ M_{R_{AB}} = -20$
as a frontier in absolute magnitude, we consider two sub-samples
corresponding to faintest and brightest objects, and analyse the
respective fraction of different spectral types. The results are
listed in Table~\ref{tab:spectra}.  While the k-spectra galaxies
dominate the ``bright'' sample, the fraction of secure k+a and k+a
candidates increase clearly in the ``faint'' one.  The luminosity
distribution of active galaxies in A3921 appears intermediate between
the MORPHS sample (Dressler \etal1999) and the Coma one (Poggianti
\etal2004). The fraction of bright k+a's ($1.9\%$ secure case, $7.7\%$
when including candidates) is lower than MORPHS but higher than
Poggianti \etal2004 who found no object at such magnitudes. In the latter
case however, the k+a distribution is dominated by a population of
dwarf galaxies (fainter than $M_{R_{AB}} =-18.4$ in our case) which is
scarcely tested by our shallower survey.  We clearly detect an
evolution of the typical luminosity of our k+a population, which is
brighter than in Coma but fainter than that detected by MORPHS.

\begin{table*}
\begin{center}
\begin{tabular}{cccccc}
\hline
\hline
Sample & ${\rm N}_{\rm tot}$  & k & k+a & k+a? & Emission line \\
\hline
Bright& 52 & 44 (84.6\%) & 1 (1.9\%) & 3 (5.8\%) & 4 (7.7\%)\\
Faint & 31 & 15 (48.4\%) & 5 (16.1\%) & 4 (12.9\%) & 7 (22.6\%)\\
Total & 83 & 59 (71.1\%) & 6 (7.2\%) & 7 (8.4\%) & 11 (13.3\%) \\
\hline
\end{tabular}
\caption{Number of spectra in different magnitude sub-samples. Bright
 and faint galaxies correspond to objects with $ M_{R_{AB}}<-20$ and $
 M_{R_{AB}}>-20$ respectively. Bright objects correspond to the same
 sampling in absolute magnitude as in the MORPHS spectroscopic
 catalogue (Dressler \etal1999) }
\label{tab:spectra}
\end{center}
\end{table*}

\subsubsection{Colours}

\begin{figure}
\resizebox{8cm}{!}{\includegraphics{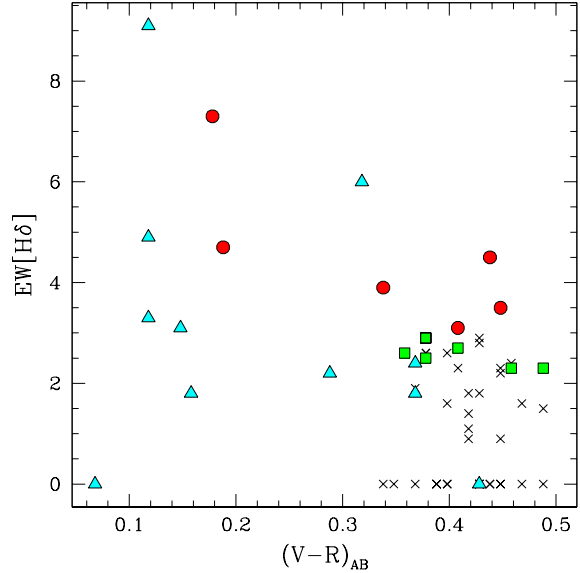}}
\hfill
\parbox[b]{8cm}{
\caption{Colour-${\rm H}\delta$ diagram for galaxy members of A3921. Crosses
indicate k type galaxies, squares k+a type galaxies, circles galaxies with
H-K inversion and triangles emission lines galaxies.}
\label{Hdcol}}
\end{figure}

However, in contrast to Coma, Fig.~\ref{cmd} shows that most of the
k+a galaxies of our spectroscopic sample have red colours, as
indicated by their position in A3921 colour-magnitude diagram (circles
in Fig.~\ref{cmd}).  This can also be seen in Fig. ~\ref{Hdcol}, which
displays the ${\rm H}\delta$ equivalent width as a function of the V-R
colour as in Poggianti \etal1999.  Among the 13 detected k+a and k+a
candidate galaxies, 11 have V-R colours spreading the same range as
the k galaxies (0.35-05), while only 2 show clearly bluer colours.
Moreover, only one k+a galaxy shows ${\rm EW(H_{\delta})}> 5 \AA$
(Fig.  ~\ref{Hdcol}), most objects displaying ${\rm EW(H_{\delta})}$
relatively low, comparable to those of the ``redder k+a'' Coma
subsample of Poggianti
\etal2004 and significantly weaker than the EWs of ${\rm H}_{\delta}$ lines
detected in their bluer k+a galaxies. We do not detect this young k+a blue
population in A3921, where galaxies with k+a spectra look rather like
resulting from an episode of star forming or star-bursting occurring $\sim 1 -
1.5 \rm ~Gyr$ ago, with redder colours and smaller Balmer lines. 

\subsubsection{Spatial and velocity distributions}

\begin{table*}
\begin{center}
\begin{tabular}{cccc}
\hline
\hline
Subsample & $N_{gal}$ & ${C}_{BI}$ & ${S}_{BI}^*$ \\
 & & [km/s] &  [km/s]  \\
\hline
\multicolumn{4}{c}{ }\\
k & 59 & $27988 ^{+86} _{-111}$ & $ 770^{+119} _{-96}$    \\
\multicolumn{4}{c}{ }\\
k+a & 6 & $27920 ^{+347} _{-386}$ & $597 ^{+86} _{-62}$  \\
\multicolumn{4}{c}{ }\\
k+a? & 7 & $27839  ^{+1402} _{-305}$ & $1188 ^{+318} _{-131}$  \\
\multicolumn{4}{c}{ }\\
k+a + k+a? & 13 & $27805  ^{+353} _{-265}$ & $964 ^{+285} _{-143}$  \\
\multicolumn{4}{c}{ }\\
e & 11 & $27941  ^{+522} _{-219}$ & $1607 ^{+314} _{-219}$ \\
\multicolumn{4}{c}{ }\\
\hline
\end{tabular}

\caption{Properties of the $cz$ distribution for the different
spectral types. $^*$ In the case of the two separate k+a and k+a?
subsamples we use the classical variance estimator for velocity
dispersion and not the biweight indicator, due to the low number of
objects (Beers \etal1990). }
\label{ROST_Tab1}
\end{center}
\end{table*}

\begin{figure}
\centering
\resizebox{8cm}{!}{\includegraphics{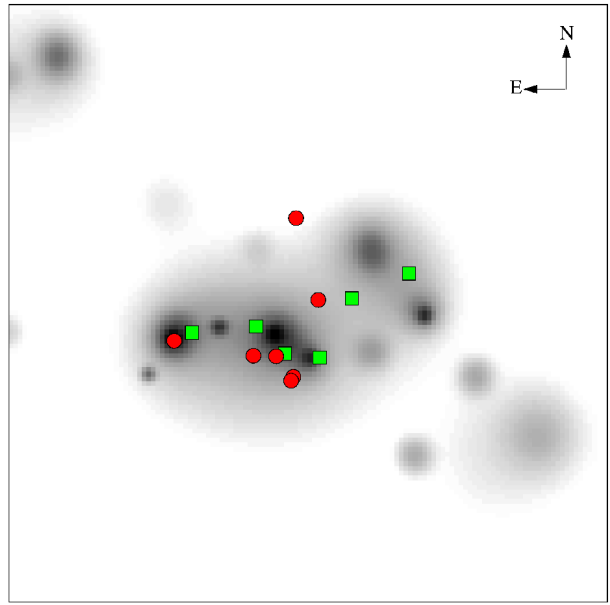}}
\parbox[b]{8cm}{
\caption{
  Projected galaxy density map (34$\times$34 ${\rm arcmin}^2$) of the
  red-sequence galaxies with $R_{\rm AB}<19$ (see Fig.\ref{IsoDMapsGR}).
  The squares represent galaxies that are members of A3921 and classified as
  k+a and the circles as "k+a?" (see text).}
\label{IsoDka}}
\end{figure}

\begin{figure*}
\centering
\resizebox{18cm}{!}
{\includegraphics*{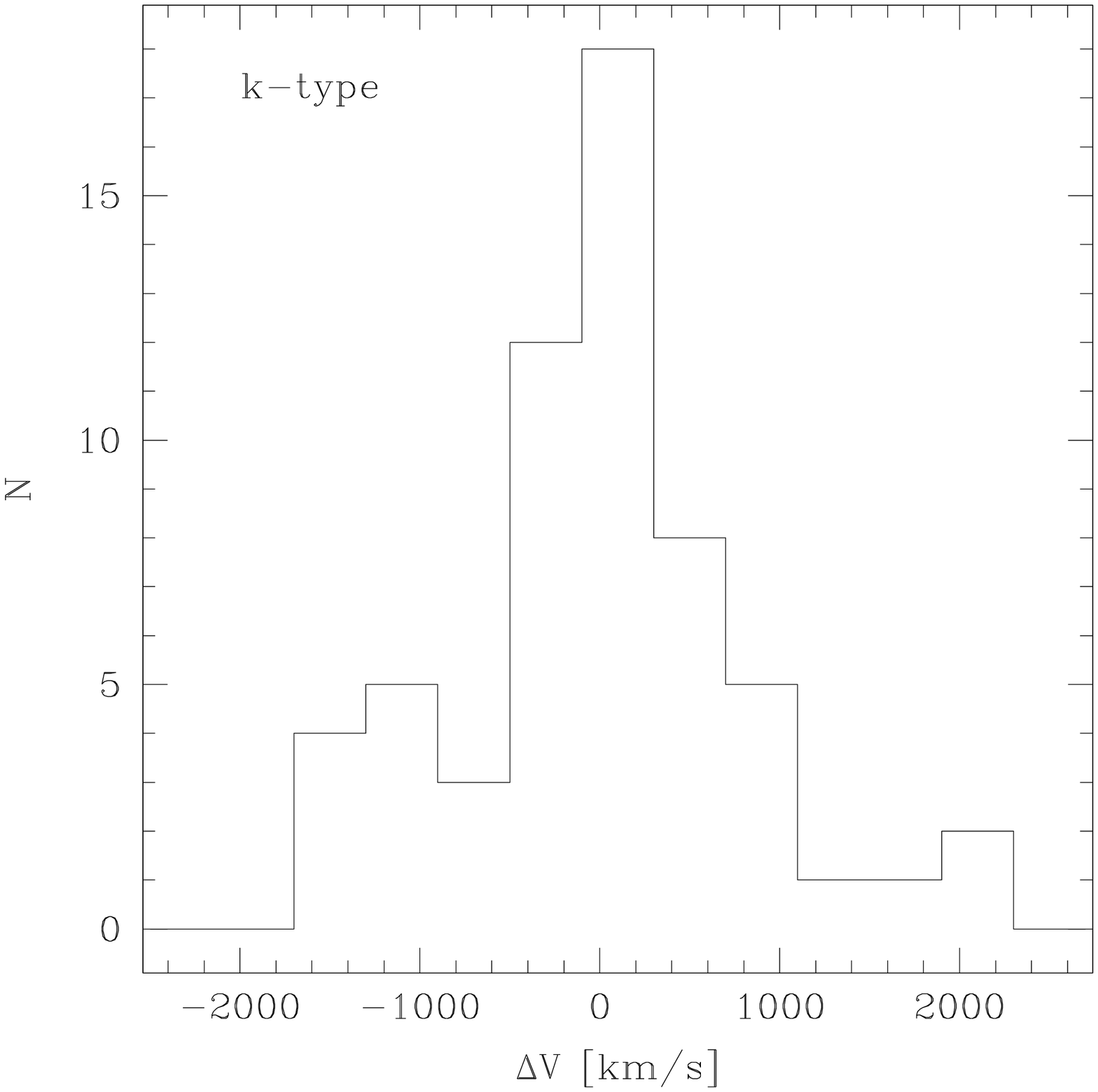}
\includegraphics*{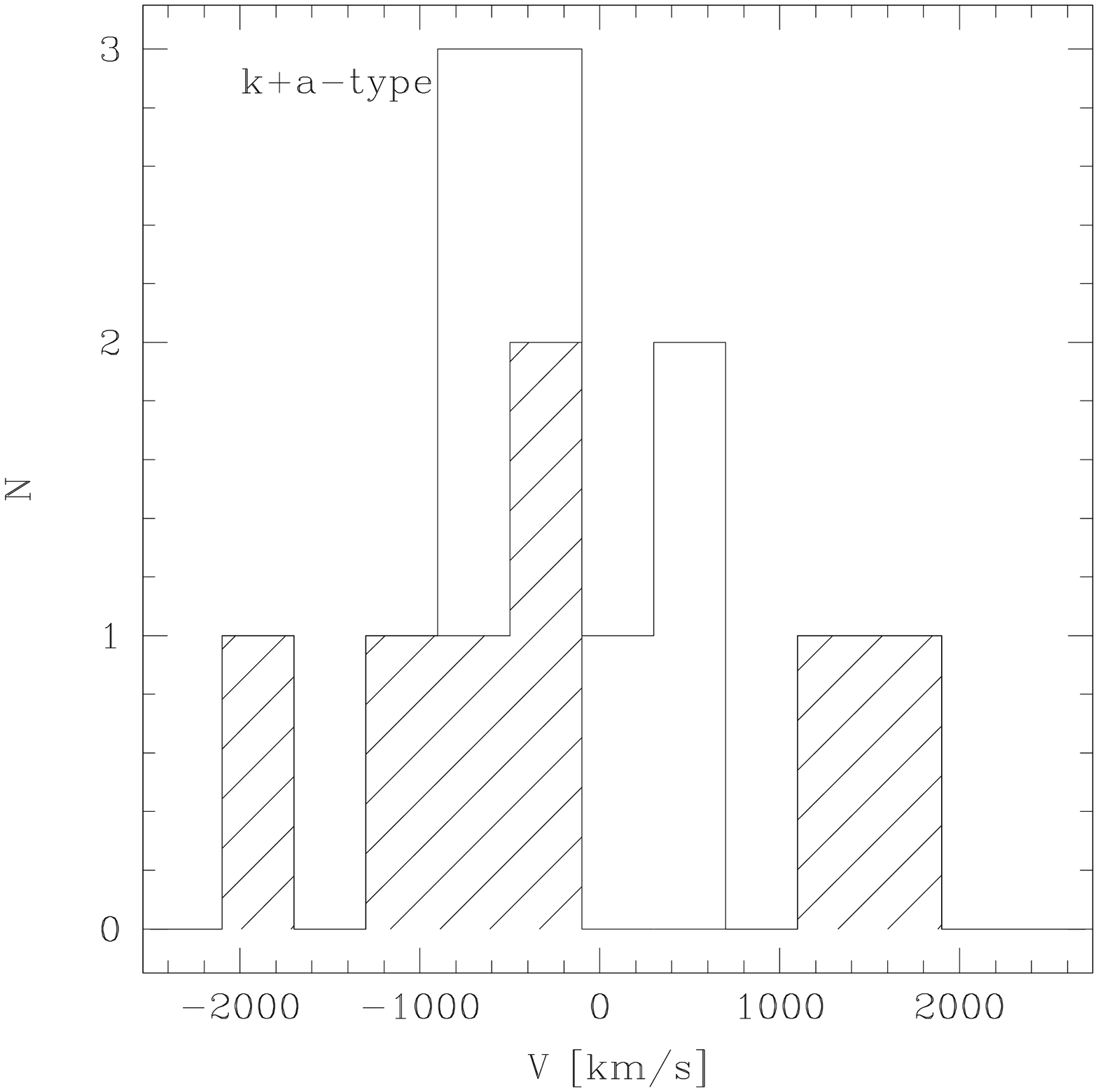}
\includegraphics*{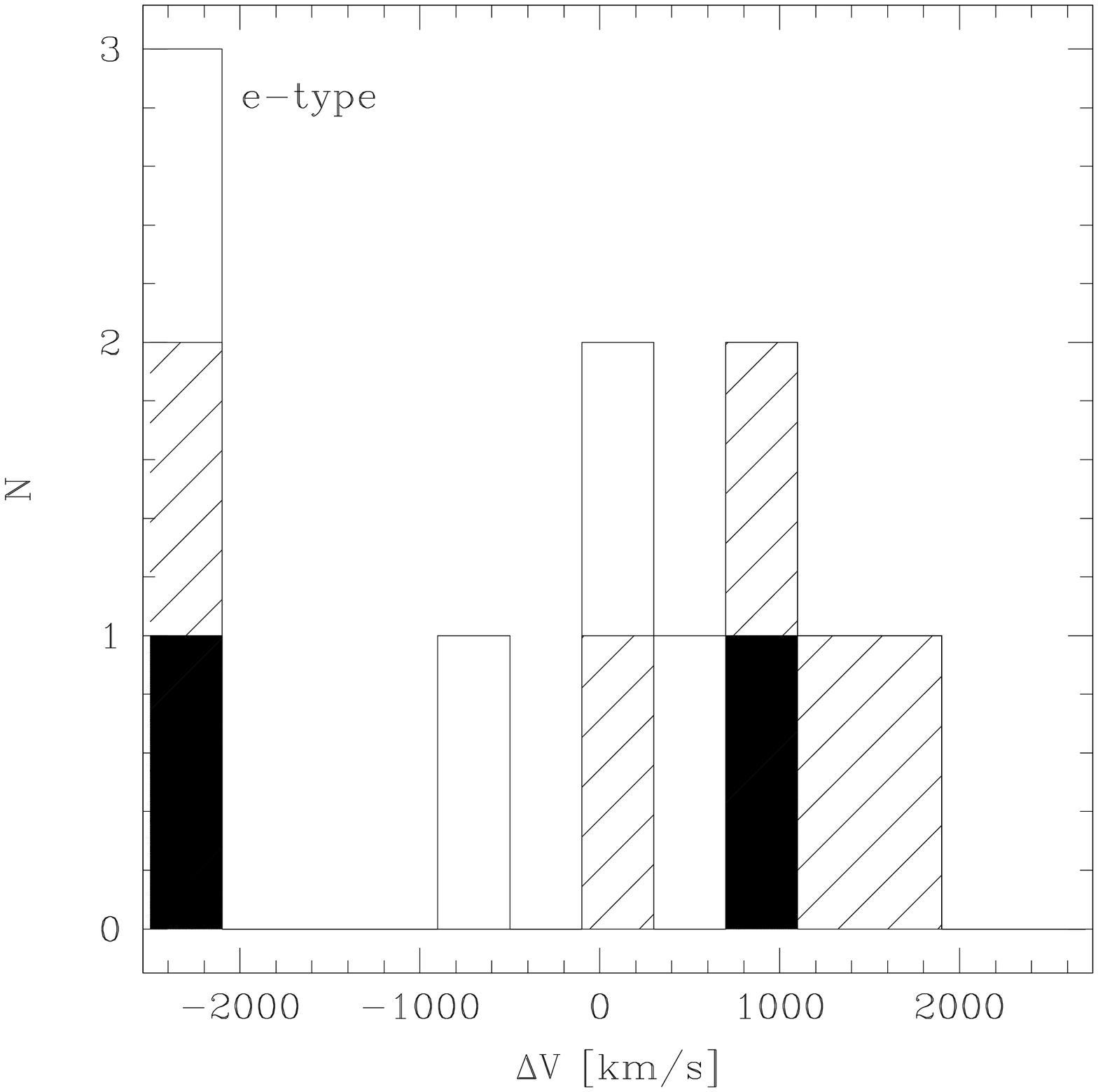}}
\parbox[b]{18cm}{
\caption{ From left to right, velocity distribution (with a binning of
  400 km/s) of k, k+a and emission line galaxies (according to the
  definition of Dressler \etal1999, see text for details).  In the
  central figure, the white component corresponds to secure k+a's and
  the shaded part to k+a candidates. In the right panel, the shaded
  component of the histogram correspond to e(a)'s, the black component
  to e(b)'s, the white part to e(c) type objects. }
\label{VelDD}}
\end{figure*}

In Fig.~\ref{IsoDka} we show the projected positions on the sky of secure
k+a's (squares) and k+a candidates (circles). From a simple visual inspection
of this figure, most of k+a galaxies seem to be concentrated in projection
around the central field of A3921-A. This is particularly clear in the case of
k+a candidates, as only one of them is locate elsewhere, in the cluster
outskirts. Confirmed k+a's are less clustered around the central field of
A3921-A. They are however concentrated in the highest density regions of the
whole cluster and roughly aligned along an East/West direction.  Applying
a bi-dimensional Kolmogorov-Smirnov test, we cannot exclude that
k+a's spatial distribution is significantly different from that of k-type
objects with a quite high significance level (87\%). An intermediate and
therefore not conclusive significance level (26\%) is on the contrary obtained
by comparing the projected positions of k+a candidates and passive galaxies.

The velocity distribution of the whole k+a sample is shown in the middle panel
of Fig.~\ref{VelDD}. According to a visual inspection and usual statistical
tests (i.e. 1D-KS, Rank-Sum, Hoel 1971, and F-tests, Press \etal1992), it is
not significantly different from the k-type's velocity distribution (left
panel). k+a candidates (shaded in Fig.~\ref{VelDD}) are characterized by a
significantly higher velocity dispersion than passive and confirmed k+a
galaxies (96\% and 93\% c.l. respectively according to an F-test), very close
to the whole cluster value (1188 km/s $vs$ 1008 km/s, see
Tables~\ref{ROST_Tab1} and~\ref{ROST_Tab}). Mean velocities of the different
sub-samples are largely comparable within 1$\sigma$ errors.

\subsection{Emission line galaxies}

\begin{figure}
\resizebox{8cm}{!}{\includegraphics{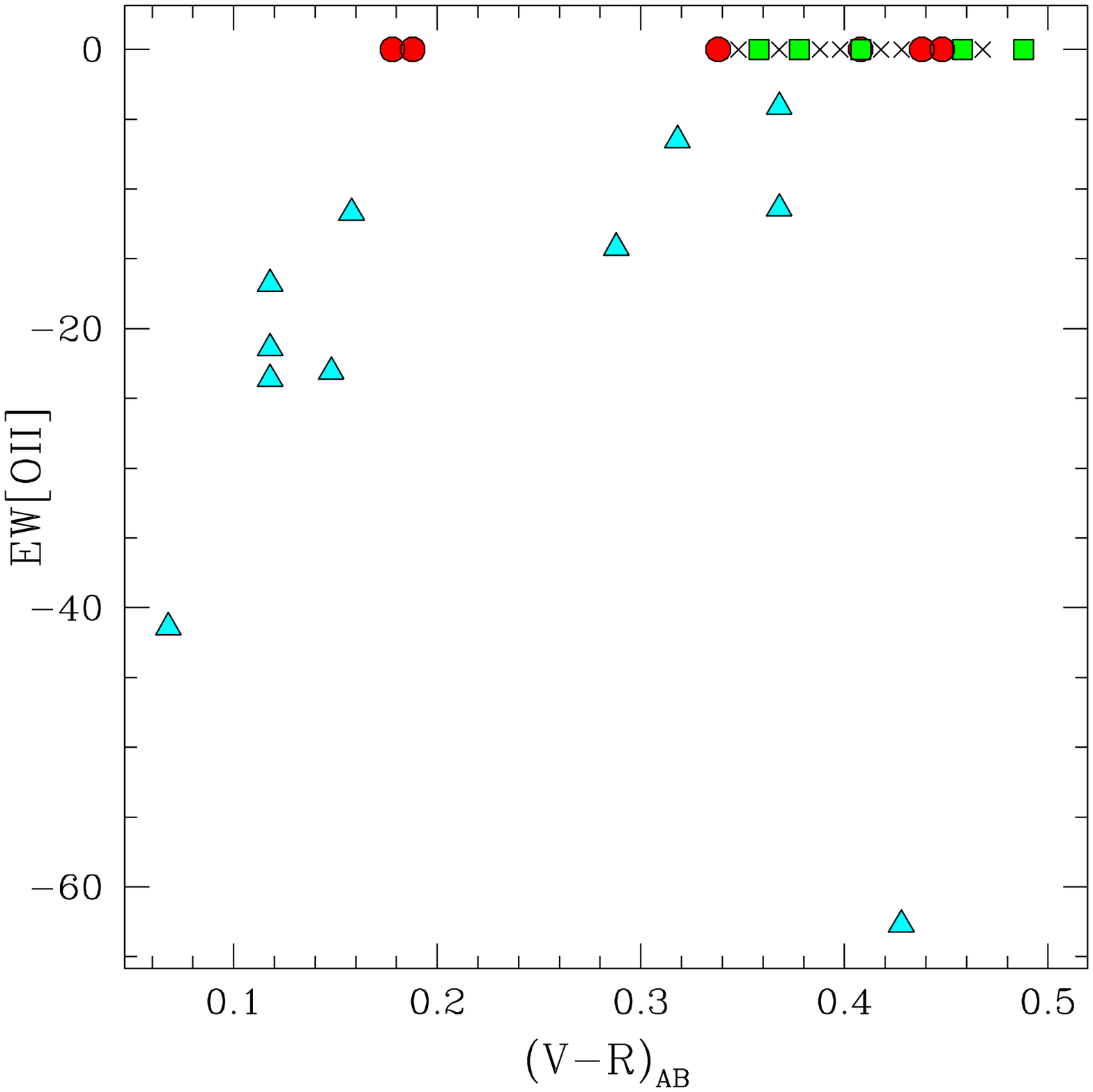}}
\hfill
\parbox[b]{8cm}{
\caption{Colour-[OII] diagram for galaxy members of A3921. Crosses
indicate k type galaxies, squares k+a type galaxies, circles galaxies with
H-K inversion and triangles emission lines galaxies.}
\label{OII_col}}
\end{figure}

The emission-line population also shows interesting features, both in
its fraction and colour distribution. We find 11 emission line
galaxies, among which 2 starbursts (e(b)), 4 classical constant rate
forming stars spirals (e(c)), and 5 e(a)-type objects, that have been
suggested to be dust-starbursts (Poggianti {\it et al.}, 1999). The
global fraction of emission line galaxies (13\%) is lower than that
estimated at high redshift from MORPHS (26.5\%), but comparable to
that measured at intermediate redshifts ($0.18<z<0.55$) in the CNOC1
survey ($\sim$16\%, Ellingson
\etal2001).  Fig.~\ref{OII_col} shows the ${\rm EW([OII])}$ as a
function of V-R colour.  The sample can be divided into two groups: the
bluest ones with $V-R < 0.3$, and typically large absolute values of
${\rm EW([OII])}$, ($ < -10 \AA )$, and the reddest ones with colours
typical of the passive population $0.3 < V-R < 0.6$, and lower
absolute values of ${\rm EW([OII])}$, $(\geq -10 \AA)$, except for one
object with an exceptionally strong [OII] line (${\rm EW([OII])} = -65
\AA$), and a red colour index $V-R\sim 0.42$, which is identified with
BG3. The presence of this red population of galaxies presenting
emission lines is quite unusual, and reminiscent of what observed in
the case of the merging cluster RX J0152.7-1357 at $z=0.837$ analysed
by Demarco \etal2004.  This suggests that an on-going episode of star formation
or bursting is occurring in an older population.

\subsubsection{Spatial and velocity distributions}\label{eml_dist}

\begin{figure}
\centering
\resizebox{8cm}{!}{\includegraphics{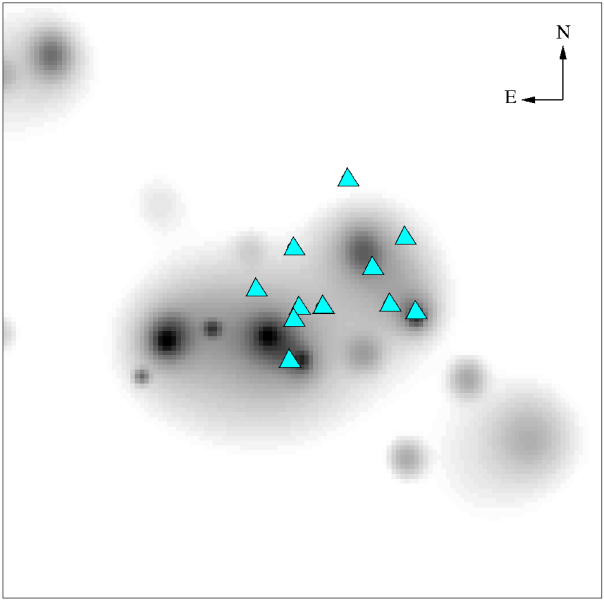}}
\parbox[b]{8cm}{
\caption{
  Projected galaxy density map (34$\times$34 ${\rm arcmin}^2$) of the
  red-sequence galaxies with $R_{\rm AB}<19$ (see Fig.\ref{IsoDMapsGR}).
  The triangles show the location of emission line cluster member galaxies
  (see text).}
\label{IsoDem}}
\end{figure}

Most (8/11) of star-forming galaxies seem to be spatially concentrated in the
central region of the whole cluster, i.e. on the NW side of A3921-A, on the SE
side and in the centre of A3921-B and between the two sub-clusters (see
Fig.~\ref{IsoDem}). A Rank-Sum test confirms this visual impression, as
star-forming objects result more concentrated than k-type galaxies around a
central position between the two subclusters with a very high c.l. (99.7\%).
Moreover, according to a bidimensional KS test, the spatial distributions of
e-type and k-type galaxies are not drawn from the same parent population (2\%
significance level).

Fig.~\ref{VelDD} shows that emission line galaxies have a much more
dispersed velocity distribution with respect to passive
objects. According to the F-test, velocity dispersion of emission line
galaxies ($1607 ^{+314}_{-219}$ km/s) is significantly higher than the
corresponding value for k-type objects ($770 ^{+119}_{-96}$ km/s),
with a $\sim$100\% confidence level.

Most of star-forming objects are therefore distributed between the two
sub-clusters, and their velocity distribution is quite more dispersed
than that of the relaxed k-type galaxies (whose velocity distribution
does not differ significantly from a Gaussian), but with a comparable
mean velocity. This suggests that, at least for a fraction of them,
the star formation episode could be the result of an interaction with
the ICM, connected with the merging process of the cluster.

\subsection{SFR from [OII] equivalent widths}

We will hereafter derive the SFR of the galaxies within A3921 from our
measurements of the [OII] equivalent widths of our galaxy spectra. The
uncertainties in the SFR estimated from [OII] are known to be larger
than when derived through ${\rm H}\alpha$. This is due to the fact
that, while flux estimates from ${\rm H}\alpha$ scale with the number
of ionizing stars, estimates from [OII] are hampered by systematics
due to large variations in excitation, implying substantial
dispersions in the [OII]/${\rm H}\alpha$ and in the [OII]/${\rm
H}\alpha$+[NII] correlations. In our case, however, [OII] is the only
tracer of SFR available in our domain of wavelength, and will be used
with caution.  Using the equation derived by Kennicutt (1992):

\begin{equation}
SFR (M\subsun {\rm yr}^{-1}) \sim
2.7~\times~10^{-12}~\frac{L_B}{L_{B\sun}}{\rm EW(OII)}E({\rm H}\alpha)
\end{equation}

\noindent the star-formation rate (SFR) is directly estimated from the
integrated broad-band B luminosity, the [OII] equivalent width,
assuming a reasonable value for the extinction correction $E({\rm
H}\alpha)$. Converting our V-band magnitudes in B-band using
$(B-V)_{AB}=0.65$ for the objects belonging to the red sequence and
$(B-V)_{AB}=0.30$ for bluer objects, and taking a solar luminosity
$M_B=5.48$, as well as the canonical value for extinction (1 mag for
${\rm H}\alpha$) used in Kennicutt 1992, we obtain an estimate of the
SFR for our 11 galaxies with [OII] emission line.  From the majority
of spectra within A3921, the SFR ranges from 0.14 to $2.40 M\subsun
{\rm yr}^{-1} $, with however one object with very high SFR ($\sim
19.2 M\subsun {\rm yr}^{-1}$) corresponding to the galaxy BG3
previously quoted.  We estimate a median SFR of $\sim 0.58$ and a mean
SFR of $\sim 0.72 M\subsun {\rm yr}^{-1}$ (BG3 excluded).  The
cumulated SFR for the whole cluster reaches $\sim 11.5 M\subsun {\rm
yr}^{-1}$ (BG3 excluded).  As expected, the values show that star
formation in A3921 is suppressed as compared to field galaxies (Tresse
and Maddox 1998). The highest optical SFR in our data apart from the
BG3, ($2.40 M\subsun {\rm yr}^{-1} $), is intermediate between the
values obtained in A114 at $z=0.32$ by Couch \etal2001 ($4 M\subsun
{\rm yr}^{-1} $), and that obtained by Duc et al. 2002 ($1.15 M\subsun
{\rm yr}^{-1} $) in Abell 1689 at $z=0.18$. However, the SFR in Couch
\etal2001 were measured with ${\rm H}\alpha$, and, as shown in Duc et
al. 2002, measurements of the SFR with [OII] can be seriously
underestimated due to dust extinction. Therefore our values can be
considered as a lower limit of the SFR in A3921.

BG3 shows an exceptionally high star formation rate, suggesting a very
active star-bursting object. The spectrum shows strong absorption
lines typical of an old stellar population (Ca K and H, G band, $4000
\AA$ break) as well as emission lines ([OII], ${\rm H}\gamma$ 
and ${\rm H}\beta$).  Additional observations are planned to further
investigate the nature of this object and to test if emission is due
to star formation or to an active galactic nucleus (only the bluer
part of the spectrum was obtained in these observations and the
equivalent width of [OIII] is necessary to perform the test).

\section{Discussion and conclusions}

\subsection{Dynamical state of the cluster}

Using our new spectroscopic and VRI-bands photometric catalogues of
the central $\sim~1.8{\times}1.2~{\rm Mpc}^2$ region of A3921, we
detect two main sub-clusters of galaxies: a South-East clump (A3921-A)
that hosts the BCG (BG1), and, at $\sim$~0.74~Mpc distance, a
North-West system (A3921-B), hosting the second brightest cluster
galaxy (BG2) and the third brightest object (BG3) in its
outskirts. From the analysis of the projected density distribution,
there are clear signs of merging events within this cluster.  The
density distribution of galaxies of A3921-A shows an elliptical shape,
while A3921-B is characterized by a very morphologically disturbed
distribution, with an extended low-density tail extending towards
South. This morphology suggests that sub-cluster B is approaching
cluster A from the North or exiting after colliding it from the
South. The fact that the sub-cluster B is well defined would suggest
that the merging has not yet occurred, but in case of a high impact
parameter, or of a low mass ratio between the secondary and the main
component, the density structure of the sub-clusters may survive.  On
the other hand, the presence of the Southern tail of subcluster B
rather suggests a post-merger event. In this case, the tail would be
made of faint galaxies lagging behind the brightest objects during
their tangential crossing of subcluster A.

\begin{figure}
\centering
\resizebox{8cm}{!}{\includegraphics{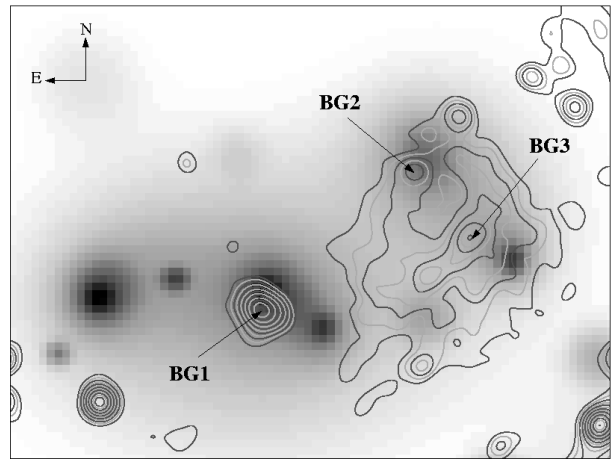}}
\parbox[b]{8cm}{
\caption{X-ray residuals after subtraction of a 2D-$\beta$ model 
(Belsole \etal2004)
  overlaid on the red sequence galaxy density map of the central part of 
A3921
($22\times 18 {\rm arcmin}^2$).  }
\label{OttX}}
\end{figure}

However, in spite of clear merging signatures on the density
distribution, the kinematical and dynamical properties both of the
whole cluster and of the two subclusters do not show strong signatures
of merging. Moreover, the two sub-clusters show a very similar mean
projected velocity. When applying a two-body dynamical model to the
two components, several solutions could explain the observed dynamics
of A3921 allowing both the pre-merging and the post-merging cases. A
comparison with the X-ray properties of the cluster can allow us to
discriminate between them. The analysis of XMM-Newton observations by
Belsole \etal(2004) reveals that the X-ray emission of A3921-A can be
modelled with a 2D$\beta$-model, leaving a distorted residual
structure toward the NW, coincident with A3921-B (see
Fig.~\ref{OttX}). The main cluster detected in X-ray is centred on the
BCG position (BG1), while the X-ray peak of the NW clump is offset
from the brightest galaxy of A3921-B (BG2) (see Belsole
\etal2004). The temperature map of the cluster shows an extended hot
region oriented parallel to the line joining the centres of the two
sub-clusters. A comparison of this image with numerical simulations by
Ricker \& Sarazin (2001) suggests that we are observing the central
phases of an off-axis merger between two unequal mass objects, where
clump A is the most massive component (Belsole \etal2004), consistent
with optical results.

By combining the signatures of merging derived both from the optical
iso-density map and from X-ray results, we can now reconsider the
solutions of the two-body dynamical model (Table~\ref{twobody}). In
the pre-merger case, the high angle solutions of cases (a1) and (a2)
would imply a very large real separation for the two sub-clusters
($\sim 6-7$ Mpc), which is very unlikely taking into account the clear
signs of interaction between the two clumps observed both in optical
and in X-ray. In the "recent" post-merger case, we can also exclude
the (c) solution, as we expect a higher value of the relative velocity
($\geq$1000 km/s) between the two clumps for obtaining a so evident
hot bar in temperature map. Finally, the comparison of observed and
simulated galaxy density and temperature maps (e.g. Schindler \&
B\"ohringer 1993, Ricker \& Sarazin 2001) clearly exclude an older
merger (e.g. $t_0$=1 Gyr), as we do not expect anymore to observe a
clear bimodal morphology in optical and we should not detect such
obvious structure in the temperature map.  Therefore, only the
solutions corresponding to the very central phases of merging ($t_0
\approx \pm 0.3$ Gyr) can explain all our observational results, implying a
collision axis nearly perpendicular to the line of sight. This is
consistent with the absence of strong merging signatures in the
observed projected velocity distribution of cluster members.

The superposition of the X-ray residuals onto the optical iso-density
map (Fig.~\ref{OttX}) shows that the bulk of X-ray emission in A3921-B
is offset towards SW from the main concentration of galaxies. As
numerical simulations show that the non-collisional component is much
less affected by the collision than the gas distribution
(e.g. Roettiger \etal1993), this offset suggests that A3921-B is
tangentially traversing A3921-A along a SW/NE direction, with its
galaxies in advance with respect to the gaseous component. The
off-axis collision geometry has probably prevented total assimilation
of the B group into the main cluster A. This off-axis collision scenario
is also consistent with the shape of the feature in the temperature
map (Belsole \etal2004).

\subsection{Has the merging event affected star formation?}

We detect very few k+a (1 secure and 3 candidates) at the bright
absolute magnitude cut of MORPHS (Dressler \etal1999). Most of our k+a
and k+a? objects lie within $R_{AB}$ absolute magnitudes in the range
[-19/-20].  This effect is very similar to the trend in luminosity
detected by Poggianti \etal2004 in Coma, but the typical luminosity of
our k+a/k+a? objects is at least one magnitude brighter. However, we
do not sample the faint end of the luminosity function as in the case
of Coma, which may contain part of the undetected population.  We
therefore confirm the increase of the typical luminosity of the k+a
population with redshift as detected in Poggianti \etal2004, comparing
Coma and MORPHS clusters. We also find a significant fraction of k+a
in our low redshift cluster lying between the fractions obtained at
low and high redshift.  The mechanism responsible of this "redshift"
effect in A3921 is not immediate. This can be a "redshift" consequence
only due to some cosmic "downsizing" effect, as it has been shown that
the star formation activity at high $z$ was more efficient for more
massive galaxies. In this case, this effect could be explained only by
the infall of the galaxies into the cluster.  On the other hand, A3921
shows very strong sub-clustering and signatures of merging, which can
be suspected to be responsible, at least in some fraction, for the
enhancement of the fraction of such objects. This will be detailed in
the following.
  
The typical EW(H$\delta$) of the k+a/k+a? galaxies detected in A3921
is moderate and most of the objects have red colours,
indistinguishable from red sequence objects. We fail to detect the
population of blue k+a with strong Balmer lines as detected in Coma by
Poggianti \etal2004, which can only be reproduced by a starburst in a
recent past. In contrast, our objects rather reflect the evolution of
galaxies having undergone starburst or starforming activity, which has
been suppressed by some physical process, and now lie in the second
half of their lifetime (typically $<1.5$ Gyr), with redder colours,
and fainter Balmer lines, before reaching a k-type spectrum.  This
population can be reproduced by simply halting star formation produced
in continuous way, without evoking a strong star burst.  The k+a?
objects are probably galaxies having undergone a still older and
fainter last episode of star formation, as they do not have Balmer
lines strong enough to enter the k+a sample, but show clear signatures
of past activity.
  
A fraction of emission line galaxies comparable to that found in
intermediate redshift clusters has also been detected. The spatial and
kinematical distributions of k+a and k+a? as well as emission line
objects have been compared to the substructures evidenced both in the
whole galaxy and gas distribution. While k+a/k+a? galaxies are mostly
distributed among the main cluster A, the emission line galaxies
mostly lie in the region of the sub-cluster B, and in the region in
between A and B. A similar spatial distribution is followed by blue
galaxies, which are more clustered in the central region of A3921-B
than in the whole field and, in particular, than in the center of the
more massive clump A (see Fig.~\ref{IsoDblue}). The comparison of the
observed distributions of blue and emission line objects to the
undergoing merging scenario presented previously ($\pm$ 0.3 Gyr)
suggests that the interaction with the ICM during the passage of the
sub-cluster B on the edge of the cluster A may have triggered
star-bursting. This hypothesis is supported by the significant
difference in the radial velocity dispersion of emission-line and
passive galaxies, suggesting that emission-line objects are a
dynamically younger population than the general cluster members. In
contrast, the k+a/k+a?  population shows a signature of an older star
formation activity which can hardly be related to the on-going merger,
but may be understood either as previously infalling galaxies, or
relics from another older merging process, but now at rest within the
main cluster, as shown from the velocity distribution.

Ongoing multi-wavelength (optical, X-ray and radio) follow-up will
allow us to study in detail the environment and star formation
properties of this cluster.

\begin{acknowledgements}   
  We thank the anonymous referee for useful suggestions that helped to
  improve this paper. We warmly thank all our collaborators working on
  the X-ray analysis of A3921: Elena Belsole, Jean-Luc Sauvageot,
  Gabriel W. Pratt and Herv\'e Bourdin. We thank B. Vandame for
  providing us his imaging reduction package "alambic" as well as his
  support to use it and G. Mars for his contribution to the
  reduction. We thank Albert Bijaoui, Andrea Biviano and Philippe
  Prugniel for fruitful discussions and helpful comments. We thank
  Christophe Adami for providing spectra of ENACS galaxies. This
  research was supported in part by Marie Curie individual fellowships
  MEIF-CT-2003-900773 (CF), by the PNC (``Programme National de
  Cosmologie'') grants in 2002 and 2003, and by the ``Conseil
  Scientifique'' of the ``Observatoire de la C\^ote d'Azur''. CF
  thanks Fran\c{c}ois Mignard and the CERGA department of the
  ``Observatoire de la C\^ote d'Azur'' for having financially
  supported part of this work. This research has made use of the
  NASA/IPAC Extragalactic Database (NED) which is operated by the Jet
  Propulsion Laboratory, California Institute of Technology under
  contract with the National Aeronautics and Space Administration.
\end{acknowledgements}

\appendix

\section{Red sequence identification}\label{RS}

The technique adopted to isolate the red sequence of the elliptical galaxies
of A3921 and calculate its parameters (slope, intercept and width) is
described in this Appendix.

The CMD of the central field of A3921 ($32{\times}32~{\rm arcmin}^2$) is shown
in Fig.~\ref{cmd}. Due to the high asymmetry of the galaxy distribution and
the heavy contamination of field objects in our CMD, a simple linear
regression plus an iterated 3$\sigma$ clip (Gladders \etal1998) does not give
satisfying results.  We thus use the median absolute deviation (MAD) as
scale indicator of the object distribution on our CMD, as it is extremely
efficient ($\sim$90\%) in the case of heavy-tailed distributions (Beers
\etal1990).  In fact, we expect that, on the CMD, the deviations of the red
sequence galaxies from the optimal linear fit follow a Gaussian
distribution, while deviations of field objects do not and populate
the tails of the distributions instead.  We moreover apply an
asymmetric clip (${n}_{i}{\cdot}{S}_{\rm MAD}$) on the two sides of
the distribution
(${n}_{1}-{n}_{2}=\sqrt[3]{skewness}/\sqrt{variance}$).  We then
follow a procedure similar to that described in Gladders {\it et al.},
iterating our linear fitting and clipping until convergence on a final
solution was obtained.  The red sequence can be described by the
linear equation $(R-I)_{\rm AB}=-0.0071I_{\rm AB}+0.5531$ with a width
of $\sigma_{\rm RS}$=0.0837 (see Fig.~\ref{cmd}).

\begin{table*}[hbp]  
\begin{center}  
\caption{\small {\rm Spectroscopic data}}  
\label {TabTOT}  
\begin{tabular}{llllllccccccc}
\hline  
\hline
\multicolumn{13}{c}{ }\\
\# &
R.A. &  
DEC. &  
$V_{\rm ab}$ &
$R_{\rm ab}$ &
$I_{\rm ab}$ &
$cz$ &  
$\Delta$$cz$ &
Run &
flag &  
EW([OII]) &
EW(${\rm H}_{\delta}$) &
Type \\  
 & (J2000) & (J2000) & & & & (km/s) & (km/s) & & & (\AA) & (\AA) & \\
\multicolumn{13}{c}{ }\\
\hline
\multicolumn{13}{c}{ }\\
1 & 22:50:19.36 & -64:25:50.7 & 17.95 & 17.52 & 17.06 & 28308.1 & 44.9 & 2001 & 1 & 0.0 & 1.8 & k \\
2 & 22:50:16.39 & -64:25:29.9 & 20.98 & 21.02 & 20.70 & -2.0 & -2.0 & 2001 & 1 & - & - & - \\
3 & 22:50:12.08 & -64:24:54.0 & 17.36 & 16.91 & 16.44 & 26858.1 & 38.1 & 2001 & 1 & 0.0 & 2.2 & k \\
4 & 22:50:10.27 & -64:24:20.8 & 18.73 & 18.04 & 17.47 & -2.0 & -2.0 & 2001 & 1 & - & - & - \\
5 & 22:50:07.64 & -64:24:28.5 & 18.86 & 18.68 & 18.39 & 27370.6 & 66.6 & 2001 & 1 & 0.0 & 7.3 & k+a \\
6 & 22:50:01.78 & -64:26:45.0 & 17.33 & 16.91 & 16.47 & 30339.4 & 36.2 & 2001 & 1 & 0.0 & 1.1 & k \\
7 & 22:49:57.52 & -64:24:44.3 & 17.71 & 17.25 & 16.78 & 27972.0 & 45.8 & 2001 & 1 & 0.0 & 2.4 & k \\
8 & 22:49:54.76 & -64:25:14.9 & 17.79 & 17.34 & 16.85 & 28023.1 & 34.7 & 2001 & 1 & 0.0 & 0.0 & k \\
9 & 22:49:52.03 & -64:26:02.6 & 19.48 & 19.14 & 18.49 & 28767.8 & 69.4 & 2001 & 1 & 0.0 & 3.9 & k+a \\
10 & 22:49:47.61 & -64:26:01.9 & 18.59 & 18.17 & 17.77 & 28129.0 & 45.3 & 2001 & 1 & 0.0 & 0.9 & k \\
11 & 22:49:42.85 & -64:25:05.9 & 19.62 & 19.55 & 19.75 & -1.0 & -1.0 & 2001 & 3 & - & - & - \\
12 & 22:49:16.07 & -64:22:48.4 & 19.19 & 19.00 & 18.69 & 27596.1 & 67.6 & 2001 & 1 & 0.0 & 4.7 & k+a \\
13 & 22:49:13.03 & -64:20:37.2 & 20.42 & 19.99 & 19.78 & 28183.9 & 74.7 & 2001 & 2 & - & - & - \\
14 & 22:49:09.93 & -64:20:06.8 & 20.37 & 20.10 & 19.81 & 26489.7 & 88.1 & 2001 & 2 & - & - & - \\
15 & 22:49:06.62 & -64:21:37.3 & 17.24 & 16.90 & 16.74 & -2.0 & -2.0 & 2001 & 1 & - & - & - \\
16 & 22:49:04.91 & -64:18:59.3 & 16.64 & 15.81 & 14.77 & -2.0 & -2.0 & 2001 & 1 & - & - & - \\
17 & 22:48:58.98 & -64:21:11.8 & 18.76 & 18.64 & 18.48 & 28141.7 & 98.3 & 2001 & 1 & -23.6 & 3.3 & e(a) \\
18 & 22:48:56.02 & -64:20:13.8 & 20.65 & 20.35 & 20.46 & 27913.5 & 87.5 & 2001 & 2 & - & - & - \\
19 & 22:48:53.06 & -64:19:20.6 & 19.98 & 18.95 & 18.09 & 26110.5 & 89.2 & 2001 & 2 & - & - & - \\
20 & 22:48:45.65 & -64:21:24.5 & 18.03 & 17.58 & 17.26 & 28714.0 & 51.8 & 2001 & 1 & 0.0 & 3.5 & k+a \\
21 & 22:48:42.86 & -64:21:17.9 & 16.88 & 16.44 & 16.02 & 27804.7 & 31.7 & 2001 & 1 & 0.0 & 0.9 & k \\
22 & 22:48:37.55 & -64:20:14.2 & 20.26 & 19.87 & 19.64 & 25078.5 & 110.0 & 2001 & 2 & - & - & - \\
23 & 22:48:35.94 & -64:20:12.3 & 18.53 & 18.08 & 17.65 & 27781.7 & 80.6 & 2001 & 1 & 0.0 & 0.0 & k \\
24 & 22:48:33.17 & -64:20:50.4 & 19.29 & 18.65 & 18.19 & 25001.0 & 110.0 & 2001 & 2 & - & - & - \\
25 & 22:49:45.36 & -64:25:14.4 & 17.92 & 17.47 & 16.99 & 27572.9 & 52.2 & 2001 & 1 & 0.0 & 0.9 & k \\
26 & 22:49:40.88 & -64:26:34.5 & 19.45 & 19.06 & 18.61 & 27548.2 & 56.3 & 2001 & 1 & 0.0 & 0.0 & k \\
27 & 22:49:38.06 & -64:26:11.8 & 17.69 & 17.24 & 16.75 & 28607.6 & 50.8 & 2001 & 1 & 0.0 & 2.3 & k \\
28 & 22:49:35.56 & -64:26:07.2 & 16.86 & 16.49 & 16.12 & 28667.0 & 30.9 & 2001 & 1 & 0.0 & 1.9 & k \\
29 & 22:49:28.71 & -64:27:20.7 & 16.27 & 16.12 & 15.98 & -2.0 & -2.0 & 2001 & 1 & - & - & - \\
30 & 22:49:27.39 & -64:24:33.2 & 18.19 & 18.06 & 17.99 & -2.0 & -2.0 & 2001 & 1 & - & - & - \\
31 & 22:49:24.24 & -64:26:15.2 & 19.84 & 19.45 & 19.10 & 29559.6 & 84.0 & 2001 & 1 & 0.0 & 0.0 & k \\
32 & 22:49:22.53 & -64:23:42.0 & 18.67 & 17.98 & 17.24 & 24970.6 & 38.1 & 2001 & 2 & - & - & - \\
33 & 22:49:18.94 & -64:26:09.7 & 20.38 & 20.12 & 19.97 & 28585.7 & 132.6 & 2001 & 2 & - & - & - \\
34 & 22:49:15.97 & -64:25:30.8 & 19.59 & 18.72 & 17.33 & 20086.4 & 97.4 & 2001 & 2 & - & - & - \\
35 & 22:49:12.36 & -64:26:54.6 & 17.10 & 16.46 & 15.95 & -2.0 & -2.0 & 2001 & 1 & - & - & - \\
36 & 22:49:07.62 & -64:26:16.7 & 17.35 & 16.92 & 16.43 & 27997.9 & 32.7 & 2001 & 1 & 0.0 & 2.8 & k \\
37 & 22:49:00.85 & -64:26:05.5 & 19.05 & 18.68 & 18.28 & 28997.1 & 100.8 & 2001 & 1 & 0.0 & 0.0 & k \\
38 & 22:50:13.53 & -64:24:41.5 & 19.76 & 19.33 & 18.88 & 28876.6 & 98.4 & 2001 & 1 & - & - & - \\
39 & 22:50:06.48 & -64:24:41.9 & 16.37 & 15.92 & 15.48 & 28038.9 & 35.8 & 2001 & 1 & 0.0 & 0.0 & k \\
40 & 22:50:04.72 & -64:26:36.3 & 17.73 & 17.38 & 17.11 & -2.0 & -2.0 & 2001 & 1 & - & - & - \\
41 & 22:49:58.18 & -64:25:48.1 & 14.95 & 14.46 & 13.97 & 28166.4 & 64.2 & 2001 & 1 & 0.0 & 1.5 & k \\
42 & 22:49:53.28 & -64:25:03.5 & 18.01 & 17.54 & 17.06 & 28483.6 & 48.1 & 2001 & 1 & 0.0 & 1.6 & k \\
43 & 22:49:48.42 & -64:27:08.5 & 19.52 & 19.14 & 18.79 & 26969.9 & 76.4 & 2001 & 1 & 0.0 & 2.5 & k+a? \\
44 & 22:49:45.45 & -64:26:40.1 & 20.16 & 20.15 & 20.30 & 30435.3 & 173.2 & 2001 & 2 & - & - & - \\
45 & 22:49:41.46 & -64:26:24.1 & 19.68 & 19.31 & 18.96 & 25576.3 & 73.1 & 2001 & 1 & -11.4 & 1.8 & e(c) \\
46 & 22:49:38.74 & -64:25:11.8 & 20.25 & 19.79 & 19.39 & -1.0 & -1.0 & 2001 & 3 & - & - & - \\
47 & 22:49:36.32 & -64:26:38.5 & 16.37 & 15.88 & 15.39 & 28061.1 & 46.8 & 2001 & 1 & 0.0 & 0.0 & k \\
48 & 22:49:32.62 & -64:26:13.9 & 19.70 & 19.26 & 18.78 & 27953.0 & 55.4 & 2001 & 1 & 0.0 & 4.5 & k+a \\
49 & 22:49:12.28 & -64:21:47.2 & 17.62 & 17.48 & 17.45 & -2.0 & -2.0 & 2001 & 1 & - & - & - \\
50 & 22:49:09.53 & -64:19:18.7 & 18.84 & 18.36 & 17.86 & 40663.7 & 107.4 & 2001 & 1 & - & - & - \\
\multicolumn{13}{c}{ }\\
\hline  
\end{tabular}  
\label{MAINTAB}
\end{center}  
\end{table*}

\begin{table*}[hbp]  
\begin{center}  
\caption{\small {\rm Spectroscopic data}}  
\begin{tabular}{llllllccccccc}
\hline  
\hline
\multicolumn{13}{c}{ }\\
\# &
R.A. &  
DEC. &  
$V_{\rm ab}$ &
$R_{\rm ab}$ &
$I_{\rm ab}$ &
$cz$ &  
$\Delta$$cz$ &
Run &
flag &  
EW([OII]) &
EW(${\rm H}_{\delta}$) &
Type \\  
 & (J2000) & (J2000) & & & & (km/s) & (km/s) & & & (\AA) & (\AA) & \\
\multicolumn{13}{c}{ }\\
\hline
\multicolumn{13}{c}{ }\\
51 & 22:49:07.26 & -64:20:56.9 & 16.62 & 15.78 & 14.71 & -2.0 & -2.0 & 2001 & 1 & - & - & - \\
52 & 22:49:04.78 & -64:20:35.6 & 15.41 & 14.91 & 14.49 & 27888.4 & 38.5 & 2001 & 1 & 0.0 & 0.0 & k \\
53 & 22:48:59.92 & -64:20:50.4 & 16.48 & 16.05 & 15.56 & 27295.5 & 42.3 & 2001 & 1 & 0.0 & 0.0 & k \\
54 & 22:48:54.08 & -64:21:21.9 & 17.57 & 17.24 & 17.07 & -2.0 & -2.0 & 2001 & 1 & - & - & - \\
55 & 22:48:50.71 & -64:21:34.4 & 21.60 & 21.21 & 21.35 & 36279.8 & 113.3 & 2001 & 2 & - & - & - \\
56 & 22:48:47.89 & -64:20:34.5 & 19.21 & 18.79 & 18.39 & 27545.7 & 67.1 & 2001 & 1 & 0.0 & 0.0 & k \\
57 & 22:48:39.66 & -64:19:22.3 & 17.00 & 16.63 & 16.28 & 27151.0 & 52.5 & 2001 & 1 & -4.1 & 2.4 & e(c) \\
58 & 22:48:36.07 & -64:22:38.4 & 17.61 & 17.20 & 16.73 & 27846.0 & 57.5 & 2001 & 1 & 0.0 & 2.7 & k \\
59 & 22:48:35.42 & -64:20:06.1 & 20.26 & 19.61 & 19.08 & 27878.7 & 94.4 & 2001 & 2 & - & - & - \\
60 & 22:48:31.10 & -64:19:51.9 & 20.67 & 19.77 & 19.23 & 27079.5 & 92.0 & 2001 & 2 & - & - & - \\
61 & 22:48:29.29 & -64:20:58.6 & 19.05 & 18.53 & 18.08 & 76835.2 & 90.7 & 2001 & 2 & - & - & - \\
62 & 22:48:26.20 & -64:20:50.1 & 17.60 & 17.51 & 17.53 & -2.0 & -2.0 & 2001 & 1 & - & - & - \\
63 & 22:49:05.15 & -64:25:03.5 & 16.61 & 16.38 & 16.19 & -2.0 & -2.0 & 2001 & 1 & - & - & - \\
64 & 22:49:01.47 & -64:25:04.0 & 18.06 & 17.63 & 17.12 & 28560.4 & 44.7 & 2001 & 1 & 0.0 & 0.0 & k \\
65 & 22:49:00.10 & -64:23:41.3 & 18.79 & 18.41 & 18.04 & 27953.0 & 58.7 & 2001 & 1 & 0.0 & 0.0 & k \\
66 & 22:48:57.61 & -64:23:59.4 & 19.54 & 18.62 & 17.91 & 32591.3 & 124.1 & 2001 & 2 & - & - & - \\
67 & 22:48:53.47 & -64:23:09.1 & 20.25 & 19.77 & 19.32 & 37746.2 & 58.5 & 2001 & 2 & - & - & - \\
68 & 22:48:51.13 & -64:23:10.1 & 20.50 & 20.34 & 19.86 & -1.0 & -1.0 & 2001 & 3 & - & - & - \\
69 & 22:48:49.07 & -64:23:12.0 & 16.21 & 15.78 & 15.33 & 29059.5 & 81.9 & 2001 & 1 & -62.7 & 0.0 & e(b) \\
70 & 22:48:44.80 & -64:22:50.2 & 17.54 & 17.17 & 16.74 & 28186.1 & 38.7 & 2001 & 1 & 0.0 & 2.5 & k \\
71 & 22:48:43.36 & -64:23:53.1 & 18.60 & 18.30 & 17.99 & 28837.7 & 101.2 & 2001 & 1 & 0.0 & 2.1 & k \\
72 & 22:48:38.08 & -64:24:14.1 & 19.84 & 19.78 & 19.78 & 83738.0 & 112.5 & 2001 & 1 & - & - & - \\
73 & 22:48:34.84 & -64:23:39.9 & 17.95 & 17.66 & 17.26 & 28505.5 & 72.0 & 2001 & 1 & -14.2 & 2.2 & e(c) \\
74 & 22:48:29.27 & -64:24:13.3 & 21.13 & 20.26 & 18.64 & 83383.1 & 120.4 & 2001 & 2 & - & - & - \\
75 & 22:48:24.27 & -64:23:42.9 & 19.46 & 18.65 & 17.49 & -2.0 & -2.0 & 2001 & 2 & - & - & - \\
76 & 22:48:22.47 & -64:21:26.7 & 16.68 & 16.49 & 16.41 & -2.0 & -2.0 & 2001 & 1 & - & - & - \\
77 & 22:48:20.13 & -64:21:58.2 & 19.39 & 18.97 & 18.65 & -1.0 & -1.0 & 2001 & 3 & - & - & - \\
78 & 22:49:48.73 & -64:21:47.2 & 15.96 & 15.71 & 15.58 & -2.0 & -2.0 & 2001 & 1 & - & - & - \\
79 & 22:49:46.31 & -64:25:12.0 & 19.15 & 18.72 & 18.30 & 27255.4 & 44.1 & 2001 & 1 & 0.0 & 0.0 & k \\
80 & 22:49:44.51 & -64:22:15.0 & 17.50 & 16.82 & 16.22 & -2.0 & -2.0 & 2001 & 1 & - & - & - \\
81 & 22:49:41.10 & -64:24:05.6 & 20.57 & 20.45 & 20.36 & 28957.4 & 46.4 & 2001 & 1 & -21.4 & 9.1 & e(a) \\
82 & 22:49:38.41 & -64:23:23.8 & 19.00 & 18.88 & 18.75 & 29659.9 & 106.8 & 2001 & 1 & -16.8 & 4.9 & e(a) \\
83 & 22:49:35.31 & -64:23:00.1 & 17.91 & 17.53 & 17.09 & 26168.7 & 34.1 & 2001 & 1 & 0.0 & 2.9 & k+a? \\
84 & 22:49:36.63 & -64:25:30.5 & 19.42 & 19.02 & 18.62 & 29173.6 & 45.9 & 2001 & 1 & 0.0 & 1.6 & k \\
85 & 22:49:32.01 & -64:21:44.1 & 19.77 & 19.41 & 19.01 & 25534.2 & 64.5 & 2001 & 2 & - & - & - \\
86 & 22:49:28.71 & -64:25:54.6 & 21.02 & 20.62 & 20.35 & 25415.1 & 92.7 & 2001 & 2 & - & - & - \\
87 & 22:49:24.15 & -64:23:01.5 & 21.63 & 21.66 & 21.49 & 15171.0 & 72.3 & 2001 & 2 & - & - & - \\
88 & 22:49:19.41 & -64:22:11.2 & 19.52 & 19.46 & 19.11 & 22355.2 & 90.2 & 2001 & 2 & - & - & - \\
89 & 22:49:17.39 & -64:22:56.0 & 20.67 & 20.49 & 20.19 & 25121.8 & 85.7 & 2001 & 2 & - & - & - \\
90 & 22:49:15.13 & -64:24:53.2 & 18.49 & 18.18 & 17.93 & -2.0 & -2.0 & 2001 & 1 & - & - & - \\
91 & 22:49:07.81 & -64:21:54.4 & 20.69 & 20.44 & 20.08 & -1.0 & -1.0 & 2001 & 3 & - & - & - \\
92 & 22:49:04.37 & -64:23:58.3 & 20.17 & 19.74 & 19.56 & 13836.7 & 98.9 & 2001 & 2 & - & - & - \\
93 & 22:50:00.10 & -64:18:43.8 & 19.88 & 19.44 & 19.18 & -1.0 & -1.0 & 2002 & 3 & - & - & - \\
94 & 22:49:58.14 & -64:20:05.8 & 18.92 & 18.50 & 18.11 & 30165.3 & 46.4 & 2002 & 1 & 0.0 & 0.0 & k \\
95 & 22:49:54.60 & -64:19:34.9 & 22.48 & 22.29 & 21.43 & -1.0 & -1.0 & 2002 & 3 & - & - & - \\
96 & 22:49:50.74 & -64:21:05.2 & 19.84 & 19.31 & 18.91 & 32423.2 & 140.8 & 2002 & 2 & - & - & - \\
97 & 22:49:47.82 & -64:19:04.4 & 15.20 & 15.12 & 15.17 & -2.0 & -2.0 & 2002 & 1 & - & - & - \\
98 & 22:49:46.01 & -64:18:14.1 & 17.60 & 17.22 & 16.90 & 27357.3 & 40.3 & 2002 & 1 & 0.0 & 2.9 & k+a? \\
99 & 22:49:43.49 & -64:20:07.5 & 20.00 & 19.60 & 19.23 & 77114.8 & 136.3 & 2002 & 1 & - & - & - \\
\multicolumn{13}{c}{ }\\
\hline  
\end{tabular}  
\end{center}  
\end{table*}

\begin{table*}[hbp]  
\begin{center}  
\caption{\small {\rm Spectroscopic data}}    
\begin{tabular}{llllllccccccc}
\hline  
\hline
\multicolumn{13}{c}{ }\\
\# &
R.A. &  
DEC. &  
$V_{\rm ab}$ &
$R_{\rm ab}$ &
$I_{\rm ab}$ &
$cz$ &  
$\Delta$$cz$ &
Run &
flag &  
EW([OII]) &
EW(${\rm H}_{\delta}$) &
Type \\  
 & (J2000) & (J2000) & & & & (km/s) & (km/s) & & & (\AA) & (\AA) & \\
\multicolumn{13}{c}{ }\\
\hline
\multicolumn{13}{c}{ }\\
100 & 22:49:41.58 & -64:19:59.0 & 18.38 & 18.31 & 18.46 & 25634.9 & 83.4 & 2002 & 1 & -41.4 & 0.0 & e(b) \\
101 & 22:49:39.50 & -64:19:35.3 & 21.42 & 21.13 & 21.13 & -1.0 & -1.0 & 2002 & 3 & - & - & - \\
102 & 22:49:36.68 & -64:18:55.4 & 18.06 & 17.79 & 17.54 & 39792.0 & 82.8 & 2002 & 1 & - & - & - \\
103 & 22:49:33.83 & -64:20:14.1 & 20.76 & 20.11 & 19.74 & 136543.4 & 137.0 & 2002 & 2 & - & - & - \\
104 & 22:49:33.39 & -64:20:15.0 & 20.45 & 20.04 & 19.58 & -1.0 & -1.0 & 2002 & 3 & - & - & - \\
105 & 22:49:30.17 & -64:20:24.2 & 17.74 & 17.61 & 17.59 & -2.0 & -2.0 & 2002 & 1 & - & - & - \\
106 & 22:49:28.53 & -64:19:46.2 & 17.40 & 16.96 & 16.57 & 28093.8 & 36.1 & 2002 & 1 & 0.0 & 0.0 & k \\
107 & 22:49:25.29 & -64:19:26.3 & 17.61 & 17.38 & 17.29 & -2.0 & -2.0 & 2002 & 1 & - & - & - \\
108 & 22:49:22.48 & -64:19:03.6 & 18.45 & 17.96 & 17.56 & 40946.8 & 90.1 & 2002 & 1 & - & - & - \\
109 & 22:49:18.63 & -64:21:12.2 & 19.15 & 18.73 & 18.36 & 27983.8 & 48.1 & 2002 & 1 & 0.0 & 1.1 & k \\
110 & 22:49:17.40 & -64:19:55.1 & 21.95 & 20.84 & 20.02 & -1.0 & -1.0 & 2002 & 3 & - & - & - \\
111 & 22:49:13.99 & -64:19:04.8 & 21.77 & 21.11 & 20.83 & -1.0 & -1.0 & 2002 & 3 & - & - & - \\
112 & 22:51:06.72 & -64:25:12.9 & - & - & - & 33771.5 & 146.7 & 2002 & 1 & - & - & - \\
113 & 22:51:05.11 & -64:25:17.8 & 18.63 & 18.42 & 18.29 & -2.0 & -2.0 & 2002 & 1 & - & - & - \\
114 & 22:51:03.17 & -64:25:03.7 & 16.98 & 16.85 & 16.80 & -2.0 & -2.0 & 2002 & 1 & - & - & - \\
115 & 22:51:01.34 & -64:24:27.6 & 14.32 & 13.88 & 13.62 & -2.0 & -2.0 & 2002 & 1 & - & - & - \\
116 & 22:50:58.45 & -64:24:51.7 & 19.89 & 19.09 & 18.28 & -2.0 & -2.0 & 2002 & 1 & - & - & - \\
117 & 22:50:56.78 & -64:24:26.1 & 20.48 & 19.90 & 19.40 & 23657.6 & 79.5 & 2002 & 2 & - & - & - \\
118 & 22:50:54.75 & -64:26:00.5 & 18.31 & 18.18 & 18.13 & -2.0 & -2.0 & 2002 & 1 & - & - & - \\
119 & 22:50:52.16 & -64:25:13.1 & 19.15 & 18.79 & 18.43 & 29843.5 & 55.2 & 2002 & 1 & 0.0 & 2.6 & k+a? \\
120 & 22:50:50.85 & -64:25:20.8 & 17.86 & 17.46 & 17.04 & 28329.6 & 33.4 & 2002 & 1 & 0.0 & 2.6 & k \\
121 & 22:50:48.31 & -64:24:57.2 & 18.07 & 17.65 & 17.22 & 27611.2 & 37.8 & 2002 & 1 & 0.0 & 1.4 & k \\
122 & 22:50:45.03 & -64:24:25.0 & 16.95 & 16.48 & 16.03 & 27793.5 & 46.1 & 2002 & 1 & 0.0 & 0.0 & k \\
123 & 22:50:44.28 & -64:24:19.4 & 16.05 & 15.88 & 15.80 & -2.0 & -2.0 & 2002 & 1 & - & - & - \\
124 & 22:50:41.83 & -64:24:46.2 & 19.27 & 18.86 & 18.48 & 27303.1 & 76.7 & 2002 & 1 & 0.0 & 3.1 & k+a \\
125 & 22:50:39.80 & -64:23:31.1 & 19.24 & 19.01 & 18.79 & -2.0 & -2.0 & 2002 & 1 & - & - & - \\
126 & 22:50:36.92 & -64:24:08.4 & 17.26 & 17.18 & 17.17 & -2.0 & -2.0 & 2002 & 1 & - & - & - \\
127 & 22:50:33.55 & -64:22:39.2 & 18.09 & 17.68 & 17.24 & 33096.0 & 77.1 & 2002 & 2 & - & - & - \\
128 & 22:50:30.31 & -64:25:46.0 & 17.60 & 17.45 & 17.40 & -2.0 & -2.0 & 2002 & 1 & - & - & - \\
129 & 22:50:26.53 & -64:24:32.0 & 18.09 & 17.69 & 17.26 & 26487.6 & 52.2 & 2002 & 1 & 0.0 & 2.3 & k \\
130 & 22:50:24.77 & -64:24:39.7 & 18.35 & 17.97 & 17.59 & 26285.7 & 66.9 & 2002 & 1 & 0.0 & 2.6 & k \\
131 & 22:50:23.61 & -64:23:06.5 & 18.73 & 18.24 & 17.77 & 28336.1 & 85.2 & 2002 & 1 & - & - & - \\
132 & 22:50:21.57 & -64:26:43.2 & 19.75 & 19.00 & 18.08 & -2.0 & -2.0 & 2002 & 1 & - & - & - \\
133 & 22:50:19.49 & -64:26:28.1 & 18.37 & 17.94 & 17.50 & 28223.9 & 40.3 & 2002 & 1 & 0.0 & 0.0 & k \\
134 & 22:50:15.19 & -64:27:53.0 & 18.68 & 18.28 & 17.89 & 26234.3 & 57.7 & 2002 & 1 & 0.0 & 0.0 & k \\
135 & 22:50:14.46 & -64:26:06.5 & 16.43 & 16.23 & 16.10 & -2.0 & -2.0 & 2002 & 1 & - & - & - \\
136 & 22:50:07.83 & -64:26:02.1 & 19.69 & 19.23 & 18.76 & 27787.5 & 64.8 & 2002 & 1 & 0.0 & 2.3 & k+a? \\
137 & 22:50:05.87 & -64:27:20.2 & 15.21 & 15.09 & 15.05 & -2.0 & -2.0 & 2002 & 1 & - & - & - \\
138 & 22:50:02.53 & -64:27:54.7 & 21.08 & 20.77 & 20.62 & 32474.3 & 47.4 & 2002 & 2 & - & - & - \\
139 & 22:49:59.65 & -64:26:40.6 & 14.76 & 14.64 & 14.60 & -2.0 & -2.0 & 2002 & 1 & - & - & - \\
140 & 22:49:58.13 & -64:26:05.4 & 18.67 & 18.18 & 17.61 & 27843.1 & 79.3 & 2002 & 1 & 0.0 & 2.3 & k+a? \\
141 & 22:49:55.09 & -64:24:41.8 & 19.07 & 18.64 & 18.22 & 27381.9 & 58.1 & 2002 & 1 & 0.0 & 0.0 & k \\
142 & 22:49:51.65 & -64:26:00.0 & 19.42 & 19.04 & 18.39 & 27507.3 & 50.8 & 2002 & 1 & 0.0 & 2.6 & k \\
143 & 22:49:48.34 & -64:27:01.1 & 20.63 & 20.22 & 19.99 & 29371.5 & 88.4 & 2002 & 1 & 0.0 & 2.7 & k+a? \\
144 & 22:49:46.78 & -64:25:38.4 & 19.67 & 19.24 & 18.78 & 28997.3 & 86.8 & 2002 & 1 & 0.0 & 2.9 & k \\
145 & 22:49:43.23 & -64:25:16.6 & 19.82 & 19.71 & 19.78 & 50679.8 & 109.3 & 2002 & 2 & - & - & - \\
146 & 22:49:41.37 & -64:26:11.9 & 22.15 & 21.27 & 19.79 & -1.0 & -1.0 & 2002 & 3 & - & - & - \\
147 & 22:49:37.90 & -64:27:09.2 & 18.51 & 18.11 & 17.67 & 28941.8 & 54.6 & 2002 & 1 & 0.0 & 0.0 & k \\
148 & 22:49:34.88 & -64:25:53.2 & 21.64 & 20.70 & 19.19 & -1.0 & -1.0 & 2002 & 3 & - & - & - \\
149 & 22:49:24.16 & -64:20:14.7 & -0.01 & 22.36 & 21.30 & 24260.9 & 112.1 & 2002 & 2 & - & - & - \\
150 & 22:49:19.12 & -64:22:10.0 & -0.01 & 21.14 & 19.83 & 76517.1 & 97.8 & 2002 & 1 & - & - & - \\
\multicolumn{13}{c}{ }\\  
\hline  
\end{tabular}  
\end{center}  
\end{table*}

\begin{table*}[hbp]  
\begin{center}  
\caption{\small {\rm Spectroscopic data}}    
\begin{tabular}{llllllccccccc}
\hline  
\hline
\multicolumn{13}{c}{ }\\
\# &
R.A. &  
DEC. &  
$V_{\rm ab}$ &
$R_{\rm ab}$ &
$I_{\rm ab}$ &
$cz$ &  
$\Delta$$cz$ &
Run &
flag &  
EW([OII]) &
EW(${\rm H}_{\delta}$) &
Type \\  
 & (J2000) & (J2000) & & & & (km/s) & (km/s) & & & (\AA) & (\AA) & \\
\multicolumn{13}{c}{ }\\
\hline
\multicolumn{13}{c}{ }\\
151 & 22:49:18.82 & -64:22:06.2 & 20.52 & 20.23 & 19.80 & 76278.4 & 68.1 & 2002 & 1 & - & - & - \\
152 & 22:49:14.86 & -64:21:59.7 & 19.34 & 18.99 & 18.75 & 45803.5 & 56.2 & 2002 & 1 & - & - & - \\
153 & 22:49:12.26 & -64:19:41.7 & 16.44 & 15.57 & 14.31 & -2.0 & -2.0 & 2002 & 1 & - & - & - \\
154 & 22:49:10.31 & -64:18:30.4 & 17.63 & 17.15 & 16.67 & 28212.5 & 70.2 & 2002 & 1 & 0.0 & 0.0 & k \\
155 & 22:49:06.83 & -64:19:33.3 & 18.50 & 18.08 & 17.67 & 28019.9 & 65.0 & 2002 & 1 & 0.0 & 0.0 & k \\
156 & 22:49:04.07 & -64:19:59.1 & 18.46 & 17.99 & 17.52 & 28299.3 & 37.5 & 2002 & 1 & 0.0 & 0.0 & k \\
157 & 22:49:02.03 & -64:20:24.9 & 16.85 & 16.14 & 15.61 & -2.0 & -2.0 & 2002 & 1 & - & - & - \\
158 & 22:48:59.87 & -64:21:24.9 & 22.26 & 21.94 & 0.49 & -1.0 & -1.0 & 2002 & 3 & - & - & - \\
159 & 22:48:56.84 & -64:22:42.1 & 21.22 & 20.82 & 20.50 & 26253.3 & 82.2 & 2002 & 1 & - & - & - \\
160 & 22:48:54.48 & -64:22:38.3 & 20.82 & 20.52 & 20.28 & 20990.9 & 64.0 & 2002 & 2 & - & - & - \\
161 & 22:48:52.09 & -64:21:07.0 & 22.08 & 21.39 & 20.51 & 11581.2 & 88.8 & 2002 & 2 & - & - & - \\
162 & 22:48:49.32 & -64:21:49.9 & 16.94 & 16.72 & 16.63 & -2.0 & -2.0 & 2002 & 1 & - & - & - \\
163 & 22:48:50.72 & -64:19:31.8 & 12.81 & 12.81 & 12.52 & -2.0 & -2.0 & 2002 & 1 & - & - & - \\
164 & 22:48:47.80 & -64:19:57.6 & 21.88 & 20.93 & 19.55 & 25499.4 & 106.2 & 2002 & 2 & - & - & - \\
165 & 22:48:44.29 & -64:19:56.7 & 19.23 & 18.76 & 18.31 & 47378.9 & 80.4 & 2002 & 1 & - & - & - \\
166 & 22:48:40.06 & -64:20:55.8 & 18.15 & 18.11 & 18.14 & -1.0 & -1.0 & 2002 & 3 & - & - & - \\
167 & 22:48:39.63 & -64:21:25.3 & 19.68 & 18.97 & 18.47 & 35278.8 & 49.3 & 2002 & 1 & - & - & - \\
168 & 22:50:04.02 & -64:22:12.1 & 19.51 & 19.04 & 18.72 & 26820.6 & 112.9 & 2002 & 2 & - & - & - \\
169 & 22:50:01.83 & -64:22:21.8 & 18.26 & 17.94 & 17.66 & 25417.1 & 65.1 & 2002 & 1 & -6.5 & 6.0 & e(a) \\
170 & 22:50:00.12 & -64:23:04.7 & 16.99 & 16.78 & 16.67 & -2.0 & -2.0 & 2002 & 1 & - & - & - \\
171 & 22:49:58.23 & -64:23:23.8 & 20.60 & 19.74 & 18.72 & 39858.5 & 84.7 & 2002 & 2 & - & - & - \\
172 & 22:49:55.83 & -64:22:37.8 & 17.44 & 17.39 & 17.41 & -2.0 & -2.0 & 2002 & 1 & - & - & - \\
173 & 22:49:54.00 & -64:23:43.1 & 19.22 & 18.91 & 18.50 & 27002.7 & 57.0 & 2002 & 1 & 0.0 & 1.6 & k \\
174 & 22:49:47.53 & -64:21:24.4 & 19.58 & 19.43 & 19.34 & 31631.0 & 86.0 & 2002 & 2 & - & - & - \\
175 & 22:49:44.99 & -64:23:00.7 & 19.70 & 19.24 & 18.88 & 28404.3 & 78.6 & 2002 & 1 & 0.0 & 0.0 & k \\
176 & 22:49:42.54 & -64:23:36.0 & 23.03 & 21.94 & 20.06 & -1.0 & -1.0 & 2002 & 3 & - & - & - \\
177 & 22:49:38.28 & -64:23:37.1 & 20.12 & 19.95 & 19.86 & -2.0 & -2.0 & 2002 & 2 & - & - & - \\
178 & 22:49:34.50 & -64:24:15.4 & 19.68 & 19.27 & 18.85 & 28796.8 & 96.5 & 2002 & 1 & 0.0 & 1.6 & k \\
179 & 22:49:29.94 & -64:21:33.7 & 19.55 & 19.38 & 19.41 & -1.0 & -1.0 & 2002 & 3 & - & - & - \\
180 & 22:49:30.53 & -64:24:14.6 & 17.15 & 16.98 & 16.85 & 38735.5 & 86.2 & 2002 & 2 & - & - & - \\
181 & 22:49:26.08 & -64:23:21.4 & 19.15 & 18.99 & 18.78 & 28360.1 & 98.0 & 2002 & 1 & -11.7 & 1.8 & e(c) \\
182 & 22:49:24.12 & -64:25:18.6 & 17.84 & 17.14 & 16.48 & -2.0 & -2.0 & 2002 & 1 & - & - & - \\
183 & 22:49:22.39 & -64:23:05.4 & 19.08 & 18.95 & 18.87 & -2.0 & -2.0 & 2002 & 1 & - & - & - \\
184 & 22:49:18.45 & -64:25:04.0 & 20.62 & 20.53 & 20.54 & 25393.9 & 108.5 & 2002 & 2 & - & - & - \\
185 & 22:49:14.39 & -64:24:43.7 & 19.84 & 19.70 & 19.62 & 29093.6 & 102.2 & 2002 & 2 & - & - & - \\
186 & 22:49:11.77 & -64:25:18.0 & 17.84 & 17.45 & 16.97 & 29982.7 & 42.6 & 2002 & 1 & 0.0 & 0.0 & k \\
187 & 22:49:07.77 & -64:24:28.6 & 20.25 & 20.07 & 19.82 & -1.0 & -1.0 & 2002 & 3 & - & - & - \\
188 & 22:49:01.75 & -64:24:39.3 & 20.72 & 20.10 & 19.55 & -1.0 & -1.0 & 2002 & 3 & - & - & - \\
189 & 22:48:58.93 & -64:23:54.6 & 18.84 & 18.63 & 18.13 & 29194.9 & 92.6 & 2002 & 2 & - & - & - \\
190 & 22:48:52.22 & -64:25:16.6 & 18.77 & 18.19 & 17.72 & -2.0 & -2.0 & 2002 & 1 & - & - & - \\
191 & 22:48:50.13 & -64:24:21.5 & 19.62 & 19.31 & 18.84 & -1.0 & -1.0 & 2002 & 3 & - & - & - \\
192 & 22:48:47.44 & -64:22:33.0 & 20.61 & 20.12 & 19.77 & -1.0 & -1.0 & 2002 & 3 & - & - & - \\
193 & 22:48:45.58 & -64:24:17.8 & 20.18 & 19.88 & 19.55 & -1.0 & -1.0 & 2002 & 3 & - & - & - \\
194 & 22:48:43.82 & -64:24:35.9 & 18.29 & 17.83 & 17.36 & 34446.7 & 84.3 & 2002 & 2 & - & - & - \\
195 & 22:48:41.31 & -64:24:06.4 & 21.10 & 20.83 & 20.61 & -1.0 & -1.0 & 2002 & 3 & - & - & - \\
196 & 22:48:39.21 & -64:22:20.8 & 13.23 & 13.30 & 13.09 & -2.0 & -2.0 & 2002 & 1 & - & - & - \\
197 & 22:48:37.35 & -64:24:06.5 & 18.51 & 18.14 & 17.76 & 26677.0 & 70.2 & 2002 & 1 & 0.0 & 0.0 & k \\
198 & 22:48:34.45 & -64:24:58.4 & 19.94 & 19.60 & 19.31 & 27698.2 & 97.0 & 2002 & 1 & 0.0 & 0.0 & k \\
199 & 22:48:33.81 & -64:23:45.6 & 18.86 & 18.48 & 18.08 & 28454.0 & 98.3 & 2002 & 1 & 0.0 & 0.0 & k \\
\multicolumn{13}{c}{ }\\
\hline
\end{tabular}
\end{center}  
\end{table*}

\begin{table*}[hbp]  
\begin{center}  
\caption{\small {\rm Spectroscopic data}}    
\begin{tabular}{llllllccccccc}
\hline  
\hline
\multicolumn{13}{c}{ }\\
\# &
R.A. &  
DEC. &  
$V_{\rm ab}$ &
$R_{\rm ab}$ &
$I_{\rm ab}$ &
$cz$ &  
$\Delta$$cz$ &
Run &
flag &  
EW([OII]) &
EW(${\rm H}_{\delta}$) &
Type \\  
 & (J2000) & (J2000) & & & & (km/s) & (km/s) & & & (\AA) & (\AA) & \\
\multicolumn{13}{c}{ }\\
\hline
\multicolumn{13}{c}{ }\\
200 & 22:48:30.44 & -64:24:04.7 & 22.20 & 21.20 & 19.62 & -1.0 & -1.0 & 2002 & 3 & - & - & - \\
201 & 22:49:35.77 & -64:29:49.2 & 20.14 & 19.50 & 18.94 & 87248.7 & 141.3 & 2002 & 2 & - & - & - \\
202 & 22:49:34.06 & -64:28:14.1 & 21.94 & 21.18 & 19.76 & -1.0 & -1.0 & 2002 & 3 & - & - & - \\
203 & 22:49:31.64 & -64:28:27.5 & 19.63 & 19.14 & 18.66 & -1.0 & -1.0 & 2002 & 3 & - & - & - \\
204 & 22:49:29.66 & -64:28:31.9 & 18.72 & 18.58 & 18.37 & 40483.6 & 78.4 & 2002 & 2 & - & - & - \\
205 & 22:49:27.93 & -64:26:24.8 & 19.31 & 18.55 & 17.32 & 12633.3 & 60.3 & 2002 & 2 & - & - & - \\
206 & 22:49:24.90 & -64:27:36.8 & 19.89 & 19.50 & 19.13 & 27849.3 & 95.7 & 2002 & 2 & - & - & - \\
207 & 22:49:20.99 & -64:28:57.9 & 18.18 & 17.74 & 17.26 & 27776.3 & 37.0 & 2002 & 1 & 0.0 & 0.0 & k \\
208 & 22:49:19.29 & -64:27:13.7 & 19.36 & 19.08 & 18.65 & -1.0 & -1.0 & 2002 & 3 & - & - & - \\
209 & 22:49:17.23 & -64:28:56.6 & 17.09 & 16.35 & 15.58 & -2.0 & -2.0 & 2002 & 1 & - & - & - \\
210 & 22:49:14.03 & -64:29:42.8 & 18.29 & 17.88 & 17.42 & 26797.2 & 44.8 & 2002 & 1 & 0.0 & 2.3 & k \\
211 & 22:49:12.50 & -64:26:43.4 & 18.16 & 17.72 & 17.31 & 27003.7 & 68.7 & 2002 & 1 & 0.0 & 0.0 & k \\
212 & 22:49:09.29 & -64:30:56.8 & 16.07 & 15.73 & 15.41 & -2.0 & -2.0 & 2002 & 1 & - & - & - \\
213 & 22:49:07.63 & -64:28:09.7 & 17.54 & 16.93 & 16.37 & -2.0 & -2.0 & 2002 & 1 & - & - & - \\
214 & 22:49:04.75 & -64:26:17.7 & 20.76 & 19.99 & 19.25 & -1.0 & -1.0 & 2002 & 2 & - & - & - \\
215 & 22:49:02.68 & -64:27:43.5 & 19.59 & 19.64 & 19.90 & 19608.5 & 77.4 & 2002 & 2 & - & - & - \\
216 & 22:49:00.93 & -64:28:11.1 & 17.89 & 17.54 & 17.10 & 26430.9 & 59.0 & 2002 & 1 & 0.0 & 0.0 & k \\
217 & 22:48:58.19 & -64:27:50.7 & 19.28 & 18.88 & 18.19 & 18467.1 & 57.0 & 2002 & 2 & - & - & - \\
218 & 22:48:56.40 & -64:29:20.7 & 16.36 & 16.23 & 16.14 & -2.0 & -2.0 & 2002 & 1 & - & - & - \\
219 & 22:48:54.53 & -64:28:35.8 & 20.55 & 20.28 & 19.32 & -1.0 & -1.0 & 2002 & 3 & - & - & - \\
220 & 22:48:52.71 & -64:29:09.1 & 19.21 & 18.80 & 18.52 & 29057.8 & 62.7 & 2002 & 2 & - & - & - \\
221 & 22:48:51.24 & -64:29:08.7 & 17.49 & 17.07 & 16.72 & 28386.4 & 34.3 & 2002 & 1 & 0.0 & 1.8 & k \\
222 & 22:49:21.73 & -64:16:08.2 & 16.48 & 16.30 & 16.22 & -2.0 & -2.0 & 2003 & 1 & - & - & - \\
223 & 22:49:19.93 & -64:13:54.9 & 18.53 & 18.22 & 17.97 & 28380.3 & 86.0 & 2003 & 2 & - & - & - \\
224 & 22:49:17.16 & -64:15:40.8 & 14.80 & 14.64 & 14.62 & -2.0 & -2.0 & 2003 & 1 & - & - & - \\
225 & 22:49:14.54 & -64:16:08.0 & 19.83 & 19.78 & 19.81 & -2.0 & -2.0 & 2003 & 1 & - & - & - \\
226 & 22:49:11.96 & -64:16:03.9 & 19.27 & 19.12 & 19.00 & 29965.4 & 129.5 & 2003 & 1 & -23.1 & 3.1 & e(a) \\
227 & 22:49:08.89 & -64:15:38.1 & 18.40 & 18.10 & 17.79 & 40723.0 & 146.7 & 2003 & 1 & - & - & - \\
228 & 22:49:07.60 & -64:15:27.3 & 17.87 & 17.35 & 16.88 & 49673.1 & 62.7 & 2003 & 1 & - & - & - \\
229 & 22:49:04.62 & -64:14:31.9 & 18.73 & 17.86 & 16.46 & -2.0 & -2.0 & 2003 & 2 & - & - & - \\
230 & 22:49:02.23 & -64:16:39.5 & 20.27 & 19.97 & 19.67 & 40621.9 & 77.9 & 2003 & 1 & - & - & - \\
231 & 22:49:00.69 & -64:14:38.1 & 19.62 & 18.90 & 18.35 & 25789.2 & 81.1 & 2003 & 2 & - & - & - \\
232 & 22:48:58.22 & -64:14:44.3 & 17.30 & 17.13 & 17.07 & -2.0 & -2.0 & 2003 & 2 & - & - & - \\
233 & 22:48:55.85 & -64:13:34.2 & 14.28 & 14.13 & 14.13 & -2.0 & -2.0 & 2003 & 2 & - & - & - \\
234 & 22:48:55.80 & -64:15:21.8 & 14.11 & 13.78 & 13.57 & -2.0 & -2.0 & 2003 & 2 & - & - & - \\
235 & 22:48:50.54 & -64:14:03.7 & 17.65 & 17.40 & 17.21 & 26343.0 & 54.0 & 2003 & 2 & - & - & - \\
236 & 22:48:48.00 & -64:15:46.3 & 21.90 & 21.13 & 20.73 & 25494.6 & 80.5 & 2003 & 2 & - & - & - \\
237 & 22:48:45.29 & -64:15:46.0 & 17.92 & 17.53 & 17.16 & 28166.1 & 41.7 & 2003 & 1 & 0.0 & 1.6 & k \\
238 & 22:48:44.06 & -64:13:43.2 & 19.35 & 19.25 & 19.15 & 31409.4 & 58.5 & 2003 & 2 & - & - & - \\
239 & 22:48:39.53 & -64:16:22.1 & 20.96 & 20.30 & 19.75 & 33496.9 & 127.2 & 2003 & 2 & - & - & - \\
\multicolumn{13}{c}{ }\\
\hline
\end{tabular}
\end{center}  
\end{table*}


\begin{thebibliography}{}

\bibitem{} {Abraham, R.G., Snecker-Hane, T.A., Hutchings, J.B.,
    Carlberg, R.G., Yee, H.K.C., Ellingson, E., Morris, S., Oke, J.B., \&
    Rigler, M., 1996, ApJ, 471, 694}
 
\bibitem{} {Arnaud, M., Maurogordato, S., Slezak, E., \& Rho, J., 2000, 
A\&A,    355, 461}
 
\bibitem{} {Arnaud, M., Rothenflug, R., B\"ohringer, H., Neumann, D., \&
    Yamashita, K., 1996, uxsa.coll, 163}

\bibitem{} {Arnouts, S., Vandame, B., Benoist, C. \etal2001, A\&A, 
379, 740}
 
\bibitem{} {Ashman, K.M., Bird, C.M., \& Zepf, S., 1994, AJ, 108, 2348}
 
\bibitem{} {Balogh, M.L., Morris, S.L., Yee, H.K.C., Carlberg, R.G., \& 
Ellingson,
E., 1999, ApJ, 527, 54}

\bibitem{} {Balogh, M.L., Shade, D., Morris, S.L., Yee, H.K.C., Carlberg,
    R.G., \& Ellingson, E., 1998, ApJL, 504, 75}

\bibitem{} {Balogh, M.L., Morris, S.L., Yee, H.K.C., Carlberg, R.G.,
    \& Ellingson, E., 1997, ApJL, 488, 75}

\bibitem{} {Baldi, A., Bardelli, S., \& Zucca, E., 2001, MNRAS, 324, 509}

\bibitem{} {Barrena, R., Biviano, A., Ramella, M., Falco, E.E., \& 
Seitz, S.,
    2002, A\&A, 386, 861B}

\bibitem{} {Bekki, K., 1999, ApJL, 510, 15}

\bibitem{} {Belsole, E., Sauvageot, J.-L., Pratt, G.W., \& Bourdin, H., 
2004,
submitted to A\&A}

\bibitem{} {Beers, T. C., Gebhardt, K., Huchra, J.P., Forman, W., Jones, 
C.,
\& Bothun, G.D., 1992, ApJ, 400, 410}

\bibitem{} {Beers, T. C., Flynn, K., \& Gebhardt, K., 1990, AJ, 100, 32}

\bibitem{} {Bertin, E. \& Arnouts, S. 1996, A\&AS, 117, 393}

\bibitem{} {Bijaoui A., \& Ru\'e F. {\it A Vision Model adapted to the
astronomical images} {\sl Signal Processing} v.46 n.3 pp. 345-362
1995}

\bibitem{} {Bird, C. M., Davis, S. D., \& Beers, T. C., 1995, AJ, 109, 920}
 
\bibitem{} {Bird, C. M., 1994, AJ, 107, 1637}

\bibitem{} {Bird, C. M., \& Beers, T. C., 1993, AJ, 105, 1596}
 
\bibitem{} {Biviano, A., Katgert, P., Mazure, A., Moles, M., den Hartog, 
R., Perea, J., \& Focardi, P., 1997, A\&A, 321, 84}

\bibitem{} {Caldwell, N., Rose, J.A., Sharples, R.M., Ellis, R.S. \& Bower, R.G.,  1993, AJ, 106, 473}

\bibitem{} {Carlberg, R.G., Yee, H.K.C., Ellingson, E., Abraham, R., Gravel,
    P., Morris, S., \& Pritchet, C.J., 1996, ApJ, 462, 32}
 
\bibitem{} {Couch, W.J. \& Sharples, R.M., 1987, MNRAS, 229, 423}

\bibitem{} {Couch, W.J., Balogh, M.L., Bower, R.G., Smail, I. Glazebrook, K.
 \& Taylor, M., 2001, ApJ, 549, 820}

\bibitem{} {Davis, D.S., Bird, C.M., Mushotzky, R.F., \& Odewahn, S.C., 
1995,
    ApJ, 440, 48}

\bibitem{} {Demarco, R., 2004, PhD thesis, University of Paris 7, France}

\bibitem{} {den Hartog, R., \& Katgert, P, 1996, MNRAS, 279, 349}
 
\bibitem{} {Donnelly, R.H., Forman, W., Jones, C., Quintana, H., 
Ramirez, A., Churazov, E., \& Gilfanov, M., 2001, ApJ, 562, 254}

\bibitem{} {Dressler, A., \& Gunn, J.E., 1983, ApJ, 270, 7}

\bibitem{} {Dressler A., Smail, I., Poggianti, B.M., Butcher, H., Couch, 
W.J., Ellis, R.S., \& Oemler, A., 1999, ApJ, 122, 51}

\bibitem{} {Dressler A., \& Shectman, S., 1988, AJ, 95, 985}  

\bibitem{} {Dressler A., 1987, in Nearly Normal Galaxies: From the 
Planck Time to the Present, ed S. Faber (New York: Springer), 276}
 
\bibitem{} {Duc, P.A., Poggianti, B.M., Fadda, D., Elbaz, D., Flores, H.,
    Chanila, P., Franceschini, A., Moorwood, A., \& Cesarsky, C., 2002, A\&A,
    382, 60}

\bibitem{} {Durret, F., Forman, W., Gerbal, D., Jones, C., \& Vikhlinin, A.,
    1998, A\&A, 335, 41}

\bibitem{} {Eke V.R., Cole S., \& Frenk C.S., 1996, MNRAS 282, 263}

\bibitem{} {Ellingson, E., Lin, H., Yee, H.K.C., \& Carlberg, R.G., 2001, ApJ, 547, 609}

\bibitem{} {Evrard, A.E., 1991, MNRAS, 248, 8}
 
\bibitem{} {Fadda, D., Slezak, E. \& Bijaoui, A., 1998, A\&AS, 127, 335}

\bibitem{} {Fadda, D., Girardi, M., Giuricin, G., Mardirossian, F., \&
    Mezzetti, M., 1996, ApJ, 473, 670}

\bibitem{} {Ferrari, C., Maurogordato, S., Cappi, A., \& Benoist, C., 2003,
    A\&A, 399, 813}

\bibitem{} {Flores, R.A., Quintana, H., \& Way, M.J., 2000, ApJ, 532, 206}  

\bibitem{} {Franx, M., 1993, PASP, 105, 1058}

\bibitem{} {Fujita, Y., Takizawa, M., Nagashima, M., \& Enoki, M., 1999, 
PASJ,
    51, L1}
 
\bibitem{} {Gavazzi, G., Cortese, L., Boselli, A., Iglesias-Paramo, J.,
    V\'ilchez, J.M., \& Carrasco, L., 2003, ApJ, 597, 210}

\bibitem{} {Girardi, M., Giuricin, G., Mardirossian, F., Mezzetti,
    M., \& Boschin, W., 1998, ApJ, 505, 74}


\bibitem{} {Gladders, M. D., Lopez-Cruz, O.,
 Yee, H. K. C., \& Kodama, T., 1998, ApJ, 501, 571}

\bibitem{} {Gregory, S.A., \& Thompson, L.A., 1984, ApJ, 286, 422}
 
\bibitem{} {Hartigan, J. A., \& Hartigan, P. M., 1985, The Annals of
Statistics, Vol.13, No. 1, 70-84}

\bibitem{} {Hoel, P.G., 1971, Introduction to Mathematical Statistics,
New York: J. Wiley \& Sons}
 
\bibitem{} {Jones, C., \& Forman, W., 1992, Clusters and Superclusters of
    Galaxies, edited by A.C. Fabian, (Kluwer, Dordrecht), pp. 49-70}

\bibitem{} {Katgert, P., Biviano, A., \& Mazure, A., 2004, ApJ, 600, 657}

\bibitem{} {Katgert, P., Mazure, A., den Hartog, R., Adami, C., Biviano, 
A., \& Perea, J., 1998, A\&AS, 129, 399}

\bibitem{} {Katgert, P., Mazure, A., Perea, J., den Hartog, R., Moles, 
M., \etal1996, A\&A, 310, 8}

\bibitem{} {Kauffmann, G., 1995, MNRAS, 274, 153}

\bibitem{} {Kauffmann, G., 1995, MNRAS, 274, 161}

\bibitem{} {Kennicutt, R.C. Jr, 1992, ApJ, 388, 310}

\bibitem{} {Landolt, A.U. 1992, AJ, 104, 340}

\bibitem{} {Lemonon, L., Pierre, M., Hunstead, R., Reid, A., Mellier, Y., \&
    B\"ohringer, H., 1997, A\&A, 326, 34}

\bibitem{} {Limber, N.D., \& Mathews, W.G., 1960, ApJ, 132, 286}

\bibitem{} {Markevitch, M., Ponman, T.J., Nulsen, P.E.J., Bautz,
    M.W., Burke, D.J., David, L.P., Davis, D., Donnelly, R.H., Forman, W.R.,
    Jones, C., Kaastra, J., Kellogg, E., Kim, D.W., Kolodziejczak, J.,
    Mazzotta, P., Pagliaro, A., Patel, S., Van Speybroeck, L., 
Vikhlinin, A.,
    Vrtilek, J., Wise, M., \& Zhao, P., 2000, ApJ, 541, 542}

\bibitem{} {Mazure, A., Katgert, P., den Hartog, R., Biviano, A., 
Dubath, P.,
    Escalera, E., Focardi, P., Gerbal, D., Giuricin, G., Jones, B., Le
    F\`evre, O., Moles, M., Perea, J., \& Rhee, G., 1996, A\&A, 310, 31}
 
\bibitem{} {McLachlan, G. J., Basford, K. E., 1988, Mixture Models (Marcel
    Dekker, New York)}
 
\bibitem{} {Moss, C., \& Whittle, M., 2000, MNRAS, 317, 667}

\bibitem{} {Muriel, H., Quintana, H., Infante, L., Lambas, D.G., \& Way, 
M.J.,
    2002, AJ, 124, 1934}
 
\bibitem{} {Navarro, J.F., Frenk, C.S., \& White, S.D.M., 1997, ApJ, 
490, 493}

\bibitem{} {Newberry, M.V., Boronson, T.A., \& Kirshner, R.P., 1990, ApJ, 350, 585}

\bibitem{} {Nikogossyan, E., Durret, F., Gerbal, D., \& Magnard, F., 1999,
    A\&A, 349, 97}

\bibitem{} {Poggianti, B.M., Bridges, T.J., Komiyama, Y., Yagi, M., 
Carter, D.,
Mobasher, B., Okamura, S., \& Kashikawa, N., 2004, ApJ, 601, 197}

\bibitem{} {Poggianti, B.M., Smail, I., Dressler, A., Couch, W.J., Barger,
    A.J., Butcher, H., Ellis, R.S., \& Oemler, A., 1999, ApJ, 518, 576}

\bibitem{} {Postman, M., Lubin, L.M., \& Oke, J.B., 1998, AJ, 116, 560}

\bibitem{} {Press, W.H, Teukolsky, S.A., Vetterling, W.T., \& Flannery, 
B.P.,
    1992, Numerical Recipes, Cambridge: Cambridge University Press}

\bibitem{} {Quintana, H., Ram\'irez, A., \& Way, M.J., 1996, AJ, 112, 36}

\bibitem{} {Ricker, P.M., \& Sarazin, C.L., 2001, ApJ, 561, 621R}

\bibitem{} {Roettiger, K., Stone, J.M. \& Mushotzky. R.F., 1998, ApJ, 
493, 62}

\bibitem{} {Roettiger, K., Burns, J., \& Loken, C., 1993, ApJ, 407, L53}

\bibitem{} {Rose, J.A., 1985, AJ, 90, 1927}

\bibitem{} {Ru\'e F., \& Bijaoui A.: {\it A Multiscale Vision Model to
Analyze field astronomical images} {\sl Experimental Astronomy} 7
pp. 129-160 1997}

\bibitem{} {Sarazin, C.L., 2003, astro-ph/0301178}
 
\bibitem{} {Sauvageot, J.L., Belsole, E., Arnaud, M., \& Ponman, T.J., 
2001,
in Clusters of galaxies and the high redshift universe observed in X-ray,
Recent results of XMM-Newton and Chandra, XXXVIth Rencontres de
Moriond, XXIst Moriond Astrophysics Meeting, Edited by D. Neumann \&
J.T.T. Van}

\bibitem{} {Schindler, S., \& B\"ohringer, H., 1993, A\&A, 269, 83}

\bibitem{} {Schindler, S., \& M\"uller, E., 1993, A\&A, 272, 137}

\bibitem{} {Schlegel, D., Finkbeiner, D., \& Davis, M. 1998, ApJ, 500, 525}
 
\bibitem{} {Tomita, A., Nakamura, F.E., Takata, T., Nakanishi, K., Takeuchi,
    T., Ohta, K., \& Yamada, T., 1996, AJ, 111, 42}

\bibitem{} {Tonry, J., \& Davis, M., 1981, ApJ, 246, 666}

\bibitem{} {Tran, K.-V.H., Franx, M., Illingworth, G., Kelson, D.D., \& van
Dokkum, P., 2003, ApJ, 599, 865}


\bibitem{} {Tresse, L., \& Maddox, S.J., 1998, ApJ, 495, 691}

\bibitem{} {Wang, Q.D., Ulmer, M.P., \& Lavery, R.J., 1997, MNRAS, 288, 702}

\bibitem{} {West, M. J., Jones, C., \& Forman, W., 1995, ApJL, 451, 5}

\bibitem{} {West, M. J., \& Bothun, G. D., 1990, ApJ, 350, 36}

\bibitem{} {Yahil, A., \& Vidal, N., 1977, ApJ, 214, 347}

\end{thebibliography}
\end{document}